\documentclass[apj, letterpaper]{emulateapj}  % for apj-lookalike
\pdfoutput=1

\usepackage{amsmath}
\def\tt{}

\begin{document}

\slugcomment{\bf}
\slugcomment{ApJ, in press}

\title{Atmospheric Circulation of Brown Dwarfs and Jupiter and Saturn-like
  planets: Zonal Jets, Long-term Variability, and QBO-type
  oscillations}

\shorttitle{Atmospheric circulation of brown dwarfs}
\shortauthors{Showman et al.}

\author{Adam P. Showman\altaffilmark{1,2}, Xianyu Tan\altaffilmark{2},
Xi Zhang\altaffilmark{3}}

\altaffiltext{1}{Department of Atmospheric and Oceanic Sciences,
Peking University, Beijing, China}
\altaffiltext{2}{Department of Planetary Sciences and Lunar and
  Planetary Laboratory, The University of Arizona, 1629 University
  Blvd., Tucson, AZ 85721 USA; showman@lpl.arizona.edu}
\altaffiltext{3}{Department of Earth and Planetary Sciences, University
  of California, Santa Cruz, CA 95064, USA}
% Abstract 
\begin{abstract}
\label{Abstract}
Brown dwarfs and directly imaged giant planets exhibit significant
evidence for active atmospheric circulation, which induces a
large-scale patchiness in the cloud structure that evolves
significantly over time, as evidenced by infrared light curves and
Doppler maps.  These observations raise critical questions about the
fundamental nature of the circulation, its time variability, and the
overall relationship to the circulation on Jupiter and Saturn.
Jupiter and Saturn themselves exhibit numerous robust zonal
(east-west) jet streams at the cloud level; moreover, both planets
exhibit long-term stratospheric oscillations involving perturbations
of zonal wind and temperature that propagate downward over time on
timescales of $\sim$4 years (Jupiter) and $\sim$15 years
(Saturn). These oscillations, dubbed the Quasi Quadrennial Oscillation
(QQO) for Jupiter and the Semi-Annual Oscillation (SAO) on Saturn, are
thought to be analogous to the Quasi-Biennial Oscillation (QBO) on
Earth, which is driven by upward propagation of equatorial waves from
the troposphere.  To investigate these issues, we here present global,
three-dimensional, high-resolution numerical simulations of the flow
in the stratified atmosphere---overlying the convective interior---of
brown dwarfs and Jupiter-like planets.  The effect of interior
convection is parameterized by inducing small-scale, randomly varying
perturbations in the radiative-convective boundary at the base of the
model.  Radiative damping is represented using an idealized Newtonian
cooling scheme. In the simulations, the convective perturbations
generate atmospheric waves and turbulence that interact with the
rotation to produce numerous zonal jets. Moreover, the equatorial
stratosphere exhibits stacked eastward and westward jets that migrate
downward over time, exactly as occurs in the terrestrial QBO, Jovian
QQO, and Saturnian SAO. This is the first demonstration of a QBO-like
phenomenon in 3D numerical simulations of a giant planet.
\end{abstract}


%%%%%%%%%%%%%%%%%%%%%%%
% Begin document body %
%%%%%%%%%%%%%%%%%%%%%%%

\section{Introduction}
\label{Introduction}

A variety of evidence now indicates the existence of a vigorous
atmospheric circulation on brown dwarfs, which are fluid hydrogen
objects thought to form like stars but with insufficient mass to fuse
hydrogen, and which resemble hot, high-gravity versions of Jupiter in
many ways.  Infrared (IR) spectra indicate the presence of clouds and
chemical disequilibrium, both of which require vertical mixing (see
reviews by \citealt{helling-casewell-2014} and
\citealt{marley-robinson-2015}).  IR variability occurring on
rotational timescales implies that the cloud and temperature patterns
are commonly patchy on regional to global length scales
(e.g., \citealt{artigau-etal-2009}; \citealt{radigan-etal-2012}; 
\citealt{apai-etal-2013}; \citealt{wilson-etal-2014}; 
\citealt{metchev-etal-2015}; \citealt{buenzli-etal-2015};
  \citealt{yang-etal-2016}; \citealt{miles-paez-etal-2017},
\citealt{apai-etal-2017}; for reviews see \citealt{biller-2017} and
\citealt{artigau-2018}).  The shapes
of IR light curves often evolve significantly over several rotation
periods, implying that the spatial patterns of clouds and temperatures
change rapidly.  Doppler imaging maps allow the surface patchiness to
be explicitly resolved \citep{crossfield-etal-2014}, and detailed IR
spectral retrievals hold similar promise for multi-wavelength light
curve observations \citep{karalidi-etal-2016}.  Moreover, comparison
of observations over longer epochs now holds the promise of placing
constraints on the long-term evolution of the cloud structure and the
underlying dynamics.  The Spitzer Storms program (PI D. Apai), for
example, has monitored six brown dwarfs at systematic intervals of up
to a year \citep[e.g.,][]{apai-etal-2017}.

These observations provide an opportunity to study how atmospheric
dynamics behaves in the rapidly rotating, high-internal heat flux
regime applicable to brown dwarfs \citep[see][]{showman-kaspi-2013}.
Brown dwarfs typically receive no external stellar irradiation, and
therefore lack the large-scale (e.g., equator-to-pole or day-night)
contrasts in stellar heating that are responsible for driving the
global circulation on hot Jupiters or solar system planets like Earth.
However, the interior of brown dwarfs convect vigorously as they lose
heat to space, and this convection is expected to perturb the
overlying, stably stratified atmosphere, generating atmospheric waves
and, potentially, a large-scale atmospheric circulation that could
consist of turbulence, vortices, storms, and zonal (east-west) jet
streams.  The rapid rotation periods of brown dwarfs ($\sim$1--10
hours) implies that rotation should play a strong role in controlling
the atmospheric dynamics, more akin to the situation on Jupiter than
on the more slowly rotating hot Jupiters \citep{showman-kaspi-2013}.
But the interior heat flux from brown dwarfs of typically
$10^3$--$10^6\rm\,W\,m^{-2}$ greatly exceeds Jupiter's interior flux
of $\sim$$7.5\rm\,W\,m^{-2}$ \citep{li-etal-2018b}, suggesting that
the convection may be far more vigorous, and the greater atmospheric
temperatures of brown dwarfs relative to Jupiter imply that the
radiative time constants are far shorter.  As yet, the atmospheric
dynamics and behavior that occur in this regime are poorly understood.
Nevertheless, variability of some Y dwarfs with effective temperature
of only a few hundred Kelvins has been detected
(\citealt{cushing-etal-2016}; \citealt{esplin-etal-2016};
\citealt{leggett-etal-2016}; see also \citealt{skemer-etal-2016} and
\citealt{morley-etal-2018}). Atmospheric circulation of these
relatively cool objects may bridge the gap between that of most
observable T and L dwarfs on the one hand, and Jupiter and Saturn on
the other.

Jupiter and Saturn themselves exhibit atmospheric circulations
dominated by numerous zonal jet streams, including a broad, fast
eastward jet at the equator, and alternating eastward and westward jet
streams in the mid-to-high latitudes (for reviews, see {\tt
  \citealt{ingersoll-etal-2004}}, \citealt{vasavada-showman-2005},
{\tt \citealt{delgenio-etal-2009}}, \citealt{showman-etal-2019}, and
{\tt \citealt{sanchez-lavega-etal-2019}}).  Wind speeds are typically
$30\rm\,m\,s^{-1}$ on Jupiter and $100\rm\,m\,s^{-1}$ on Saturn in the
mid-to-high latitudes, but reach faster speeds in the equatorial
jet---approximately $100\rm\,m\,s^{-1}$ on Jupiter and
$400\rm\,m\,s^{-1}$ on Saturn.  The zonal jet structure is associated
with latitudinal temperature variations of $\sim$3--5$\rm\,K$, a
zonally banded cloud pattern, and a wealth of eddies, ranging from
coherent vortices like Jupiter's Great Red Spot to smaller, highly
time variable storms, vortices, and turbulence.  Additionally, both
planets exhibit oscillations in the stratospheric jet and temperature
structure at low latitudes, in which vertically stacked eastward and
westward jets---and associated temperature anomalies---slowly migrate
downward over time.  On Jupiter, this oscillation has a period of
$\sim$4 years and has been dubbed the Quasi-Quadrennial Oscillation or
QQO \citep{orton-etal-1991, leovy-etal-1991}, whereas on Saturn it has
a period of $\sim$15 years and is called the Saturn Semi-Annual
Oscillation or SAO (\citealt{orton-etal-2008};
\citealt{fouchet-etal-2008}; \citealt{guerlet-etal-2011};
\citealt{guerlet-etal-2018}; for a review see
\citealt{showman-etal-2019}).\footnote{Other notation has been adopted
  as well, particularly for the Saturnian oscillation.
  \citet{guerlet-etal-2018} adopt the term Saturn Equatorial
  Oscillation or SEO, while \citet{fletcher-etal-2017} adopt the more
  general phrase Saturnian Quasi-Periodic Oscillation (QPO).  For the
  Saturnian oscillation, we maintain consistency with earlier
  literature by using the phrase Saturnian SAO; for the general
  phenomenon regardless of planet or period, we adopt the phrase
  ``QBO-like oscillation,'' to emphasize the links to the dynamics of
  the QBO, which remains much better studied than any of the other
  oscillations.}  These oscillations are thought to be analogous to
the well-studied Quasi-Biennial Oscillation (QBO) on the Earth, which
is driven by the absorption in the stratosphere of upwardly
propagating, convectively generated waves from the troposphere, and
which exerts a variety of influences on global climate
\citep{baldwin-etal-2001}.

Only a few studies of the atmospheric circulation of brown dwarfs have
yet been performed.  \citet{freytag-etal-2010} presented
two-dimensional calculations of convection in a local box and its
interaction with an overlying stably stratified layer.  These models
generally ignored rotation.  \citet{showman-kaspi-2013} presented the
first global models of interior convection, demonstrating the
importance of rotation in the dynamics, and constructed a theory for
the characteristic wind speeds and horizontal temperature differences
in the stratified atmosphere.  \citet{zhang-showman-2014} performed
global calculations of the atmospheric flow using a ``one-and-a-half''
layer shallow-water model, in which an active atmospheric layer
overlies a deeper layer that represents the interior and was assumed to
be quiescent.  Convection was parameterized with a small-scale
forcing, and radiation with a simple damping scheme.  These
simulations showed that conditions of strong forcing and/or weak
radiative damping lead to a zonally banded pattern, while weak forcing
and/or strong damping lead to a pattern of horizontally isotropic
turbulence with no banding. \citet{tan-showman-2017} explored the
dynamical effect of latent heating associated with condensation of
silicates and iron in idealized 3D models, but did not include any
representation of the (dry) convection expected to occur throughout
the convection zone, which should exert significant effects on the
overlying radiatively stratified atmosphere.

By comparison, numerical simulations of the global circulation on
Jupiter and Saturn have a much longer history, although many aspects
remain poorly explored.  Such models have shown that small-scale
turbulence can interact with the planetary rotation to generate zonal
jets (for reviews, see for example \citealt{vasavada-showman-2005},
{\tt \citealt{dritschel-mcintyre-2008}, \citealt{delgenio-etal-2009},}
and \citealt{showman-etal-2019}).  This line of inquiry started with
two-dimensional (one-layer) models in which convection was
parameterized with small-scale sources of vorticity or mass randomly
injected into the layer \citep[e.g.,][]{williams-1978,
  nozawa-yoden-1997a, showman-2007, scott-polvani-2007}.  Thick-shell
spherical convection models show how interior convection can induce
the formation of zonal jets in the deep interior, although such models
usually do not include a representation of the overlying atmosphere
\citep[e.g.][]{christensen-2002, heimpel-etal-2005, kaspi-etal-2009,
  duarte-etal-2013}.  Three-dimensional models of the circulation in
the stratified atmosphere have focused on jet formation by baroclinic
instabilities associated with latitudinal variations in solar heating
\citep{williams-1979, williams-2003a, lian-showman-2008,
  schneider-liu-2009, liu-schneider-2010, liu-schneider-2011,
  young-etal-2019} or by storm eddies associated with latent heat
release \citep{lian-showman-2010}.  {\tt Yet other studies impose the
  observed zonal-jet profile near the bottom of the domain, and
  examine how the overlying stratospheric circulation responds to
  realistic radiative forcing \citep[e.g.][]{friedson-moses-2012}.}
Note that the jet-formation processes emphasized in many of these
studies---baroclinic instabilities and latent heating---are likely not
relevant to brown dwarfs, where the interior (dry) convective heat
flux is large, the latent heating is relatively weak, and where no
external irradiation gradient exists to provide externally imposed
baroclinicity.  The third obvious possibility---that interior dry
convection directly perturbs the stably stratified atmosphere, causing
the formation of zonal jets and other aspects of the atmospheric
circulation \citep{dritschel-mcintyre-2008}---has not previously been
explored in a 3D model, either for Jupiter or brown dwarfs.

To date, only a handful of studies have explored the dynamics of the
Jovian QQO or Saturnian SAO \citep{friedson-1999, li-read-2000,
  cosentino-etal-2017}.  All of these are two-dimensional or
quasi-two-dimensional models\footnote{These investigations all solve a
  2D system for the time evolution of the zonal-mean zonal wind versus
  latitude and height subject to parameterized wave forcing in this 2D
  meridional plane; 3D considerations of wave dynamics were used to
  determine the form this wave parameterization would take, under
  the assumption that the wave forcing results from a small set of
  specified wave modes.}  that represent the latitudinal and vertical
structure of the flow, but allow no longitudinal variation.  Because
the waves that drive these QBO-like phenomena exhibit propagation and
oscillatory behavior in both longitude and height, they cannot be
fully simulated in a purely 2D model, but rather must be
parameterized.  As yet there exist no full 3D models of giant planets
demonstrating the emergence of a QBO- or QQO-like oscillation, in
which the wave generation, propagation, and absorption that can drive
the oscillation are explicitly represented.

Here, we present three-dimensional (3D), global simulations of the
atmospheric circulation on brown dwarfs and Jupiter- and Saturn-like
planets to explore the extent to which convection interacting with a
stably stratified atmosphere can drive a circulation in the
atmosphere.  {\tt The simulations are idealized, in that the
  perturbing effects of convection near the base of the stratified
  atmosphere are parameterized in a highly idealized manner, and the
  radiative heating/cooling is represented using a simple Newtonian
  relaxation scheme.  These features provide a clean, simple
  environment in which the dynamical processes can be more fully
  diagnosed and understood.}  We wish to ascertain the fundamental
nature of the circulation, including the existence/absence and roles
of zonal (east-west) jet streams, vortices, waves, and turbulence,
determine the typical wind speeds and horizontal temperature
differences, and characterize the temporal variability, including that
over long timescales.  We show that zonal jet formation and QBO-like
oscillations can occur under appropriate conditions, and we determine
the sensitivity to radiative time constant and other parameters.
Section~\ref{model} presents our model, Section~\ref{results}
describes our results, and Section~\ref{conclusions} concludes.

\section{Model}
\label{model}

We solve the global, spherical 3D primitive equations in pressure
coordinates.  These are the standard dynamical equations for a
stratified atmosphere with horizontal length scales greatly exceeding
the vertical length scales, as appropriate to the global-scale
atmospheric flow on brown dwarfs, Jupiter, and Saturn (for reviews,
see \citealt{vallis-2006} or \citealt{showman-etal-2010}).  To
represent the effect of convection and radiation on the stratified
atmosphere, we introduce source terms in the thermodynamic energy
equation:
\begin{equation}
{q\over c_p} =  {T_{\rm eq}(p) - T(\lambda,\phi,p,t)\over 
\tau_{\rm rad}} + S
\label{thermo} 
\end{equation}
where $q$ is the specific heating rate ($\rm W\rm\,kg^{-1}$), $c_p$
is the specific heat, $\lambda$ is longitude, $\phi$ is latitude,
$p$ is pressure, and $t$ is time.

The first term on the righthand side of (\ref{thermo}) represents
radiative heating/cooling, which we parameterize with a Newtonian
heating/cooling scheme that acts to relax the local temperature
$T$ toward a prescribed radiative equilibrium temperature, $T_{\rm
  eq}$, over a prescribed radiative timescale, $\tau_{\rm rad}$.  On
irradiated planets, day-night gradients in incident stellar flux would
cause $T_{\rm eq}$ to vary spatially (e.g., being hotter on the
dayside than the nightside), but for an isolated brown dwarf, the
radiative-equilibrium temperature of the stratified atmosphere is
determined solely by the upwelling IR radiation coming from below.
Because the convection zone should exhibit minimal horizontal entropy
variations, this radiative-equilibrium temperature should be nearly
independent of longitude or latitude \citep[see][]{showman-kaspi-2013}.
Therefore, we take $T_{\rm eq}$ to be a function of pressure only.
In this context, radiation acts to {\it remove} horizontal temperature
differences---and available potential energy---and thus damps the
flow.  

In the observable atmosphere, radiative time constants are expected
to vary greatly with temperature, and near the photosphere can be
represented approximately by \citep{showman-guillot-2002}
\begin{equation}
\tau_{\rm rad} \sim {P c_p\over 4 g \sigma T^3}
\label{tau-rad}
\end{equation}
where $P$ is the photospheric pressure, $g$ is gravity, $c_p$ is the
specific heat at constant pressure, $\sigma$ is Stefan-Boltzmann
constant, and $T$ is temperature.  Inserting appropriate values for
Jupiter ($P\sim 0.5\rm\,bar$, $c_p=1.3\times10^4\rm
\,J\,kg^{-1}\,K^{-1}$, $g=23\rm\,m\,s^{-1}$ and $T=130\rm\,K$) yields
$\tau_{\rm rad}\approx 6 \times10^7\rm\,s$.  This is similar to values
estimated from sophisticated radiative transfer calculations {\tt
  providing explicit calculations of the radiative time constant as a
  function of pressure}, which indicate that $\tau_{\rm rad}$ ranges
between $10^8$ and $10^7\rm\,s$ from 1 bar to 1 mbar
\citep{kuroda-etal-2014, li-etal-2018}.  {\tt Nevertheless, some other
  studies have suggested longer radiative time constants between
  $10^8$ and $10^9\rm\,s$ \citep{conrath-etal-1990,
    guerlet-etal-2014}.}  In contrast, at the higher temperatures of
brown dwarfs, the above scaling predicts much shorter radiative time
constants in the range $\sim$$10^4$--$10^5\rm\,s$
\citep[see][]{showman-kaspi-2013, robinson-marley-2014}.  To capture
this range, we vary the radiative time constant from
$10^4$--$10^8\rm\,s$.  For simplicity, we take $\tau_{\rm rad}$
constant with pressure.

%%%%%%%%%%%%%%%%%%
% FIGURE 1
%%%%%%%%%%%%%%%%%
\begin{figure*}
\begin{minipage}[c]{0.47\textwidth}
\includegraphics[scale=0.51, angle=0]{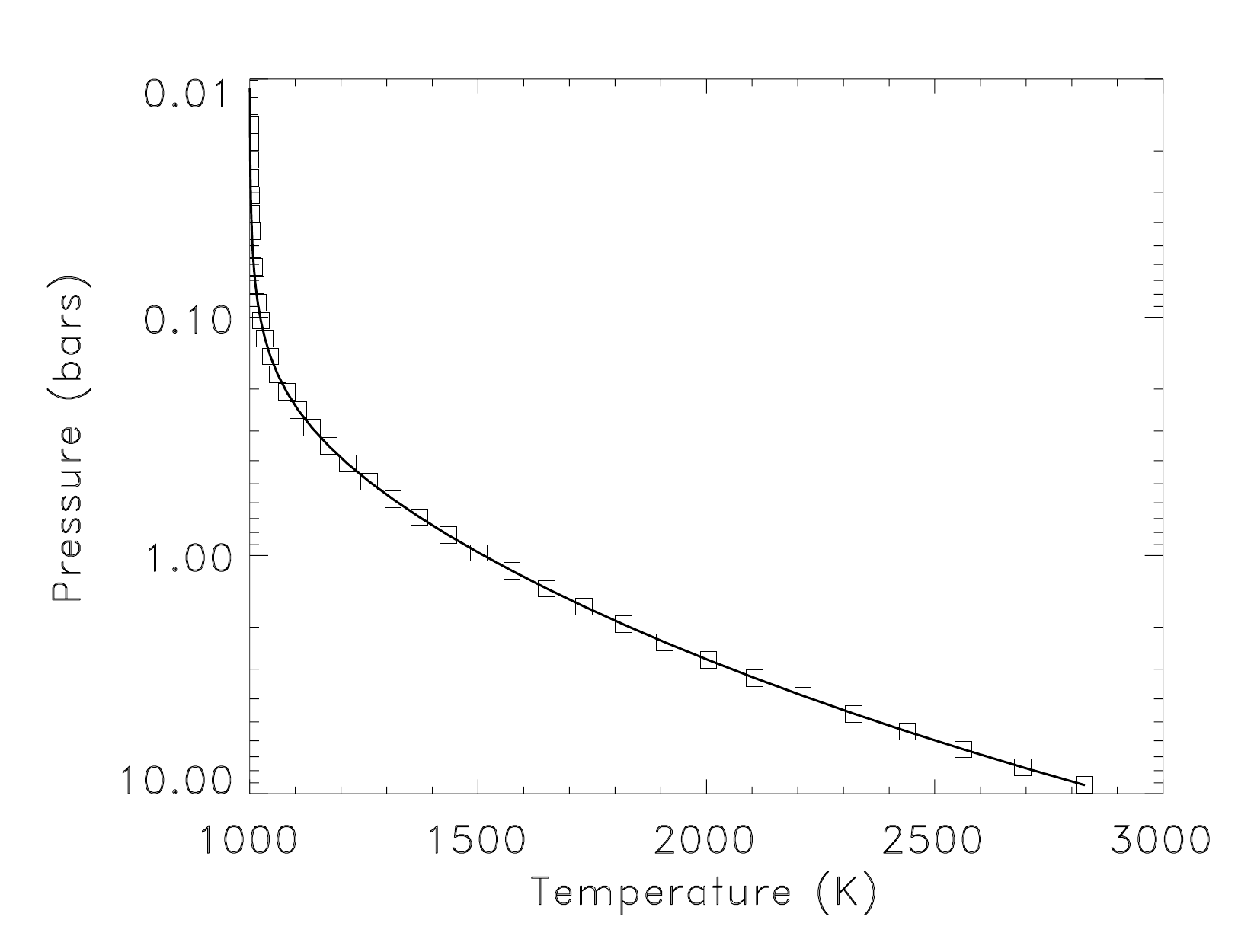}
\put(-186.,173.){(a)}\\
\end{minipage}
\begin{minipage}[c]{0.3\textwidth}
\includegraphics[scale=0.72, angle=0]{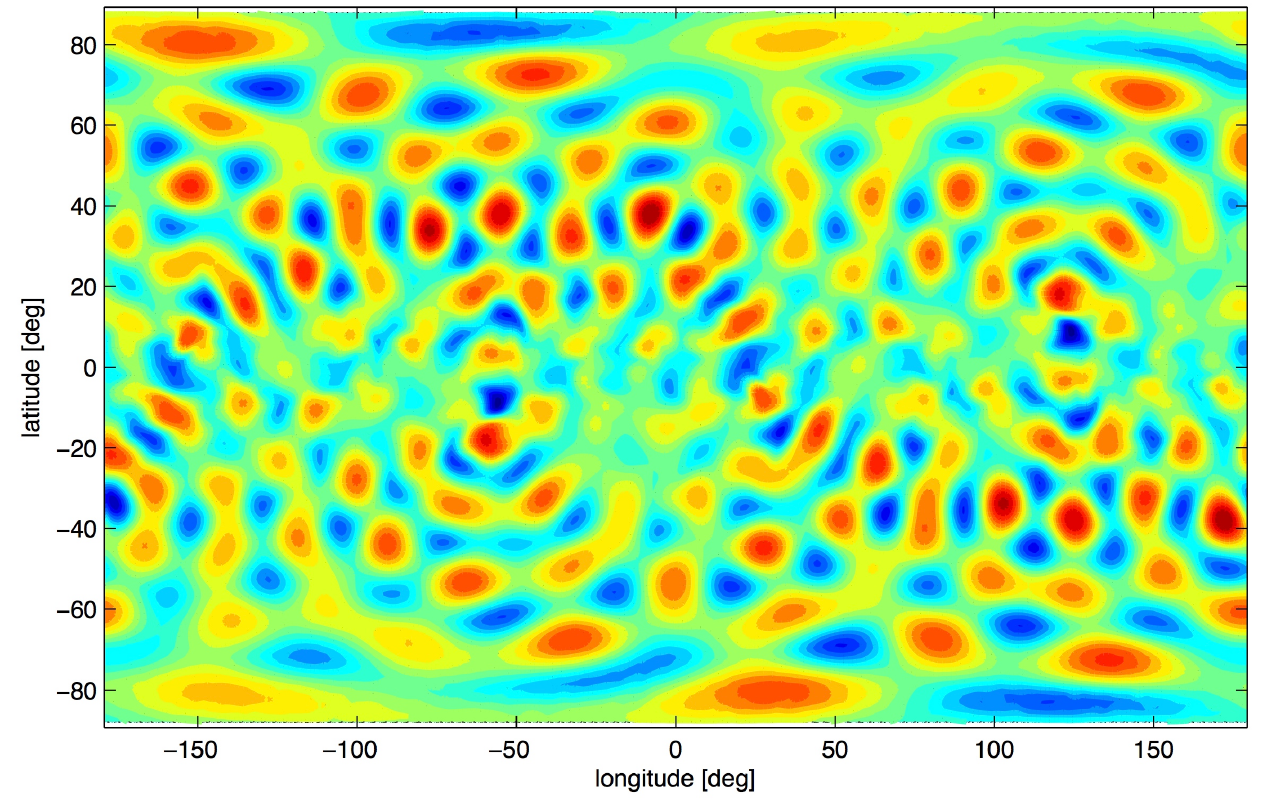}
\put(-230.,170.){(b)}\\
\end{minipage}
\caption{(Left): Radiative-equilibrium temperature-pressure profile
  $T_{\rm eq}(p)$ (Eq.~\ref{teq}) adopted in the numerical
  experiments.  The profile transitions from an adiabat at depth to an
  isotherm aloft, here with $T_{\rm iso}=1000\rm\,K$ and $\theta_{\rm
    ad}=1500 \,\rm K$. The squares show the pressures of the cell
  centers adopted in the 3D simulations for the example of a 40-level
  model.  (Right): a realization of the random, horizontally isotropic
  forcing pattern, $S_h$, used in the simulation, here shown for a
  total forcing wavenumber $n_f=20$.}
\label{setup}
\end{figure*}

Sophisticated 1D radiative-transfer models show that the temperature
profile on brown dwarfs transitions from an adiabatic interior to a
nearly isothermal atmosphere, with a transition at typically
a few bars \citep[e.g.][]{burrows-etal-2006}. To mimic this behavior,
we adopt a radiative-equilibrium profile
\begin{equation}
T_{\rm eq}(p)= (T_{\rm
  iso}^n + T_{\rm ad}^n)^{1/n},
\label{teq}
\end{equation}
where $T_{\rm iso}$ is a constant representing the isothermal
atmosphere, and $T_{\rm ad}(p)$ is an adiabatic temperature-pressure
profile representing the interior, which can be expressed as $T_{\rm
  ad} = \theta_{\rm ad}(p/p_0)^{R/c_p}$, where $\theta_{\rm ad}$ is
the potential temperature of the interior (constant for an adiabat),
$R$ is the specific gas constant, and $p_0=1\rm\,bar$ is a reference
pressure.  Here, we set $T_{\rm iso}=1000\rm\,K$ and
$\theta_{ad}=1500\rm\,K$, repsenting conditions of a typical dwarf
near the L/T transition, although we emphasize that the results of our
3D simulations are insensitive to the precise values.  $n$ is a
smoothing parameter, which controls the vertical scale over which the
temperature profile transitions between adiabatic and isothermal; we
here take it equal to $7$, which causes the transition to occur
smoothly across a vertical distance comparable to a scale height.
Figure~\ref{setup} shows the profile $T_{\rm eq}(p)$ resulting from
these choices.

The second term on the righthand side of (\ref{thermo}), $S$,
represents the effect of convection on the atmosphere, specifically
the perturbation of the radiative-convective boundary (RCB) by
convection, as well as possible convective overshoot and mixing across
the RCB.  Both of these processes should occur in a highly variable
manner.  Global convection models could in principle provide
constraints on the dominant length and timescales of this process, but
such models can only be performed with unrealistic parameter settings,
rendering any such predictions uncertain \citep{showman-etal-2011,
  showman-kaspi-2013}.  Instead, we choose to parameterize the
forcing, giving us full control over the forcing length and
timescales, and allowing us to determine how they affect the
atmospheric circulation.  In keeping with a long history of turbulence
studies, we parameterize this forcing as a random, homogeneous,
isotropic source/sink in (horizontal) space, which evolves randomly in
time via a Markov process \citep[e.g.][]{lilly-1969, williams-1978,
  nozawa-yoden-1997a, scott-polvani-2007}.  The forcing is confined
near the bottom of the domain near the RCB.  Because we envision that
convective plumes will push the RCB up and down, the thermal
perturbations (on isobars) should remain vertically coherent near the
RCB, and thus we adopt a forcing function of the form
$S(\lambda,\phi,p,t)=S_v(p)S_h(\lambda,\phi,t)$.  Here, $S_v(p)$
represents the (nondimensional) vertical structure of the forcing,
which we assume to vary linearly in log-pressure from one at the base
of the domain to zero one scale height above.  The quantity
$S_h(\lambda,\phi,t)$ represents the spatial and temporal distribution
of the forcing across the globe, with units of $\rm K\,s^{-1}$.  In
keeping with standard procedure, this is represented as a Markovian
process
\begin{equation}
S_h(\lambda,\phi,t+\delta t)= r S_h(\lambda,\phi,t) +
\sqrt{1-r^2} F(\lambda,\phi,t)
\label{markov}
\end{equation}
where $r$ is a dimensionless memory coefficient, $F$ represents the
random modifier of $S_h$, and $\delta t$ represents the model
timestep.  This formulation causes the forcing to vary smoothly from
one spatially random pattern to another over a characteristic
decorrelation timescale $\tau_{\rm for}$.  The two limits of
white-noise forcing and time-constant forcing would be represented as
$r=0$ and $r=1$, respectively.  For correlation timescales long
compared to a timestep, the memory coefficient is
\begin{equation}
r = 1 - {\delta t \over \tau_{\rm for}}.
\end{equation}

The forcing wavenumber $n_f$ represents the characteristic total
horizontal wavenumber on which convection perturbs the RCB.
Individual convective plumes are expected to be small-scale and cannot
be resolved in a global model; for example, the local, 2D box
simulations of \citet{freytag-etal-2010} suggest that individual
convective plumes are typically $\sim$10~km across (corresponding to a
spherical wavenumber of 22,000 in a global model with Jupiter's
radius!).  Despite the impossibility of resolving the individual
convective plumes in a global model, it is likely that the convection
will exhibit organization across a wide range of scales (as is common
in Earth's tropics, for example), and in our models, the forcing
wavenumber represents the supposed wavenumber of this large-scale
organization.  The largest possible forcing wavenumber that we can
resolve numerically is $N/4$, where $N$ is the total spherical
wavenumber corresponding to the model resolution.  For our nominal
resolutions of C128 (see below), $N\approx 170$, implying that the
largest forcing wavenumber that we can numerically resolve is
$n_f=42$.  We explore values of 40 and 20.  The characteristic
timescale on which the large-scale convective organization varies is
unknown but is presumably longer than the timescale associated with
individual convective plumes (associated with overshoot, for example).
We perform most simulations with $\tau_{\rm for}=10^5\rm\,s$ but also
explore $10^4\rm\,s$ and $10^6\rm\,s$ in a few integrations.

We represent the spatial structure of the forcing as a horizontally
isotropic superposition of spherical harmonics of a characteristic
total forcing wavenumber, $n_f$:
\begin{equation}
F = f_{\rm amp} \sum_{m=1}^{n_f}  N_{n_f}^m(\sin\phi)\cos[m(\lambda + \psi_m)]
\label{random}
\end{equation}
where $N_n^m(\sin\phi)$ are the normalized associated Legendre
polynomials, $m$ is the zonal wavenumber, $n$ is the total wavenumber,
$f_{\rm amp}$ is the forcing amplitude in units of $\rm K\,s^{-1}$,
and $\psi_m$ is a randomly chosen phase, different for each mode.  New
random phases $\psi_m$ are chosen each time (\ref{markov}) is
evaluated, meaning that, statistically, there is no correlation in the
spatial pattern of $F$ between one timestep and the next. {\tt The
  appendix presents presents simple arguments on the values of $f_{\rm
    amp}$ that are appropriate for a given heat flux and other
  parameters.  These estimates suggest that $f_{\rm amp}$ of a few
  $\times10^{-6}\rm\,K\,s^{-1}$ to a few $\times10^{-5}\rm\,K\,s^{-1}$
  are generally appropriate for giant planets and brown dwarfs.
  Although the detailed formulations differ, our overall approach of
  adding thermal perturbations near the bottom of the domain is
  similar to a scheme employed in the 3D Saturn stratospheric GCM of
  \citet{friedson-moses-2012}, who introduced randomly fluctuating
  thermal perturbations near the tropopause as way of producing
  stratospheric eddies.}

Because the temperature structure is nearly adiabatic at the bottom of
our domain, any winds generated there would penetrate deeply into the
planetary interior following the Taylor-Proudman theorem
\citep[e.g.][]{vasavada-showman-2005}, and at great depths, Lorentz
forces may act to brake these columnar flows
\citep[e.g.][]{busse-2002, liu-etal-2008, duarte-etal-2013}.  We
parameterize this process by introducing a frictional drag scheme near
the bottom of the domain \citep[cf][]{schneider-liu-2009,
  liu-showman-2013}.  The drag is represented in the momentum
equations as $-k_v{\bf v}$, where $k_v(p)$ is a pressure-dependent
drag coefficient, and ${\bf v}$ is horizontal velocity.  The drag
coefficient is zero (meaning no drag) at $p\le p_{\rm drag,top}$ and
varies linearly in $p$ from zero at $p_{\rm drag,top}$ to $\tau_{\rm
  drag}^{-1}$ at the bottom of the domain, where the drag is
strongest.  Here, $\tau_{\rm drag}$ represents the timescale of this
drag at the bottom of the domain.  We generally adopt $p_{\rm drag,
  top}=4\rm\,bars$, which is just below the RCB in our models.  
$\tau_{\rm drag}$ is considered to be a free parameter, which we vary
over a wide range.

The domain extends from 10 bars at the bottom (below the RCB) to 0.01
bars at the top.  All models adopt Jupiter's radius and a rotation
period of 5~hours, which is typical for brown dwarfs
\citep{reiners-basri-2008}, $c_p=13000\rm\,J\,kg^{-1}\,K^{-1}$, and
$R/c_p=2/7$, appropriate to an H$_2$ atmosphere.  The gravity is set
to either $23\rm\,m\,s^{-2}$ or $500\rm\,m\,s^{-2}$, representing
objects of 1 and $\sim$20 Jupiter masses,
respectively.\footnote{Interestingly, however, for our particular
  model formulation, the entire system is independent of the value of
  gravity.  When written in pressure coordinates, the gravity never
  explicitly appears in the primitive equations.  The gravity
  indirectly enters the system via the geopotential, which appears in
  the horizontal pressure gradient force, $-\nabla\Phi$.  The
  geopotential, which is a dependent variable in the
  pressure-coordinate version of the primitive equations, is
  determined by vertically integrating the hydrostatic balance
  equation $\partial\Phi/\partial \ln p = -RT$ with respect to
  pressure (where here we have assumed ideal gas).  If we specify the
  3D temperature structure $T(\lambda,\phi,p)$ {\it as a function of
    pressure} (not as a function of height), then the geopotential
  $\Phi(\lambda,\phi,p)$ is independent of
  gravity. 
  %Essentially, the tendency of stronger gravity to amplify
  %the pressure gradient force is cancelled out by the fact that
  %stronger gravity atmospheres have smaller scale heights, so that the
  %vertical extent of an atmosphere (of a given pressure range) is
  %smaller when the gravity is higher.  
  In more realistic models,
  gravity would typically enter the system via the radiative
  transfer---for a given opacity, the photosphere pressure, and more
  generally the detailed radiative heating/cooling rates, depend on
  the gravity. For a given atmospheric mass, the basal pressure of an
  atmosphere likewise depends on gravity.  Thus, we would expect
  gravity to affect the outcome in more realistic models that include
  radiative transfer.  In our formulation, however, the domain depth
  and the forcing and damping formulations (Eq.~\ref{thermo}) are
  functions of pressure, and therefore the temperature structure---and
  indeed the entire simulation---is independent of gravity.  We
  verified this by comparing otherwise identical models with gravities
  of 23 and $500\rm\,m\,s^{-1}$; such models behave identically.}  The
latter represents a typical brown dwarf and, with the former, brackets
the range of gravities expected on directly imaged giant planets. The
models are initialized from rest using $T_{\rm eq}(p)$ as an initial
temperature-pressure profile, and are integrated until a
(time-fluctuating) statistical equilibrium is reached.

Given our forcing and damping setup, our model can be
considered a three-dimensional generalization of the
forced-dissipative one-layer turbulence models commonly used to
explore jet formation on giant planets, and in geophysical fluid
dynamics (GFD) more generally \citep[e.g.][]{zhang-showman-2014,
  scott-polvani-2007, sukoriansky-etal-2007, nozawa-yoden-1997a}.

We have kept our model formulation as simple as possible to provide a
well-defined, clean environment in which to study jet and QBO-like
dynamics.  In particular, we do not include sub-grid-scale
parameterizations of damping due to small-scale, numerically
unresolved gravity waves; rather, the jets and QBO-type oscillations
here result solely from the explicitly resolved wave dynamics and
their interaction with the mean flow.  While small-scale gravity waves
likely contribute to the momentum budgets of the actual QQO and SAO
\citep[e.g.][]{cosentino-etal-2017}, parameterizing such waves would
introduce numerous assumptions involving the directions, phase speeds,
amplitudes, and spectra of the waves being parameterized, and would
significantly complicate an understanding of the resulting dynamics.
In our view, it is essential to first understand the behavior of
resolved, idealized systems such as the one we present; only once such
idealized models are understood should gravity wave parameterizations
be added.

Although modest horizontal resolution is adequate for hot Jupiters
\citep[e.g.][]{liu-showman-2013}, brown dwarfs---like
Jupiter---exhibit small Rossby deformation radii and require higher
resolution.  Away from the equator, the deformation radius in the
stratified atmosphere is given by $L_D = NH/f$, where $N$ is the
Brunt-Vaisala frequency, $H$ is the scale height, and
$f=2\Omega\sin\phi$ is the Coriolis parameter.  Adopting a rotation
period of 5 hours implies $\Omega=3.5\times 10^{-4}\rm\,s^{-1}$.
For a vertically isothermal temperature profile, $NH=R\sqrt{T/c_p}$.
Under brown dwarf conditions, these numbers imply $L_D \approx
1500$--$3000\rm\,km$ depending on latitude.  Ideally, the numerical grid
should resolve this scale, as well as the forcing scale.

We solved the equations using the MITgcm \citep{adcroft-etal-2004} in
cubed-sphere geometry.  Motivated by the above considerations, we
generally adopt C128 (i.e., $128\times128$ finite-volume cells on each
cube face), corresponding to an approximate global resolution of
$512\times256$ in longitude and latitude, or $0.7^\circ$ per grid
point (corresponding to a resolution of $800\rm\,km$ for an object of
Jupiter radius).  We also performed several simulations at C256,
corresponding to global resolutions of approximately $1024\times 512$
in longitude and latitude ($0.35^\circ$, or $\sim$$400\rm\,km$ per
grid point), to confirm that our qualitative results are unchanged.
The vertical resolution is $N_r=40$, 80, or 160 levels.  The models
are generally integrated up to $5,000$--$10,000$ Earth days ($25,000$
to $50,000$ brown dwarf rotation periods).  The timestep was
$100\rm\,s$ in most cases ($50\rm\,s$ in a handful of the
highest-resolution models).  A standard, fourth-order Shapiro filter
was used to control noise at the grid scale.

\section{Results}
\label{results}

\subsection{Basic flow regime}
\label{flow-regime}

\begin{figure*}
\centering
\begin{minipage}[c]{0.32\textwidth}
\includegraphics[scale=0.4, angle=0]{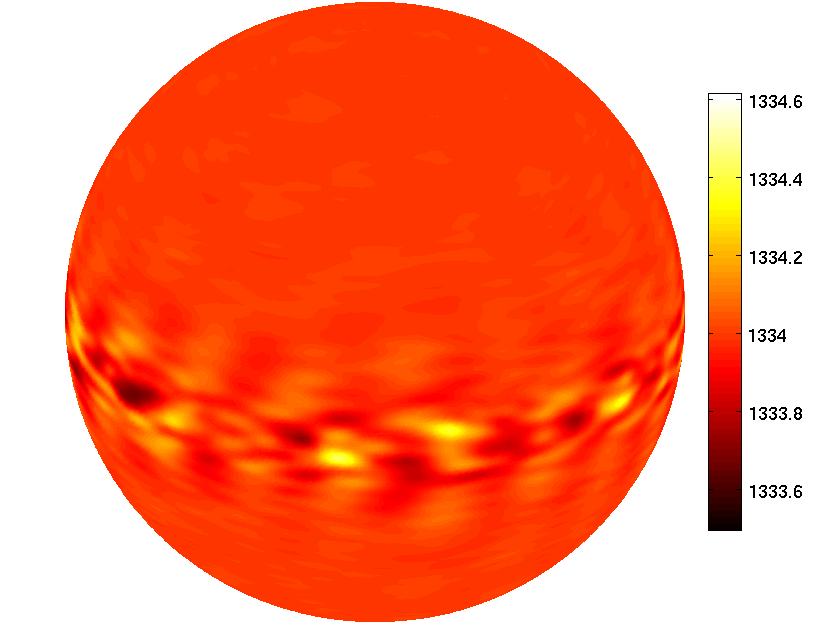}
\includegraphics[scale=0.4, angle=0]{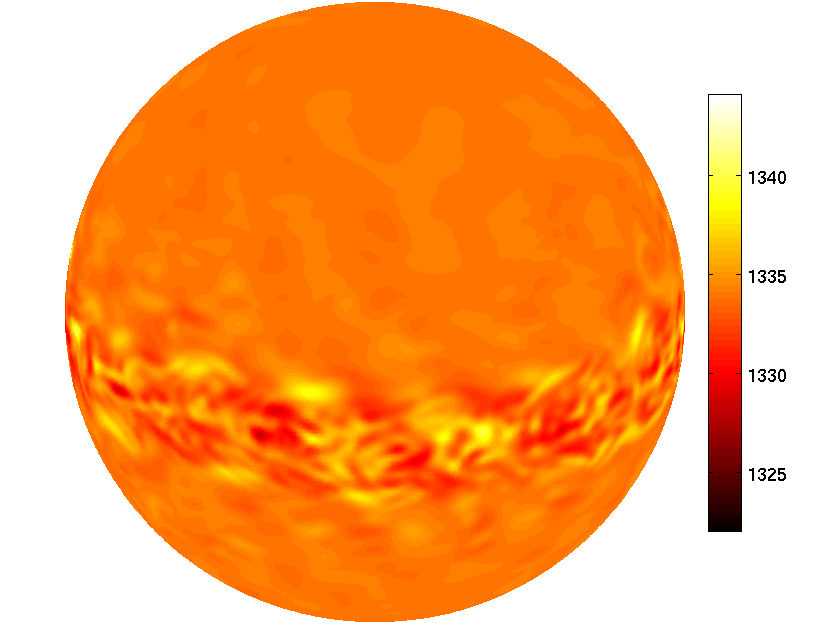}
\includegraphics[scale=0.4, angle=0]{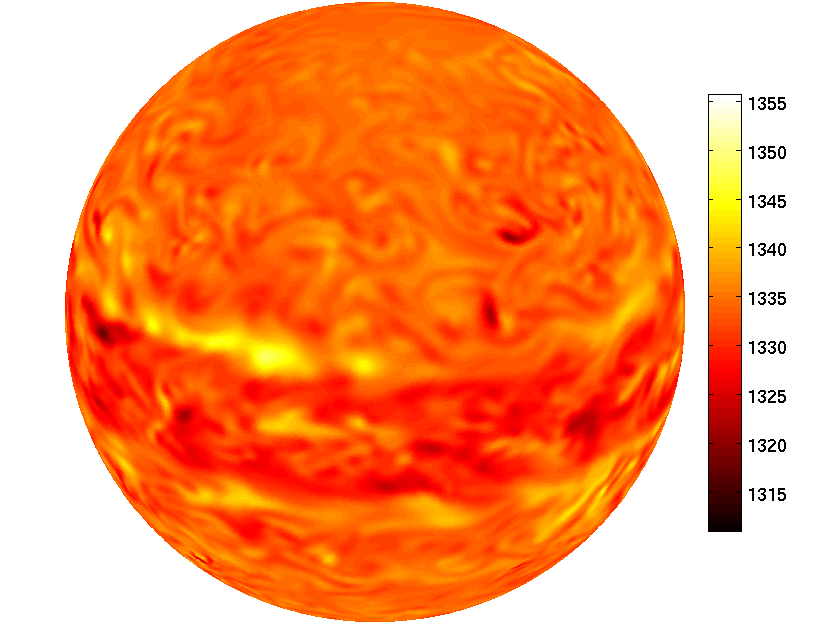}
\includegraphics[scale=0.4, angle=0]{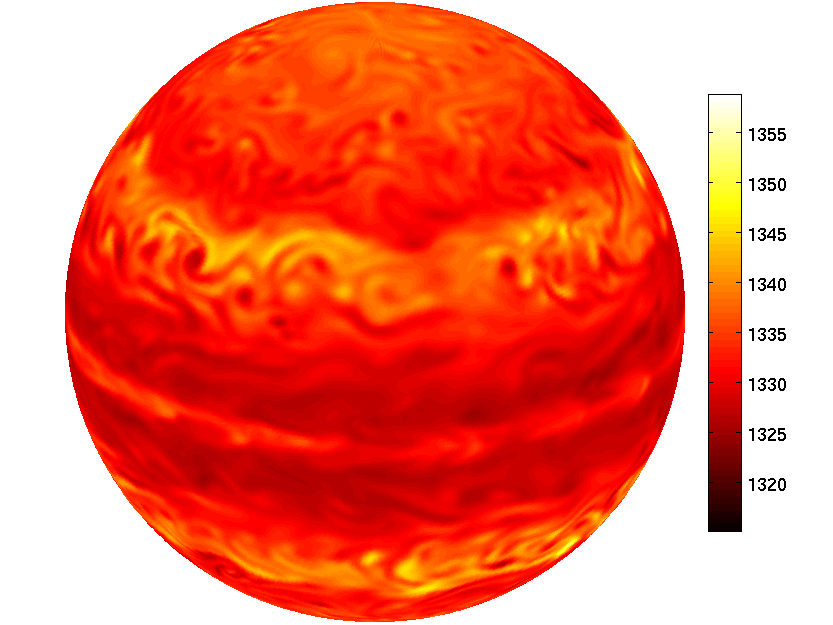}
\end{minipage}
\put(-140.,240.){(a)}
\put(-140.,115.){(b)}
\put(-140.,-10.){(c)}
\put(-140.,-130.){(d)}
\begin{minipage}[c]{0.3\textwidth}
\includegraphics[scale=0.4, angle=0]{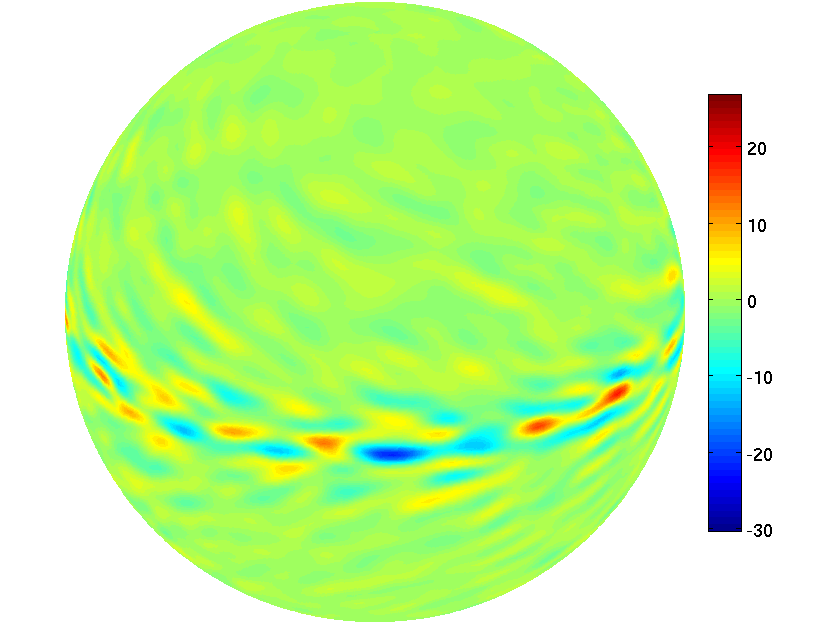}
%\put(-90.,120.){\tiny C\ofast}\\
\includegraphics[scale=0.4, angle=0]{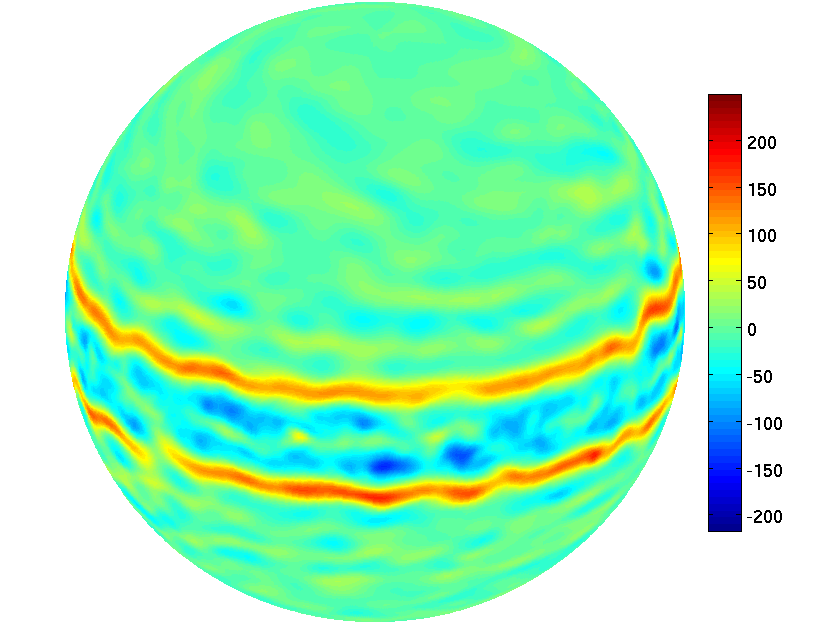}
%\put(-90.,120.){\tiny W\ofast}\\
\includegraphics[scale=0.4, angle=0]{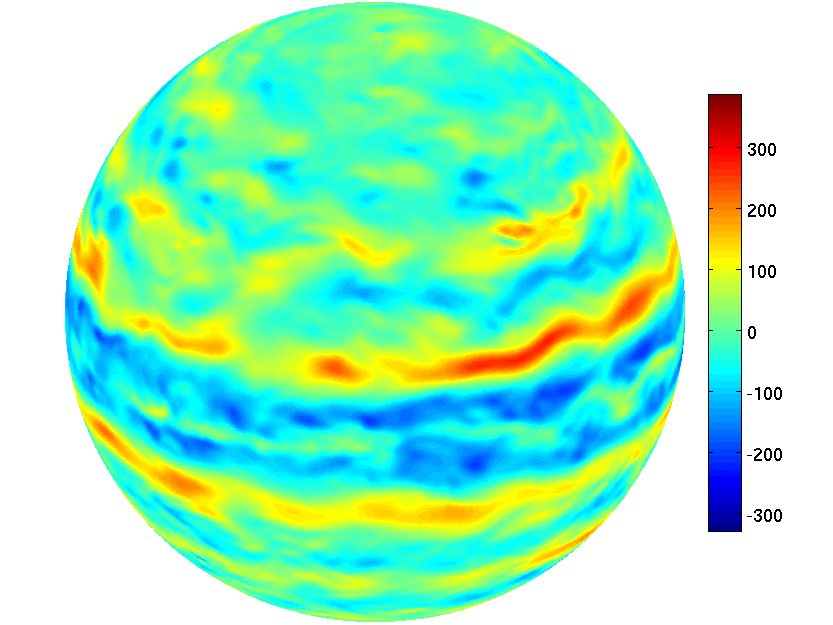}
%\put(-90.,120.){\tiny H\ofast}
\includegraphics[scale=0.4, angle=0]{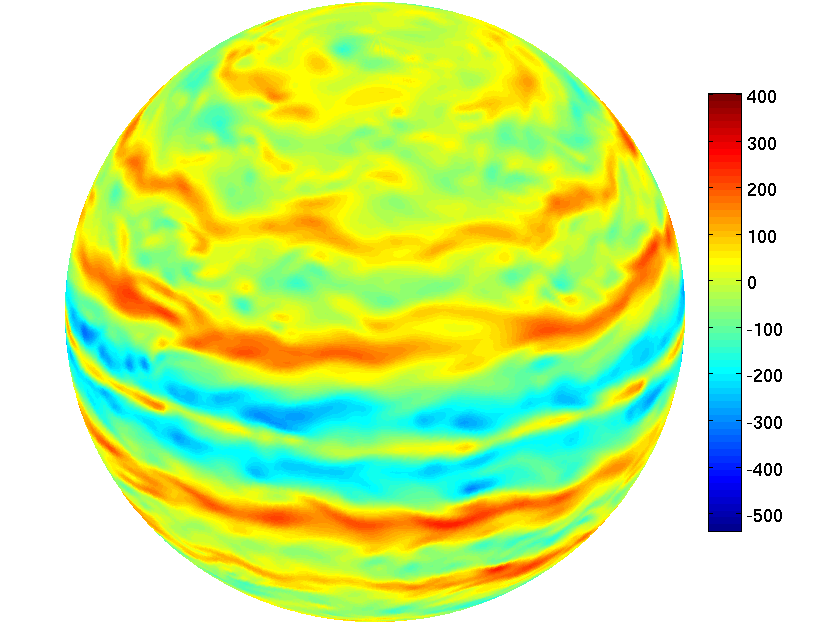}
\end{minipage}
\caption{Temperature (K, left) and zonal winds ($\rm m\,s^{-1}$,
  right) for four simulations showing the emergence of zonal jets and
  eddies in high-resolution 3D models of brown dwarfs.  These are
  snapshots shown once the flow reaches a statistical equilibrium.
  The structure is shown at a pressure of 0.6 bars, close to a typical
  IR photosphere pressure for a brown dwarf. Each row shows a
  different simulation.  The four simulations are identical except for
  the radiative time constant, which is $10^4$, $10^5$, $10^6$, and
  $10^7\rm\,s$ in rows (a), (b), (c), and (d), respectively.
  Resolution is C128 (corresponding to a global grid of $512\times256$
  in longitude and latitude) with 160 vertical levels.  The forcing
  amplitude is $f_{\rm amp}=5\times 10^{-5}\rm\,K\,s^{-1}$, rotation
  period is 5 hours, gravity is $500\rm\,m\,s^{-2}$, drag time
  constant is $10^6\rm\,s$, convection decorrelation timescale
  $\tau_{\rm for}=10^5\rm\,s$, forcing wavenumber $n_f=20$, and other
  parameters are as described in the text. }
\label{globes}
\end{figure*}

Our key result is that, although the forcing and damping are
horizontally isotropic, the interaction of the turbulence with the
planetary rotation leads to a zonally banded flow pattern and the
emergence of robust zonal jets over a wide range of parameters.
Figure~\ref{globes} illustrates this phenomenon for a series of four
high-resolution models that vary the radiative time constant from
$10^4\rm\,s$ (a, top row) to $10^7\rm\,s$ (d, bottom row).  All four
models are shown at long times after the flow has reached a
statistical equilibrium.  Temperature and zonal wind are depicted in
the left and right columns, respectively, at a pressure of 0.6 bars,
close to a characteristic IR photosphere pressure for typical brown
dwarfs.\footnote{The photosphere---namely, the pressure level where
  the majority of radiation escapes to space---depends greatly on
  wavelength, as well as on the properties of any clouds that may be
  present.  In spectral windows like J band, it commonly probes to
  several bars, whereas in absorption bands such as those of water
  vapor, it can be a bar or less.}

Although dynamical activity and zonally elongated structures occur in
all of the models (Figure~\ref{globes}), the specific jet structure,
wind speeds, and temperature differences vary significantly depending
on the radiative time constant.  In particular, when the radiative
time constant is long ($10^7\rm\,s$, panel d), zonal jets occur at all
latitudes from the equator to nearly the poles.  The low-latitude jets
exhibit faster speeds than the high-latitude jets, a phenomenon that
occurs also on Jupiter and Saturn, but the differences are not
great: wind speeds for the highest-latitude jets remain a significant
fraction of those for the low-latitude jets.  On the other hand, at
progressively shorter radiative time constants, the strong dynamical
activity is confined progressively closer to the equator, and the high
latitudes lack prominent zonal jets.  Specifically, when $\tau_{\rm
  rad}=10^6\rm\,s$, zonal banding and significant zonal winds occur
from the equator to at least $60^\circ$ latitude but weaken
poleward of that latitude.  The low-latitude confinement becomes even
more prominent at shorter radiative time constants; strong jets
and horizontal temperature differences occur primarily equatorward of
$\sim$$20^\circ$ latitude when $\tau_{\rm rad}=10^5\rm\,s$ and
$10^\circ$ latitude when $\tau_{\rm rad}=10^4\rm\,s$.  Despite this
overall trend, weak dynamical activity---particularly in the wind
field---nevertheless occurs at all latitudes even in the models with
short radiative time constant.  This can be seen in the righthand
panels of Figure~\ref{globes}(a) and (b).  Especially at the shortest
radiative time constant, this structure takes the form of weak
alternating eastward and westward structures that are phase tilted
northwest-southeast in the northern hemisphere and southwest-northeast
in the southern hemisphere.  We identify them as barotropic Rossby
waves that are generated at low latitudes by the convective forcing
and propagate poleward to higher latitudes (compare to the expected
phase relations in Figure~6 of \citealt{showman-etal-2013b}).

%%%%%%%%%%%%%%%%%%
% FIGURE 2
%%%%%%%%%%%%%%%%%

The equilibrated wind speeds and horizontal temperature differences
also vary systematically with the radiative time constant.
As described previously, our convective parameterization represents
forcing that adds energy to the model, while radiation represents
damping which removes it.  The statistical steady state occurs when
there exists a (statistical) balance between the forcing and damping.
We would thus expect the equilibrated winds and horizontal temperature
differences to be greater when the convective forcing is stronger
and/or the radiative damping is weaker, and the winds and temperature
differences to be smaller when the convective forcing is weaker and/or
the radiative damping is stronger.  Our results confirm this
expectation.  In the models shown in Figure~\ref{globes}, peak-to-peak
wind speeds and temperature differences are $\sim$$1\rm\,K$ and
$\sim$$50$--$100\rm m\,s^{-1}$ when $\tau_{\rm rad}$ is extremely
short ($10^4\rm\,s$), but rise to $\sim$$30\rm\,K$ and
$900\rm\,m\,s^{-1}$ when $\tau_{\rm rad}$ is $10^7\rm\,s$.
Interestingly, the trend between these extremes is not smooth.  The
wind speeds and temperature differences increase by nearly an order of
magnitude as $\tau_{\rm rad}$ increases from $10^4$ to $10^5\rm\,s$
but the trend weakens for further increases in $\tau_{\rm rad}$; the
winds and temperature differences increase by less than a factor of
two as $\tau_{\rm rad}$ rises from $10^5$ to $10^6\rm\,s$ and then by
another similar factor as $\tau_{\rm rad}$ rises from $10^6$ to
$10^7\rm\,s$.  This apparent regime shift in part likely reflects a
decrease in the importance of radiative damping relative to frictional
drag as an overall energy-loss process as $\tau_{\rm rad}$ becomes
large.  But there is also likely a change in the underlying dynamics
as the turbulence becomes more strongly nonlinear as the flow
amplitude increases.

The equatorial confinement of the zonal flow under conditions of
strong damping (Figure~\ref{globes}) can be qualitatively understood
using simple dynamical arguments (cf \citealt{tan-showman-2017}).  The
fast rotation rates and large length scales on brown dwarfs ensure
that the large-scale flow is close to geostrophic balance
\citep{showman-kaspi-2013}.  As a rule, rapid rotation and the
resulting geostrophy tend to weaken the horizontal divergence relative
to that expected otherwise.  For a geostrophically balanced flow, the
horizontal divergence is
\begin{equation}
\nabla\cdot{\bf v} = {\beta\over f}v = {v\over a\tan\phi}
\label{div}
\end{equation}
where ${\bf v}$ is the horizontal wind vector, $v$ is the meridional
(northward) velocity, $a$ is the planetary radius, $\phi$ is latitude,
$f=2\Omega\sin\phi$ is the Coriolis parameter, and $\beta$ is the
gradient of the Coriolis parameter with northward distance, given on
the sphere by $\beta=2\Omega\cos\phi/a$.  This equation approximately
applies where the Rossby number, $Ro=U/fL \ll 1$, where $U$ and $L$
are the characteristic horizontal wind speed and lengthscale of the
flow, respectively. For conditions relevant to our simulations ($U\sim
100\rm\,m\,s^{-1}$ and $L$ of a few thousand km), we expect $Ro\sim 1$
at latitudes of a few degrees; any latitude significantly poleward
will exhibit $Ro\ll 1$, and thus the flow should be close to
geostrophy.  Equation~(\ref{div}) implies that for flows close to
geostrophic balance, the horizontal divergence maximizes at low
latitudes and becomes zero at the poles.  In addition to the
divergence allowed by Equation~(\ref{div}), the ageostrophic component
of the horizontal velocity, which is typically a fraction $Ro$ of the
geostrophic component, may be associated with significant horizontal
divergence, and again this component of the horizontal divergence is
likely to be greater at low latitudes due to the latitude dependence
of the Coriolis parameter $f$.  The greater horizontal divergence at
low latitudes permits greater vertical motions there, allowing greater
vertical entropy advection and therefore greater horizontal
temperature differences than at higher latitudes (Figure~\ref{globes},
(a) and (b)).  Moreover, the greater horizontal divergence leads to
efficient generation and radiation of Rossby waves at low latitudes
(see discussion in \citealt{schneider-liu-2009}), which promotes zonal
jet formation and helps to explain the predominance of jets at low
latitudes in our short-$\tau_{\rm rad}$ models.  (The greater value of
$\beta$ at low latitudes than high latitudes also makes it easier to
generate Rossby waves at low latitudes, but by itself this factor does
not depend on $\tau_{\rm rad}$).

%%%%%%%%%%%%%%%%%%
% FIGURE 3
%%%%%%%%%%%%%%%%%
\begin{figure}
\centering
\begin{minipage}[c]{0.32\textwidth}
\includegraphics[scale=0.4, angle=0]{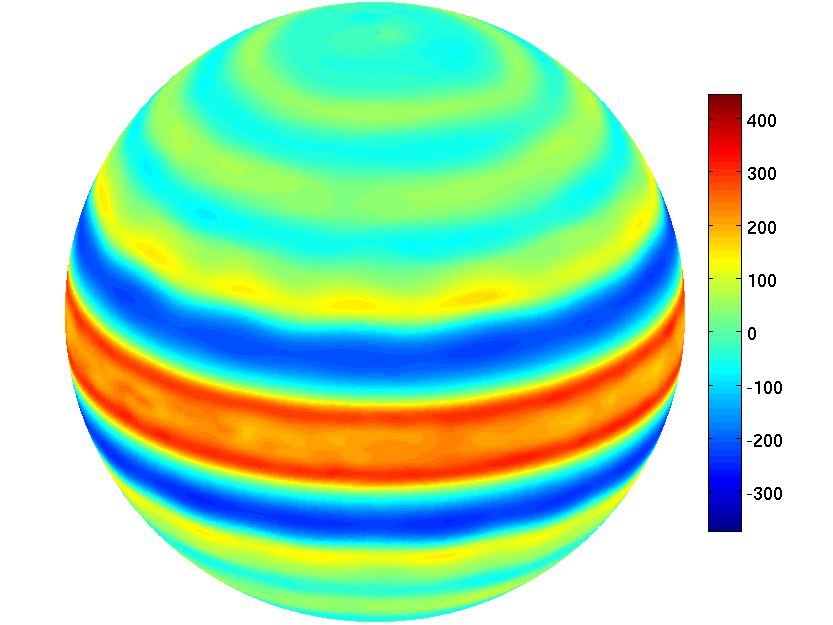}
\includegraphics[scale=0.4, angle=0]{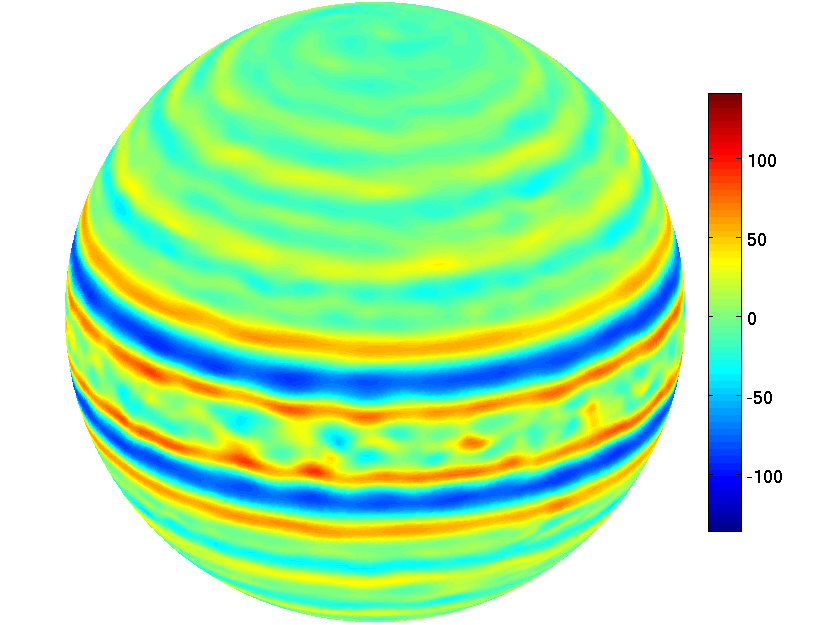}
\includegraphics[scale=0.4, angle=0]{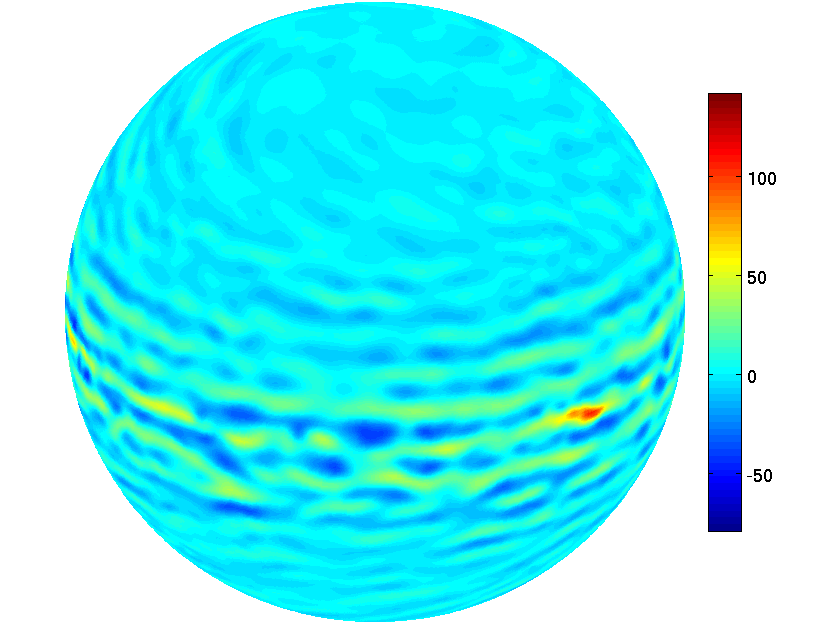}
\includegraphics[scale=0.4, angle=0]{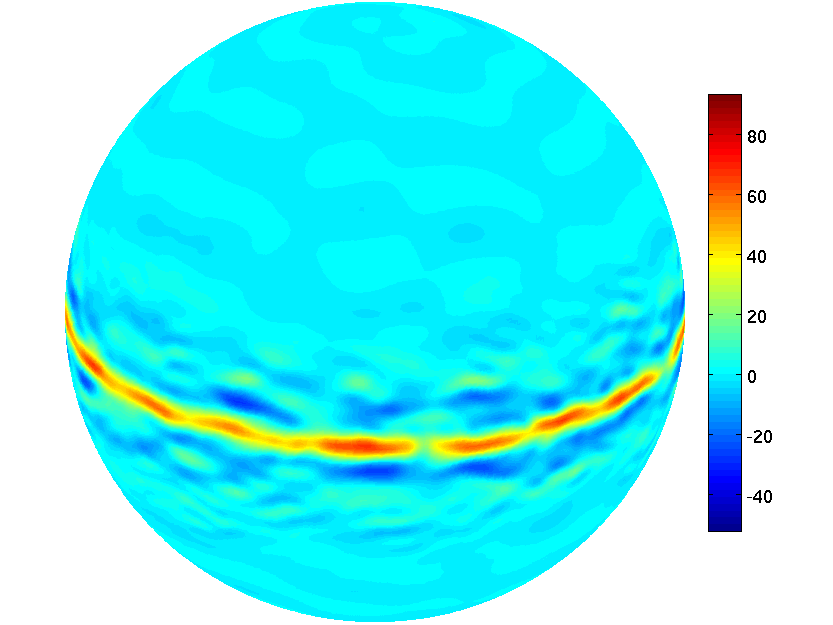}
\end{minipage}
\put(-140.,240.){(a)}
\put(-140.,115.){(b)}
\put(-140.,-10.){(c)}
\put(-140.,-130.){(d)}
\caption{Zonal winds ($\rm m\,s^{-1}$) for four simulations
  illustrating the dependence of the jets on bottom drag.  These are
  snapshots shown once the flow reaches a statistical equilibrium.
  The structure is shown at a pressure of 0.77 bars, close to a
  typical IR photosphere pressure for a brown dwarf. Each row shows a
  different simulation.  The four simulations are identical except for
  the frictional drag time constant, which is $10^\infty$ (meaning no
  basal drag), $10^7$, $10^6$, and $10^5\rm\,s$ in (a), (b), (c), and
  (d), respectively.  Resolution is C128 (corresponding to a global
  grid of $512\times256$ in longitude and latitude) with 160 vertical
  levels.  The forcing amplitude is $f_{\rm amp}=5\times
  10^{-6}\rm\,K\,s^{-1}$, rotation period is 5 hours, gravity is
  $500\rm\,m\,s^{-2}$, radiative time constant is $10^6\rm\,s$,
  convection decorrelation timescale $\tau_{\rm for}=10^5\rm\,s$,
  forcing wavenumber $n_f=20$, and other parameters are as described
  in the text. }
\label{drag}
\end{figure}

The zonal jet speeds and properties in our models also depend
significantly on the strength of frictional drag imposed at the bottom
of the model.  Figure~\ref{drag} shows four models with, respectively,
$\tau_{\rm drag}$ of $\infty$ (meaning no basal drag), $10^7\rm\,s$,
$10^6\rm\,s$, and $10^5\rm\,s$, from top to bottom.  All other
parameters are identical, including a radiative timescale of
$10^6\rm\,s$ and a forcing amplitude of
$5\times10^{-6}\rm\,K\,s^{-1}$.  Note that in all of these cases, the
frictional drag is applied only at pressures greater than 4 bars near
the bottom of the model; the upper atmosphere remains free of
large-scale drag (though the model's Shapiro filter is applied
everywhere).  The zonal jets are extremely well developed in the
weak-drag models (Figure~\ref{drag}a and b), and well-developed but
weaker in the intermediate-drag model (Figure~\ref{drag}c).  In the
model with strongest drag, however, a zonal jet is prominent only at
the equator.  These results imply that, as with radiative damping,
strong frictional drag can have the effect of confining robust zonal
jets to low latitudes.  Poleward of this critical latitude
($\sim$$45^\circ$ and $\sim$$10^\circ$ in Figures~\ref{drag}c
and d, respectively), prominent flow structures nevertheless occur.
These structure exhibit a preferential northwest-southeast tilt in the
northern hemisphere and southwest-northeast tilt in the southern
hemisphere, again reminiscent of Rossby-wave propagation, similar to
the behavior in our models with strong radiative damping (compare
Figure~\ref{drag}c and d to Figure~\ref{globes}a).

The wind speeds also vary significantly with drag strength; they range
from over $400\rm\,m\,s^{-1}$ in the drag-free case to
$\sim$$80\rm\,m\,s^{-1}$ when drag is strongest (still, this is a
modest variation considering the drag amplitude varies by orders of
magnitude across this range).  Interestingly, weak drag promotes
strong zonality of the zonal jets, in the sense that the jets are
relatively zonally symmetric, and exhibit a zonal-mean zonal wind that
is comparable to or significantly greater than the wind amplitude of
the small-scale eddies that coexist with the jets.  For example, the
weak-drag model in Figure~\ref{drag}a exhibits similar wind speeds to
the stronger-forced but stronger-drag model from Figure~\ref{globes}c;
the zonal jets are much more regular and zonally symmetric in the
former model than the latter.  This result is readily understood
  from the fact that the feedbacks that reorganize the turbulence into
  a banded structure take time to operate; when forcing and damping
  are weak, the timescales on which they modify the flow properties
  are long, and so the dynamical reorganization of the turbulence into
  a banded pattern can occur relatively unimpeded; in contrast, strong
  forcing and damping tend to modify the flow properties on short
  timescales, partially disrupting the dynamical ``zonalization'' process
  as it occurs.

The jet speeds and properties also depend significantly on the forcing
amplitude.  An example can be seen by comparing Figures~\ref{globes}c
and \ref{drag}c.  The former adopts a forcing amplitude ten times
greater than the latter ($5\times10^{-5}$ versus
$5\times10^{-6}\rm\,K\,s^{-1}$); all other parameters are identical
between the simulations, including radiative and drag timescales that
are both $10^6\rm\,s$.  At greater forcing, the typical jet speeds
reach $\pm100\rm\,m\,s^{-1}$, with peak speeds exceeding
$300\rm\,m\,s^{-1}$.  With a forcing ten times weaker, the
characteristic speeds are just tens of $\rm m\,s^{-1}$, and
exceed $50\rm\,m\,s^{-1}$ in just a few locations.  The prominent
low-latitude jets that exist equatorward of $\sim$$45^\circ$ latitude
are also significantly narrower, more closely spaced in latitude,
and appear to exhibit greater zonal symmetry in the weaker-forcing
model.

\subsection{Vorticity and implications for mechanisms of jet pumping}

%%%%%%%%%%%%%%%%%%%%%%%%%%%%%%%%%%%%%%%%%%%%%%%%%%%%%%%%%%%%%%%%%%%%%
% FIGURE 4: globes of zeta+f and zeta, and line plots of zonal means 
% for run 62
%%%%%%%%%%%%%%%%%%%%%%%%%%%%%%%%%%%%%%%%%%%%%%%%%%%%%%%%%%%%%%%%%%%%%
\begin{figure*}
\centering
\begin{minipage}[c]{0.32\textwidth}
\includegraphics[scale=0.4, angle=0]{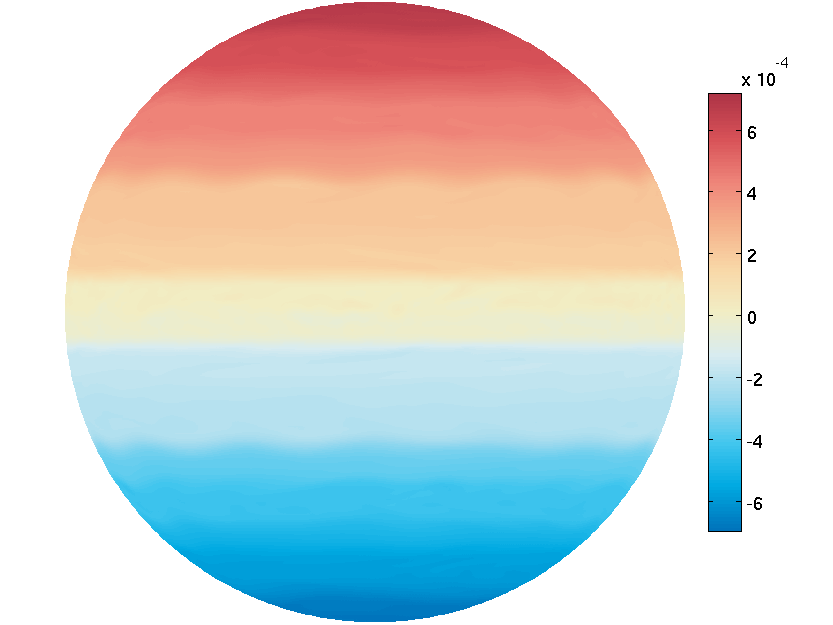}
\end{minipage}
%\put(-140.,240.){(a)}
%\put(-140.,115.){(b)}
%\put(-140.,-10.){(c)}
%\put(-140.,-130.){(d)}
\begin{minipage}[c]{0.3\textwidth}
\includegraphics[scale=0.4, angle=0]{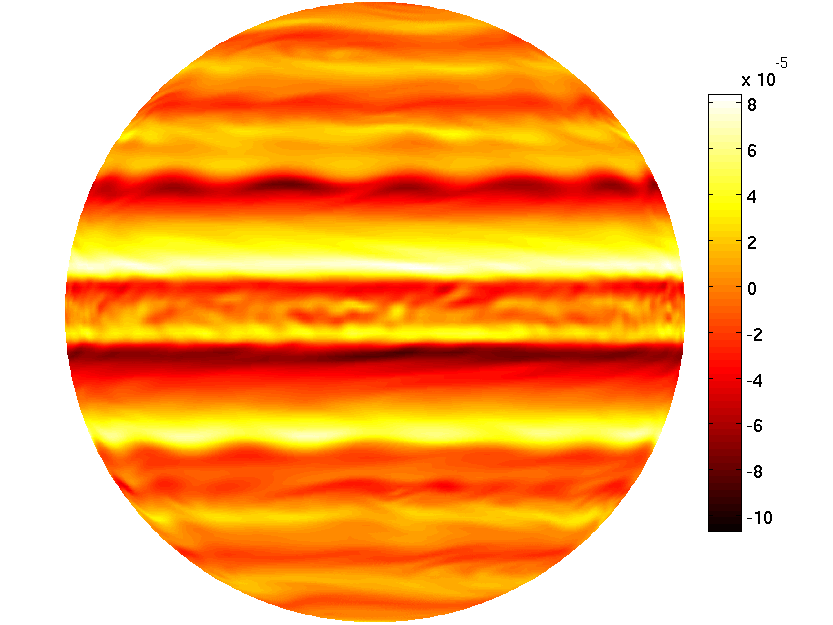}
\end{minipage}
\begin{minipage}[c]{0.3\textwidth}
\includegraphics[scale=0.4, angle=0]{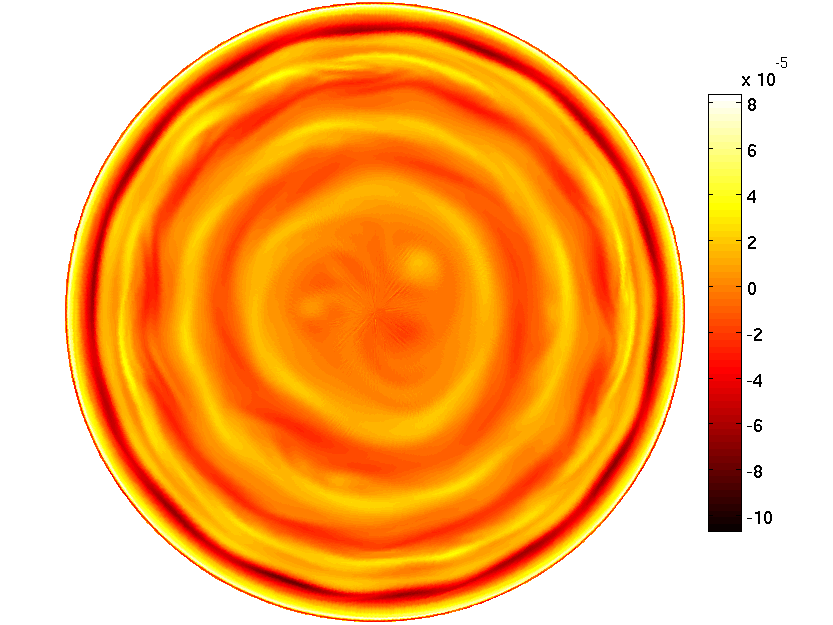}
\end{minipage}
%%%%%%%%%%%%%%%%%%%%%%%%%%%%%%%%%%%%%%%%%%%%%%%%%%%%
\begin{minipage}[c]{0.32\textwidth}
\includegraphics[scale=0.4, angle=0]{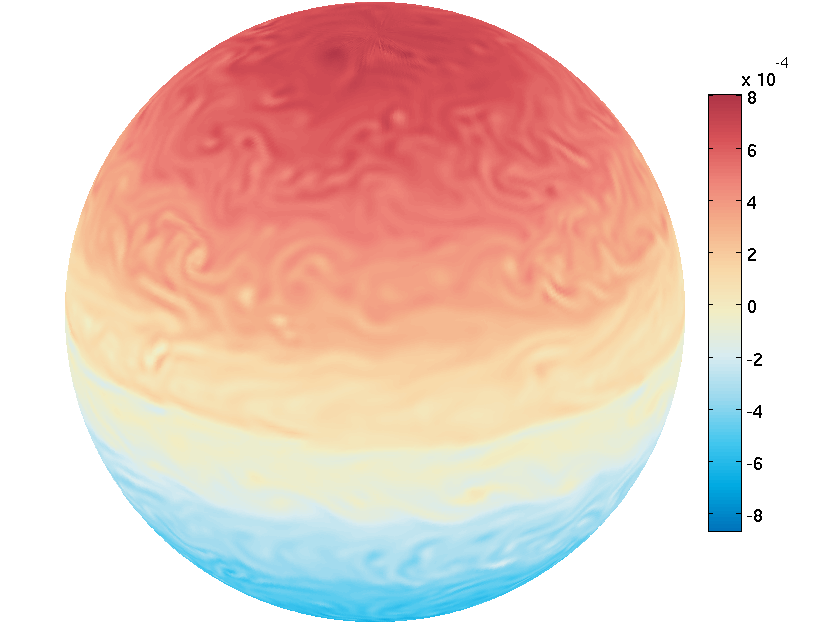}
\end{minipage}
\begin{minipage}[c]{0.3\textwidth}
\includegraphics[scale=0.4, angle=0]{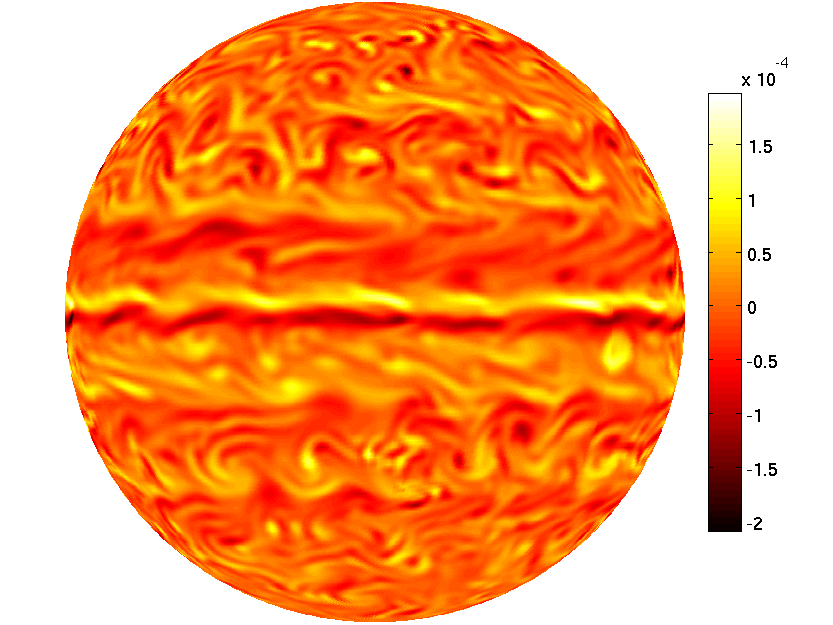}
\end{minipage}
\begin{minipage}[c]{0.3\textwidth}
\includegraphics[scale=0.4, angle=0]{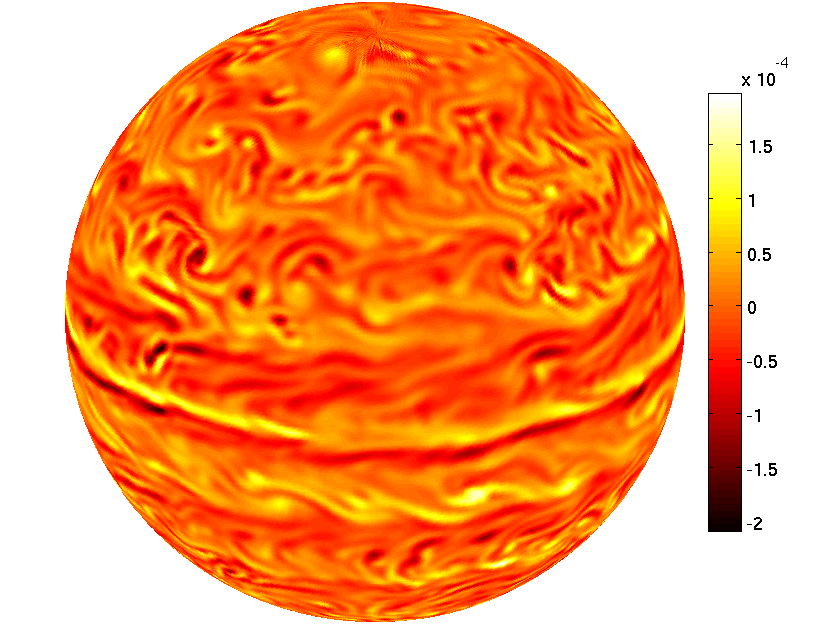}
\end{minipage}
\caption{Vorticity structure illustrating the development of banding
  in two models---a weakly forced, weakly damped model in the top row,
  and a more strongly forced, strongly damped model in the bottom row.
  The left panels shows the absolute vorticity $\zeta+f$ (with
  equatorward and oblique viewing angles in the top and bottom rows,
  respectively).  The middle and righthand panels show the relative
  vorticity $\zeta$.  The middle column shows the view from the
  equatorial plane.  On the right, the top panel shows the view
  looking down over the north pole, and the bottom panel shows an
  oblique view.  Note the organization into strips of nearly constant
  absolute vorticity, and the existence of turbulent filamentary
  structures and edge waves between the strips.  The top row is the
  same model as in Figure~\ref{drag}a, with $\tau_{\rm
    rad}=10^6\rm\,s$, $\tau_{\rm drag}=\infty$, and $f_{\rm
    amp}=5\times 10^{-6}\rm \,K\,s^{-1}$.  The bottom row is the same
  model as in Figure~\ref{globes}d, with $\tau_{\rm rad}=10^7\rm\,s$,
  $\tau_{\rm drag}=10^6\rm\,s$, and $f_{\rm
    amp}=5\times10^{-5}\rm\,K\,s^{-1}$.}
\label{vorticity-globes}
\end{figure*}

We next turn to examine the vorticity, which provides additional
information on jet formation.  Figure~\ref{vorticity-globes} shows the
absolute and relative vorticity for two different models illustrating
the development of banded structure under different conditions.
Absolute vorticity is defined as $\zeta+f$, where $\zeta={\bf k}\cdot
\nabla\times{\bf v}$ is the relative vorticity, and ${\bf k}$ is the
local upward unit vector on the sphere.  Under certain conditions, the
absolute vorticity provides an approximation to the potential
vorticity (PV), which is a materially conserved quantity under
frictionless, adiabatic conditions \citep{vallis-2006}.  In a
stratified atmosphere, the potential vorticity is defined as the
absolute vorticity, $\zeta+f$, divided by a measure of the vertical
spacing between isentropes, which can vary spatially and temporally
due to atmospheric dynamics.  However, in highly stratified regions
with modest horizontal temperature perturbations, the primary
contribution to PV variations are due to the variations in absolute
vorticity rather than thermal structure.

\begin{figure*}
\centering
\includegraphics[scale=0.85, angle=0]{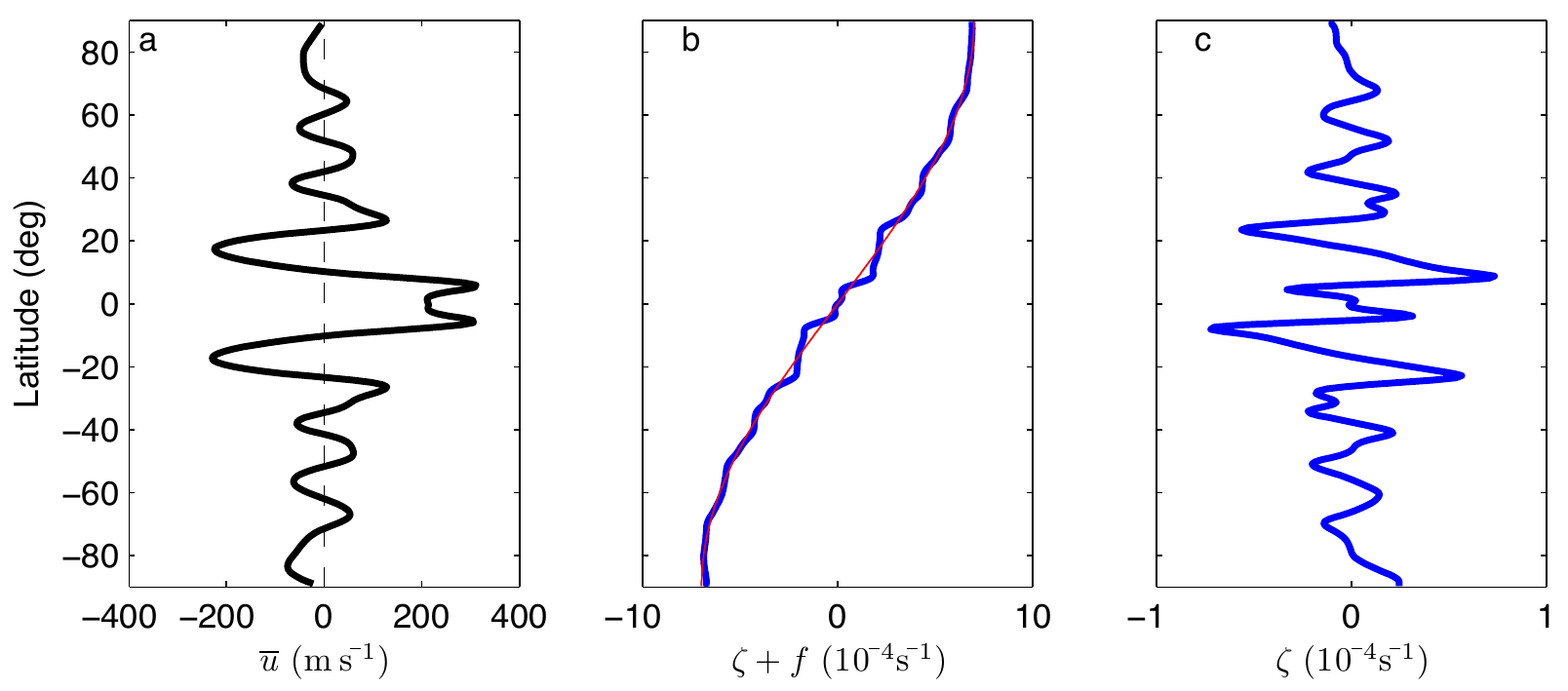}
\caption{Zonal-mean zonal wind, absolute vorticity, and relative
  vorticity at 1.5~bars for a model exhibiting a flow field
  qualitatively similar to Jupiter and Saturn, with strong equatorial
  superrotation and numerous higher-latitude eastward and westward
  jets.  Note that in this case the equatorial superrotation is a
  steady feature.  The absolute vorticity becomes nearly homogenized
  in strips, whose edges occur at the latitudes of the eastward jets.
  Relative vorticity exhibits the characteristic sawtooth pattern one
  expects for a flow exhibiting such absolute-vorticity
  homogenization.  This is the same model as in the top rows of
  Figures~\ref{drag} and \ref{vorticity-globes}.}
\label{zonal-winds-vorticity}
\end{figure*}

In situations when robust zonal jets emerge, our models tend toward a
state where the absolute vorticity, $\zeta+f$, becomes nearly
homogenized within zonal strips, with relatively sharp meridional
gradients in absolute vorticity at the edges of adjacent strips.  This
is clearly illustrated in Figure~\ref{vorticity-globes} (leftmost
panels); the top row represents a weakly forced, weak-friction model
whose resulting jets are nearly zonally symmetric with relatively
modest eddy activity; the bottom row represents a more strongly
forced, strong-friction model with stronger eddies (these are just the
two models from Figures~\ref{globes}d and \ref{drag}a, respectively).
In both cases in Figure~\ref{vorticity-globes}, the homogenization of
absolute vorticity in zonal strips manifests clearly as discrete bands
of differing colors, ranging from blue in the southern hemisphere
(where absolute vorticity is negative), to red in the northern
hemisphere (where it is positive).  At the boundaries between the
strips, the absolute vorticity gradients are relatively sharp.  The
strips are more prominent in the weakly forced, weak-friction model
than in the strongly forced, strong-friction model, a result also seen
in highly idealized one-layer turbulence studies
\citep[e.g.,][]{scott-dritschel-2012}.  The relationship between
vorticity and winds implies that the boundaries between strips
correspond to the latitudes of the eastward zonal jets, whereas the
interior of the strips themselves (where PV approaches a homogenized
state) correspond to to the latitudes of westward jets \citep[see,
  e.g., the discussion in][]{dritschel-mcintyre-2008}. The zonal-mean
zonal winds, absolute vorticity, and relative vorticity for the former
model are illustrated in Figure~\ref{zonal-winds-vorticity}, where the
absolute vorticity staircase is evident, and the correspondence
between the latitudes of the jets and the staircase steps can be seen.
{\tt The tendency to develop PV staircases, analogous to those in our
  models, has also been observed on Jupiter and Saturn
  \citep[e.g.][]{read-etal-2009b}.}

In some models, a robust eastward equatorial jet---atmospheric
superrotation---develops.  If this eastward jet were sharp (with a
single, sharp local maximum of zonal wind at its core, centered at the
equator) then this would be associated with a sharp jump
(discontinuity) in absolute vorticity centered on the equator.  In
cases such as Figure~\ref{zonal-winds-vorticity}, however, the
superrotation exhibits ``cusps,'' wherein the zonal wind within the
superrotating jet maximizes on either side of the equator, with a
shallow local minimum of zonal wind right on the equator.  In this
case, the region between the cusps corresponds to a zonal strip of
(partially) homogenized absolute vorticity, centered on the equator,
as seen in Figure~\ref{vorticity-globes} and
\ref{zonal-winds-vorticity}b.  This strip tends to be narrower in
latitudinal extent than the zonal strips of homogenized absolute
vorticity at higher latitudes.

The boundaries between strips meander more
substantially in longitude in the more strongly forced model, and
numerous turbulent, filamentary structures can be seen in that case
(Figure~\ref{vorticity-globes}, bottom row).  These represent the role
of PV mixing due to breaking Rossby waves, which in some cases can
breach the PV barriers and cause mixing between the strips, leading to
filamentary structures with local minima or maxima of PV within any
given strip.  The relative vorticity structure
(Figure~\ref{vorticity-globes}, middle and righthand panels) likewise
shows a prominent banded structure with superposed eddies.  Both
models show quite strikingly the existence of oscillatory wave
structures at the boundaries between the zonal strips of
nearly-constant absolute vorticity; these meanders can be thought of
as ``edge'' Rossby waves whose restoring force results in large part
from the quasi-discontinuous jump in PV from one strip to the next.

%%%%%%%%%%%%%%%%%%%%%%%%%%%%%%%%%%%%%%%%%%%%%%%%%%%%%%%%%%%%%%%%%%%%%
% FIGURE 5: Zonal-mean wind and vorticity line plots 
% for run 62
%%%%%%%%%%%%%%%%%%%%%%%%%%%%%%%%%%%%%%%%%%%%%%%%%%%%%%%%%%%%%%%%%%%%%

The organization of the flow into strips of nearly constant absolute
vorticity suggests that Rossby wave breaking plays a critical role in
zonal jet formation in these models.  Rossby-wave breaking tends to
occur preferentially in regions of weaker meridional PV gradient, and
the wave breaking causes mixing that tends to decrease the
(zonal-mean) meridional PV gradient still further.  This leads to a
positive feedback: given modest initial variations of meridional PV
gradient as a function of latitude, Rossby-wave breaking will
preferentially occur at the latitudes of weaker PV gradient, and the
lessening of the PV gradient due to the wave breaking in those regions
allows wave breaking to occur even more easily at those latitudes,
lessening the PV gradient still further.  The end state is a flow
where PV is almost totally homogenized in strips, with sharp PV
discontinuities in between (for reviews of the mechanism see
\citealt{dritschel-mcintyre-2008} and \citealt{showman-etal-2013b}).
Research using idealized, high-resolution one-layer models shows that
this idealized limit is achieved most readily under conditions of weak
forcing and damping; in contrast, stronger forcing and damping can
cause PV sources/sinks, and mixing between the strips, that partially
smooths the staircase pattern and prevents complete PV homogenization
within the strips \citep[e.g.][]{scott-dritschel-2012}.  This helps
explain why the staircase pattern is more prominent in the weakly
forced/damped simulation than the strongly forced/damped simulation in
Figure~\ref{vorticity-globes}.  As an initial flow slowly
self-organizes into a banded state due to this mechanism, the defining
relationship between PV and winds will imply the existence of eddy
momentum fluxes that transport eastward momentum out of westward jets
into the cores of eastward jets---helping to generate the jets and
(once they are equilibrated) maintain them against frictional and
radiative damping.

\subsection{Low-latitude QBO-like oscillations}
\label{qbo-like}

%%%%%%%%%%%%%%%%%%%%%%%%%%%%%%%%%%%%%%%%%%%%%%%%
% FIGURE 6: equatorial jet flips sign over time
%%%%%%%%%%%%%%%%%%%%%%%%%%%%%%%%%%%%%%%%%%%%%%%%
\begin{figure*}
\centering
\begin{minipage}[c]{0.32\textwidth}
\includegraphics[scale=0.4, angle=0]{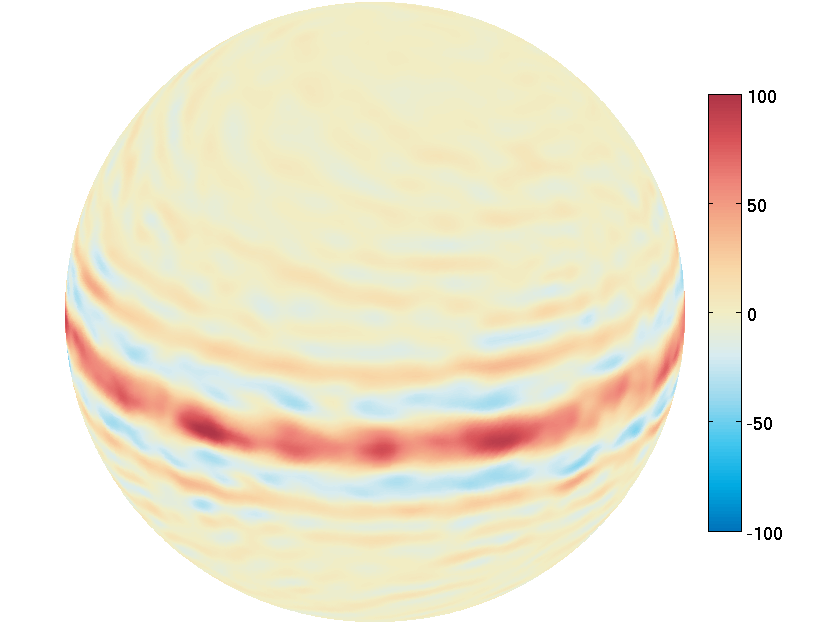}
\put(-140.,110.){(a)}
\end{minipage}
\begin{minipage}[c]{0.32\textwidth}
\includegraphics[scale=0.4, angle=0]{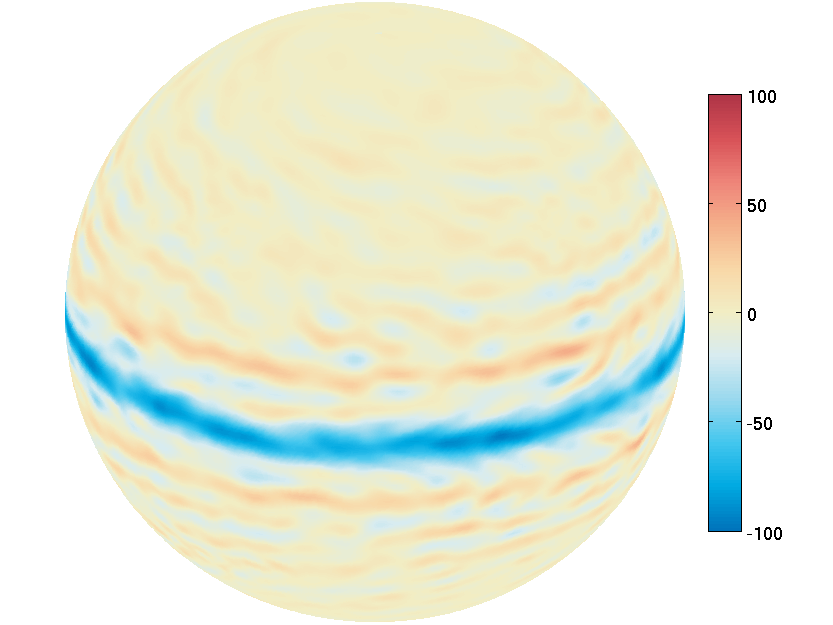}
\put(-140.,110.){(b)}
\end{minipage}
\begin{minipage}[c]{0.32\textwidth}
\includegraphics[scale=0.4, angle=0]{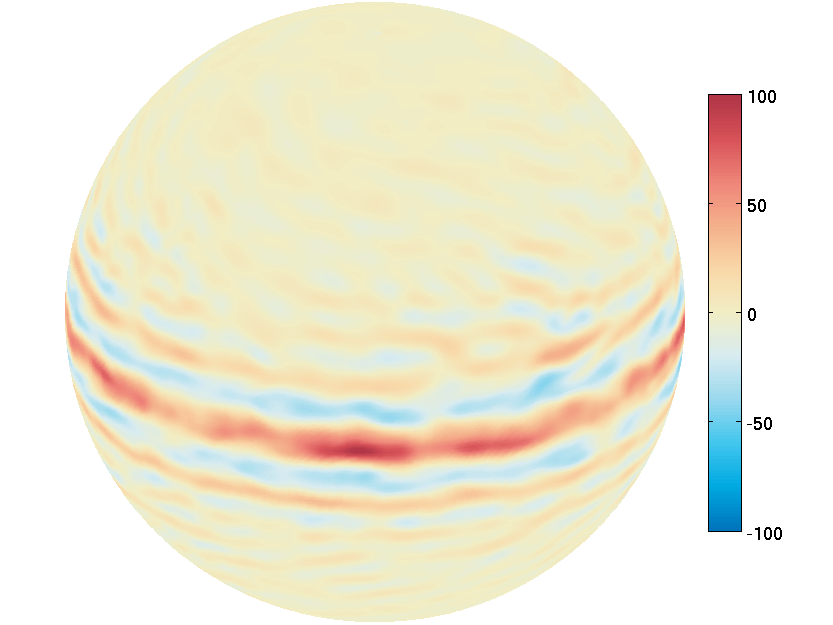}
\put(-140.,110.){(c)}
\end{minipage}
\begin{minipage}[c]{0.32\textwidth}
\includegraphics[scale=0.4, angle=0]{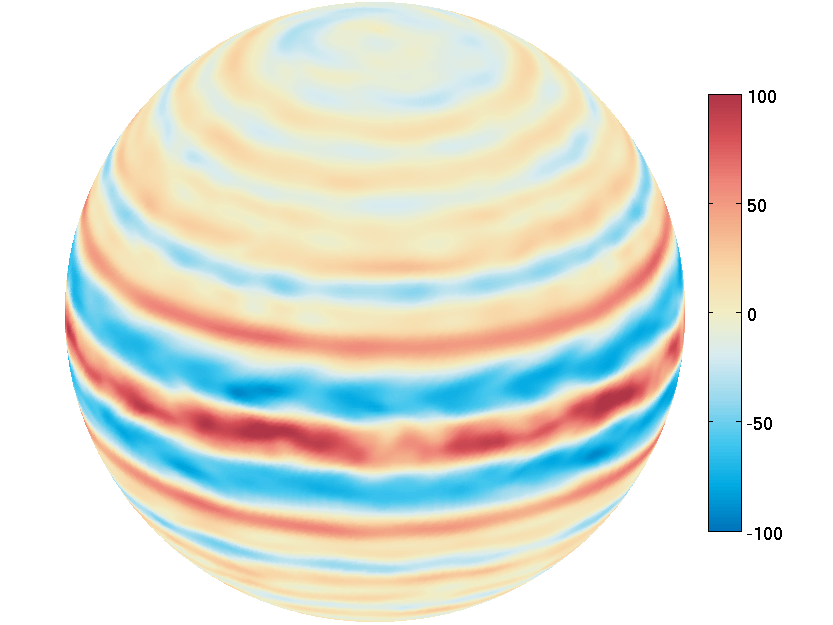}
\put(-140.,110.){(d)}
\end{minipage}
\begin{minipage}[c]{0.32\textwidth}
\includegraphics[scale=0.4, angle=0]{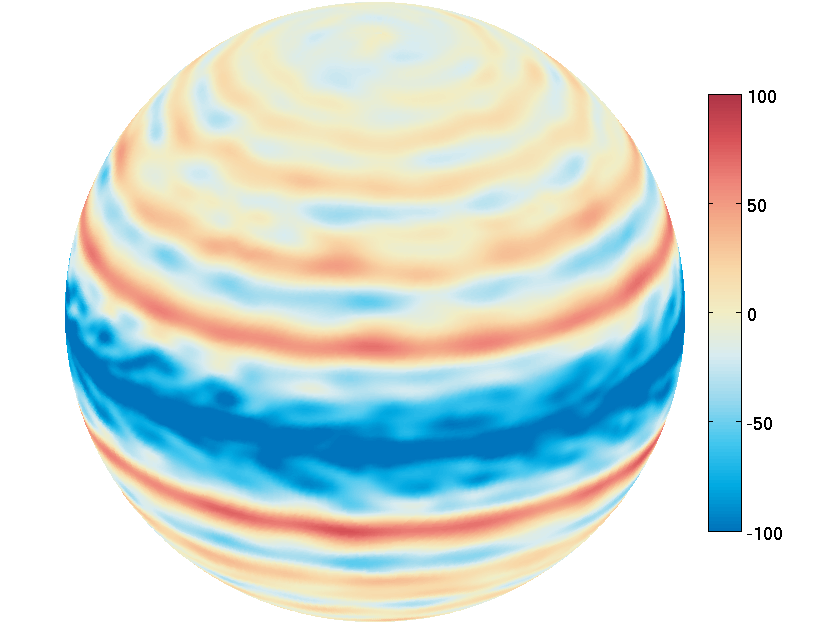}
\put(-140.,110.){(e)}
\end{minipage}
\begin{minipage}[c]{0.32\textwidth}
\includegraphics[scale=0.4, angle=0]{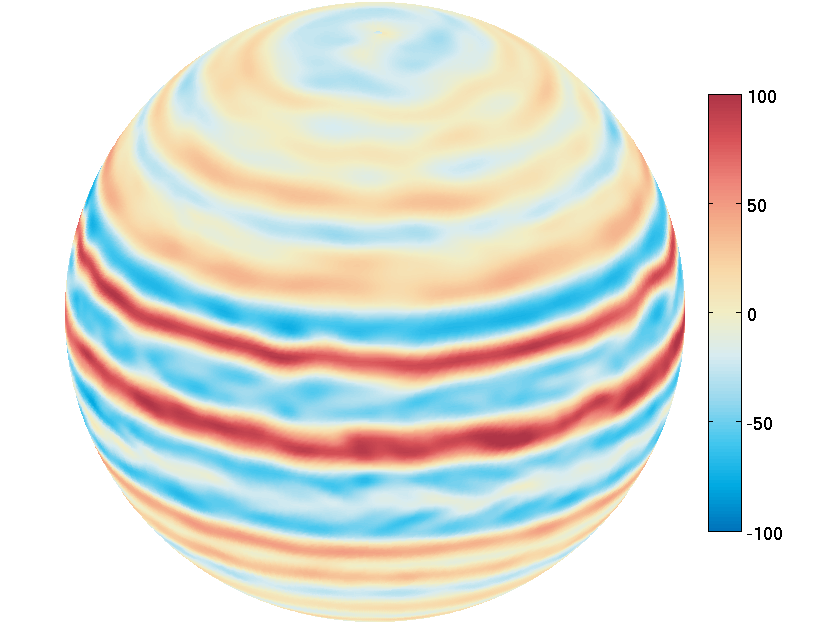}
\put(-140.,110.){(f)}
\end{minipage}
\caption{Zonal winds ($\rm m\,s^{-1}$) over time in two models,
  illustrating a long-period oscillation in the low-latitude wind
  structure.  Each row represents one integration with time increasing
  from left to right.  The top row depicts a simulation with weaker
  jets, due to stronger drag, $\tau_{\rm drag}=10^6\rm\,s$.  The
  snapshots are depicted at 1389, 3588, and 5787 Earth days in (a),
  (b), and (c), respectively (corresponding to 6666, 17,222, and
  27,778 rotation periods).  The bottom row represents a simulation
  with stronger jets due to weaker drag, $\tau_{\rm drag}=10^7\rm\,s$,
  and are shown at 2200, 6600, and 11,000 Earth days in (d), (e), and
  (f), respectively (corresponding to $\sim$10,500, 32,000, and 53,000
  rotation periods).  Note how the equatorial jet shifts from an
  eastward jet (equatorial superrotation), to a westward jet, and back
  again.  The structure is shown at a pressure of 0.2 bars, close to a
  typical IR photosphere pressure for a brown dwarf. Resolution is
  C128 (corresponding to a global grid of $512\times256$ in longitude
  and latitude) with 160 vertical levels.  In both models, the forcing
  amplitude is $f_{\rm amp}=5\times 10^{-6}\rm\,K\,s^{-1}$, rotation
  period is 5 hours, gravity is $500\rm\,m\,s^{-2}$, radiative time
  constant is $10^6\rm\,s$, forcing wavenumber $n_f=20$, convective
  decorrelation timescale is $\tau_{\rm for}=10^5\rm\,s$, and other
  parameters are as described in the text.}
  %[NOTE: second row, run 63, is 2200, 6600, and 11,000 Earth
  %days, which are output steps 1900000, 5700000, and 9500000.]
\label{qbo1}
\end{figure*}

Our simulations commonly show the emergence of a long-term,
multi-annual oscillation in the low-latitude zonal-jet and temperature
structure, analogous to the terrestrial QBO, the Jovian QQO, and the
Saturnian SAO.  Figure~\ref{qbo1} shows examples of this oscillation
for two simulations, one (top row) with stronger basal drag
($\tau_{\rm drag}=10^6\rm\,s$) and the other (bottom row) with weaker
basal drag ($\tau_{\rm drag}=10^7\rm\,s$); both simulations adopt a
radiative time constant of $10^6\rm\,s$.  The figure shows the zonal
wind over the globe on a constant-pressure surface of 0.2 bars at
three snapshots at different times, increasing from left to right
within each row.  The times of these snapshots are not at regular
intervals but rather are chosen to illustrate the extrema of the
oscillation over one cycle.  At the beginning of the depicted
oscillation cycle, the equatorial jet is eastward, with a speed of
about $100\rm\,m\,s^{-1}$ (left panels, Figure~\ref{qbo1}a and d).  At
intermediate times, thousands of days later, the equatorial jet has
reversed direction entirely at 0.2 bars, and now flows westward, with
a zonal wind speed of about $-100\rm\,m\,s^{-1}$ (middle panels,
Figure~\ref{qbo1}b and e).  At even later times, however, the
equatorial jet has flipped back to its eastward configuration (right
panels, Figure~\ref{qbo1}c and f), and the overall structure resembles
that in Figure~\ref{qbo1}a and d.  The total oscillation period is
approximately 4400 and $\sim$7000 days (12 and 19
years) in these two models, respectively.

%%%%%%%%%%%%%%%%%%%%%%%%%%%%%%%%%%%%%%%%%%%%%%%%%%%%%%%%%%%%%%
% FIGURE 7: QBO zonal wind meridional structure versus time
%%%%%%%%%%%%%%%%%%%%%%%%%%%%%%%%%%%%%%%%%%%%%%%%%%%%%%%%%%%%%%
\begin{figure*}
\centering
%\begin{minipage}[c]{0.32\textwidth}
\includegraphics[scale=0.6, angle=0]{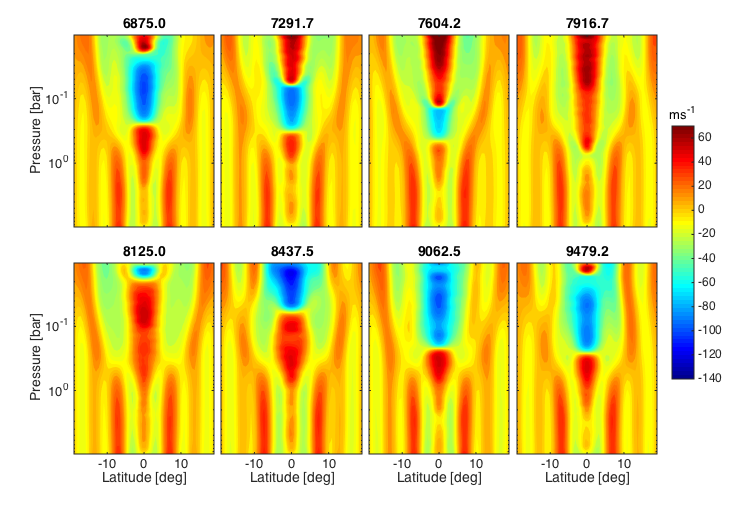}
%\end{minipage}
\caption{Time evolution of zonal-mean zonal wind ($\rm m\,s^{-1}$)
  versus latitude and pressure, showing the occurrence of a
  multi-annual variation of the structure of the low-latitude zonal
  jets. Each panel depicts the structure at a specific time, shown in
  Earth days above the panel.  This is the same model as shown in the
  top row of Figure~\ref{qbo1}.}
\label{qbo-merid-u}
\end{figure*}

In models where this oscillation occurs, the low-latitude regions
generally contain {\it both} an eastward zonal jet {\it and} a
westward zonal jet at any given time---in a vertically stacked
configuration, with one jet overlying the other---and the entire
structure migrates downward over time.  This is illustrated for a
particular case in Figure~\ref{qbo-merid-u}, which shows the vertical
and latitudinal structure of the zonal-mean zonal wind as a function
of latitude and pressure. In the simulation shown in
Figure~\ref{qbo-merid-u}, at $\sim$6875 Earth days, there exists an
eastward equatorial jet at $p\lesssim 0.03$~bars, a westward
equatorial jet from 0.05--0.5 bars, and an eastward equatorial jet
deeper than 0.5 bars.  Over time, the eastward jet near the top of the
domain deepens vertically, the underlying westward jet shrinks in
vertical extent, and the transition between them slowly shifts
downward.  Eventually, at $\sim$8000~days, a new westward equatorial
jet emerges at the top of the domain, deepening gradually in vertical
extent.  All of this means that, at some times during the oscillation,
the structure at $p\lesssim 1$~bar comprises an eastward jet on top of
a westward jet, but at other times it comprises a westward jet on top
of an eastward jet.  At some times there are three or even four
stacked jets, of alternating sign.  Figure~\ref{qbo-merid-u} makes
clear that, as these jets migrate downward, any given isobar
successively experiences either eastward or westward flow, alternating
in time---as seen previously in Figure~\ref{qbo1}.

\begin{figure*}
\centering
%\begin{minipage}[c]{0.32\textwidth}
\includegraphics[scale=0.6, angle=0]{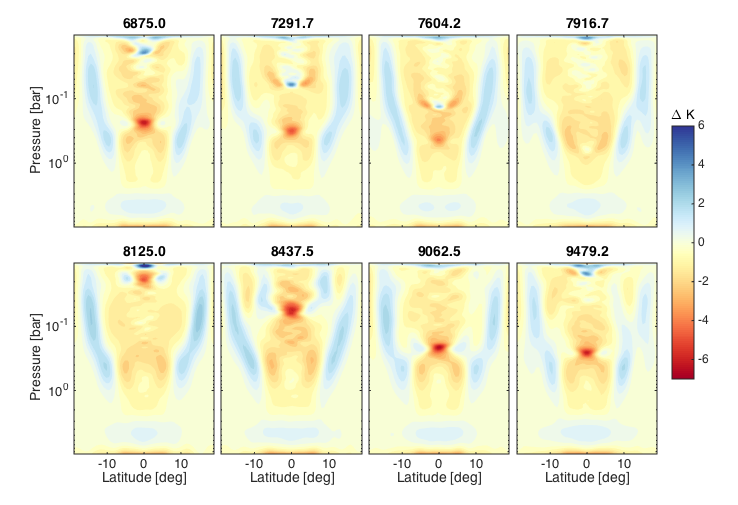}
%\end{minipage}
\caption{Time evolution of zonal-mean temperature anomaly (K) versus
  latitude and pressure, showing the occurrence of a multi-annual
  variation of the structure. Each panel depicts the structure at a
  specific time, listed above each panel in days.  This is the same
  model shown in the top row of Figure~\ref{qbo1}.}
\label{qbo-merid-temp}
\end{figure*}

The oscillation affects not only wind but also the thermal structure.
Perturbations of the temperature are correlated with the wind
structure and also migrate downward over time with the same period.
This is shown in Figure~\ref{qbo-merid-temp}, which depicts the
temperature anomalies (defined here as deviations of the local,
zonal-mean temperature from the reference temperature profile, $T_{\rm
  eq}(p)$) versus latitude and pressure for the same simulation as in
Figure~\ref{qbo-merid-u}.  It can be seen that the typical temperature
perturbations reach about $\sim$$10\rm\,K$, and that they migrate
downward over time.  {\tt Figure~\ref{qbo-merid-temp} shows that
  the temperature extrema at the equator tend to be anticorrelated to
  those occurring off the equator, such that (for example) locally
  warm regions at the equator are flanked by locally cooler regions
  immediately north and south.}  At the fast rotation rates of brown
dwarfs, the background thermal structure is close to geostrophic
balance, which implies that the thermal-wind equation approximately
holds \citep[e.g.,][p.~82]{holton-hakim-2013}:
\begin{equation}
{\partial u\over\partial \log p} = {R\over f} \left({\partial T
\over \partial y}\right)_p
\label{thermal-wind}
\end{equation}
where $u$ is zonal wind, $y$ is northward distance on the sphere, $R$
is the specific gas constant, and the horizontal derivative is taken
at constant pressure.  This equation only holds where the Rossby
number is small, implying poleward of several degrees on a brown
dwarf.  But if $u$ and $T$ in Equation~(\ref{thermal-wind}) are taken
as their zonal means, then the equation holds considerably closer to
the equator, and can thus provide
guidance. Equation~(\ref{thermal-wind}) implies that regions with
significant vertical shear of the zonal wind should likewise exhibit
significant latitudinal temperature gradients, and this is in fact
what we see in Figures~\ref{qbo-merid-u}--\ref{qbo-merid-temp}.

\begin{figure}
\centering
%\begin{minipage}[c]{0.32\textwidth}
\includegraphics[scale=0.1, angle=0]{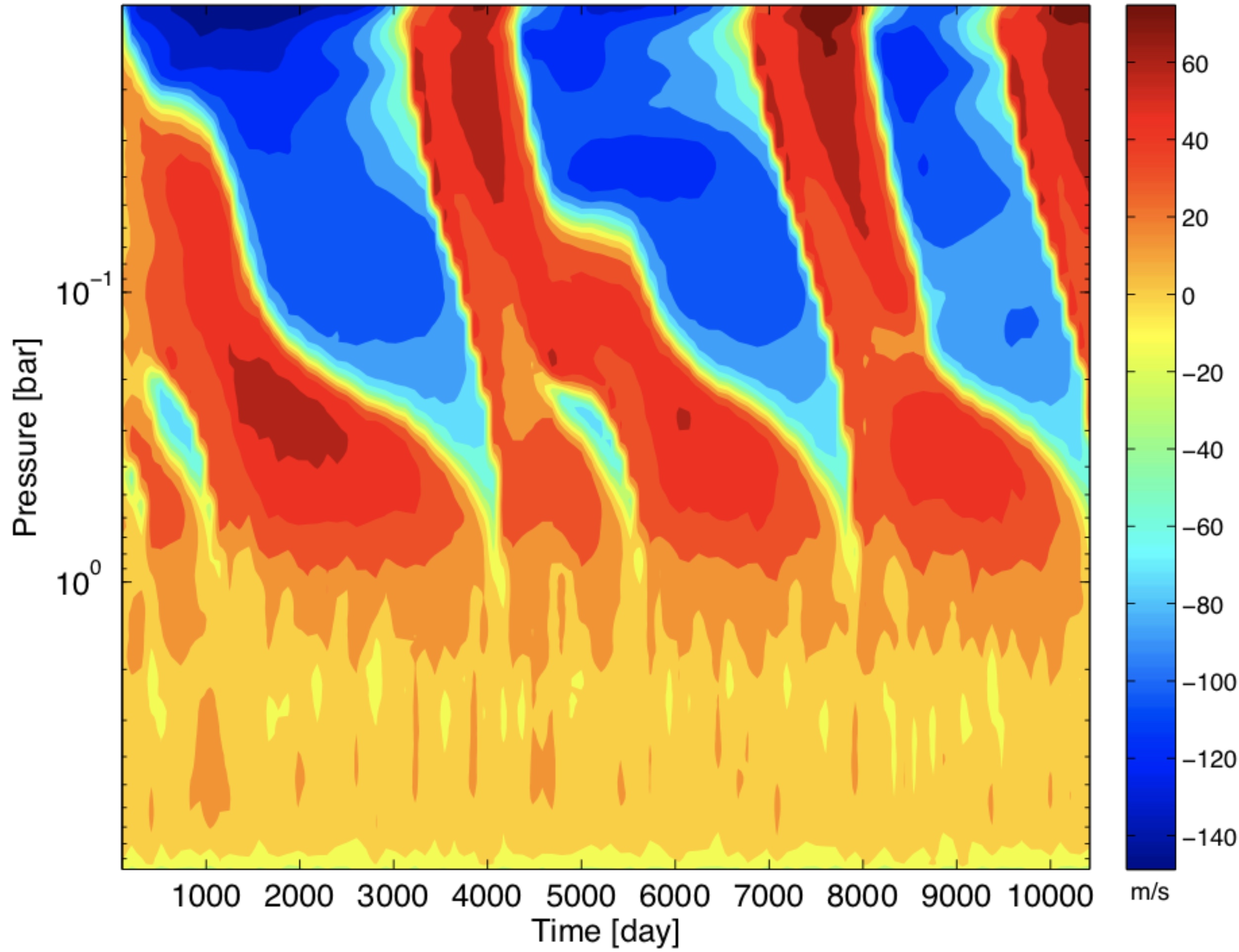}
%\end{minipage}
\caption{Zonal-mean zonal wind at the equator versus pressure and time
  (in Earth days) from a model with high horizontal resolution (C128)
  and 160 vertical levels.}
\label{qbo3}
\end{figure}

Figure~\ref{qbo3} provides another view of the vertical structure,
showing how the zonal-mean zonal wind, as a function of height,
evolves over time.  The structure of the vertically stacked eastward
and westward jets, and their downward evolution over time, is
striking.  The period is about $\sim$4400 days in this model.  Despite
the periodicity, the behavior at any given pressure is not sinusoidal
throughout the cycle but exhibits a more complex structure.  In
particular, at low pressures, a significant fraction of the cycle is
occupied by westward phase, with the eastward phase representing
significantly less than 50\% of the oscillation period---but the
reverse is true at greater pressures (with a transition around
$\sim$0.2 bars).  However, the detailed nature of the structure is
sensitive to parameter values and differs between models.

%%%%%%%%%%%%%%%%%%%%%%%%%%%%%%%%%%%%%%%%%%%%%%%%%%%%%%%%%%%%%%%
% FIGURE 8: QBO temperature meridional structure versus time
%%%%%%%%%%%%%%%%%%%%%%%%%%%%%%%%%%%%%%%%%%%%%%%%%%%%%%%%%%%%%%%

%%%%%%%%%%%%%%%%%%%%%%%%%%%%%%%%%%%%%%%%%%%%%%%%%%%%%%%%%
% FIGURE 9: equatorial slice of QBO structure over time
%%%%%%%%%%%%%%%%%%%%%%%%%%%%%%%%%%%%%%%%%%%%%%%%%%%%%%%%%

As foreshadowed earlier, the phenomenon depicted in
Figures~\ref{qbo1}--\ref{qbo3} is extremely similar to the well-known
oscillation in the Earth's stratosphere called the Quasi-Biennial
Oscillation (QBO).  The QBO likewise involves vertically stacked
eastward and westward stratospheric, equatorial zonal jets---and
associated temperature anomalies---which migrate downward with a
period of approximately 28 months (for reviews, see
\citealt{baldwin-etal-2001} or
\citealt[][pp.~313-331]{andrews-etal-1987}).  A similar oscillation
has been detected on Jupiter with a period of $\sim$4 years and is
called the Quasi Quadrennial Oscillation or QQO
\citep{leovy-etal-1991, friedson-1999, simon-miller-etal-2006}, and on
Saturn with a period of $\sim$15 years, called the Saturn Semi-Annual
Oscillation or SAO (\citealt{orton-etal-2008};
\citealt{fouchet-etal-2008}; \citealt{guerlet-etal-2011}; 
\citealt{guerlet-etal-2018}; for a review
see \citealt{showman-etal-2019}).  A wide range of theoretical and
modeling studies have been conducted to identify the dynamical
mechanisms of the QBO, and similar dynamics are believed to control
the behavior of the Jovian QQO and Saturnian SAO.

\begin{figure}
\centering
%\begin{minipage}[c]{0.32\textwidth}
\includegraphics[scale=0.5, angle=0]{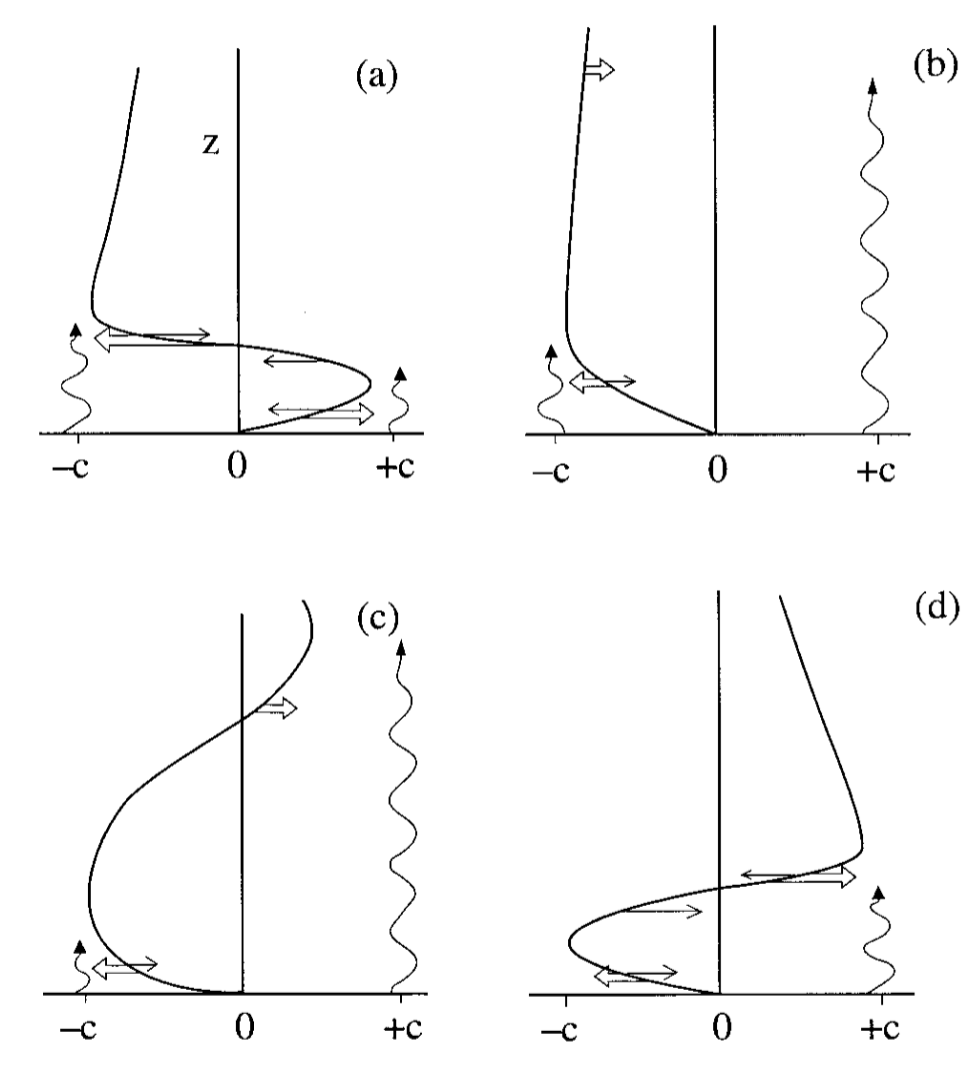}
%\end{minipage}
\caption{Schematic of the QBO mechanism.  This depicts half of a
  cycle, from the idealized model by \citet{plumb-1977}.  The curvy
  solid line represents the background zonal-mean zonal flow.
  \citet{plumb-1977} imagined eastward and westward propagating waves,
  depicted by $+c$ and $-c$ respectively.  Preferential absorption of
  these waves leads to zonal accelerations that are shown by the thick
  double arrows.  The peaks in these accelerations occur below the
  peaks in the jets themselves, causing the jet structure to migrate
  down over time.  From \citet{baldwin-etal-2001}, after
  \citet{plumb-1984}.}
\label{qbo-schematic}
\end{figure}

It is now well accepted that these QBO-type oscillations result from a
wave-mean-flow interaction involving the upward propagation of
equatorial waves generated in the lower atmosphere, and their damping
and absorption in the middle and upper atmosphere
\citep{lindzen-holton-1968, holton-lindzen-1972}.  The fundamental
mechanism is illustrated in Figure~\ref{qbo-schematic}, based on a
simple model of the phenomenon due to \citet{plumb-1977}.  Imagine the
presence of upwardly propagating waves with both eastward and westward
phase speeds.  These waves are associated with a vertical flux of
zonal momentum.  In the absence of damping, such waves would propagate
upward without causing any significant alteration to the background
flow.  If they are damped or absorbed, however, the resulting decrease
in wave amplitude causes a divergence in the upward zonal-momentum
flux, implying that the waves induce a zonal acceleration of the
background flow.  This allows for the emergence of zonal flows in
response.

%%%%%%%%%%%%%%%%%%%%%%%%%%%%%%%%%%%%%%%%%
% FIGURE 10: Schematic of QBO mechanism
%%%%%%%%%%%%%%%%%%%%%%%%%%%%%%%%%%%%%%%%%

Crucially, the presence of a background zonal flow spatially organizes
the wave absorption, allowing the emergence of a coherent zonal-jet
structure that evolves in time.  Generally, absorption of an eastward-
(westward) propagating wave causes an eastward (westward) acceleration
of the background flow.  Simultaneous damping of both eastward- and
westward-propagating waves in a single location leads to counteracting
accelerations that partly cancel out.  But in the presence of zonal
flow, eastward and westward-propagating waves damp at differing rates,
causing a preferential net acceleration that is either east or west.
For example, in the presence of a weak eastward zonal flow,
eastward-propagating waves exhibit reduced vertical group velocity,
allowing them to be more easily damped than westward propagating
waves---thereby promoting a net eastward acceleration.
\citet{plumb-1977} showed that, because of this mechanism, a state
with initially zero zonal-mean zonal flow is unstable and will, over
time, develop vertically stacked eastward and westward jets.  Once
these jets are sufficiently strong, upward propagating waves can
encounter critical levels on the lower flanks of these jets: layers
where the background zonal jet speed equals the zonal phase speed of
the upwardly propagating wave.  Waves that encounter critical levels
tend to be absorbed at an altitude close to the critical level,
causing a zonal acceleration of the same sign as the background wind.
Because such critical levels are located on the bottom flank of the
jet, the peak zonal acceleration will occur {\it below} the maximum of
eastward zonal wind, and this causes the jet to migrate downward over time.
As shown in Figure~\ref{qbo-schematic}a, an eastward jet will absorb
eastward-propagating waves but tends to be transparent to
westward-propagating waves, which propagate through the eastward jet
until they reach an overlying westard jet, where they can be
preferentially damped or absorbed.  In this way, vertically stacked
jets of both eastward and westward phases propagate downward over time
(Figure~\ref{qbo-schematic}b through d).

In Earth's atmosphere, the waves responsible for driving the QBO
result primarily from tropospheric convection and exhibit a range of
length scales and periods.  The dominant wave types include
large-scale modes such as the eastward propagating Kelvin wave and
westward propagating mixed Rossby-gravity and Rossby waves, as well as
smaller-scale inertia-gravity and gravity waves of both eastward and
westward phases \citep[e.g.][]{baldwin-etal-2001}.  The equatorial
confinement of the main jet structure within the QBO arises from the
fact that many of the waves responsible for driving it are
equatorially trapped, but also the fact that in the extratropics,
wave-induced accelerations are to large degree cancelled out by
Coriolis accelerations due to a mean-meridional circulation triggered
by the wave forcing, whereas in the tropics, where the Coriolis force
is weaker, a greater fraction of the wave torques are able to cause a
net acceleration of the zonal wind \citep{baldwin-etal-2001}.  In our
models, the convective forcing has sufficiently low wavenumber
($\lesssim40$) that it will primarily trigger the large-scale class of
wave modes, and we have intentionally excluded any subgrid-scale
parameterizations of small-scale, numerically unresolved gravity
waves.  The resolved waves that drive the QBO-like oscillations in our
models are therefore large-scale waves such as the Kelvin waves,
Rossby waves, and mixed Rossby-gravity waves.  {\tt We provide
a detailed analysis of the wave modes and how they influence the 
mean flow in Section~\ref{diag}.}

%To investigate the above picture, we performed an analysis of the
%zonal-mean zonal flow acceleration (eddy momentum convergence) caused
%by the waves.  At individual snapshots, the results are noisy due to
%the stochastic nature of the wave forcing.  In cases when QBO-like
%oscillations occur, however, the time averaged results confirm that
%the peak eastward acceleration caused by the waves occurs on the lower
%flank of the equatorial eastward jets, whereas the peak westward
%acceleration caused by the waves occurs on the lower flank of the
%equatorial westward jets.  As mentioned above, this must occur if the
%jets are to migrate downward over time, and it is consistent with the
%above dynamical picture of QBO-like dynamics.  We defer a detailed
%quantitative diagnosis of the specific wave modes for future work.

In most of our simulations exhibiting QBO-type phenomena, the
oscillation period is between $\sim$1000 and 10,000 Earth days,
depending on parameters.  One expects that the characteristic
timescale, wavenumber, and amplitude of the convective forcing will
control the population of upwardly propagating waves, which will
strongly shape the oscillation period of the QBO-type phenomenon
(e.g., stronger convective forcing amplitude could lead to stronger
wave amplitudes and shorter QBO oscillation periods).  Similarly,
radiative and frictional damping can affect both the jet
structure---modifying the speed and meridional structure of the
jets---and the convectively generated waves themselves, and thus
should likewise influence the QBO oscillation period, as well as
whether the oscillation occurs at all.  Although we have run a variety
of simulations exhibiting QBO-type oscillations, the computational
expense of these high-resolution models makes it difficult to
systematically characterize how all these parameters quantitatively
influence the QBO properties, a task we leave to the future.

We also note the importance of vertical resolution, as the
convectively generated waves that drive the QBO have short vertical
wavelengths {\tt near their critical levels}, which requires high
vertical resolution to capture adequately.  Terrestrial GCMs with
coarse vertical resolution generally fail to generate a QBO, and only
once vertical resolution exceeds 40--50 levels does a realistic
QBO-like oscillation emerge, albeit not necessarily with the detailed
properties of the actual terrestrial QBO
\citep[e.g.][]{takahashi-1996,takahashi-1999}.  Interestingly, in the
parameter regimes where the QBO-type oscillation can occur, our models
with 40, 80, and 160 vertical levels all exhibit such QBO-type
oscillations, but the period and other details of the oscillation
differ between these models.  {\tt For example, in one set of
  otherwise identical models, we found that the oscillation period
  changed from 4.1 to 5.7 to 7.1 years as the resolution was increased
  from 40 to 80 to 160 vertical levels.  The sensitivity of the
  oscillation amplitude appears to be much more modest, however; the
  eastward and westward phases of the oscillation in these models have
  peak zonal wind speeds of $\sim$$70\rm m\,s^{-1}$ and
  $-120\rm\,m\,s^{-1}$, essentially independent of resolution, across
  this entire set.  The vertical-resolution sensitivity of the
  oscillation period may result in part from overly coarse resolution
  of the critical-level wave-absorption regions, but also overly
  coarse resolution of the parameterization of convective
  perturbations to the RCB at the bottom of the model, when too few
  levels are used.  The basal forcing region is only a few levels
  thick in the lowest-resolution models; it is possible that changes
  to the vertical resolution are influencing the effective model
  forcing amplitude, which in turn would influence the resulting wave
  amplitudes and momentum fluxes driving the QBO-like oscillation.}
Regardless of the details, these results demonstrate that vertical
resolutions of 40 and 80 levels are insufficient, and even for 160
levels we cannot as yet guarantee full numerical convergence.  (In
contrast, the properties of the off-equatorial jets---poleward of
$\sim$$10^\circ$ latitude---appear to be largely similar over the full
range of vertical resolutions we explored, from 40 to 160 levels.)
Future detailed work will be necessary to explore the resolution
sensitivity at extreme vertical resolutions.

The QBO-like oscillations captured here are an example of an emergent
property that results from the nonlinearity of the system.  In the
simulations in Figures~\ref{qbo1}--\ref{qbo3}, all the explicit
timescales in the system---the radiative timescale, drag timescale,
and convective forcing timescale---are short, of order tens of days or
less.  Yet the oscillation that emerges has timescales of thousands of
days.  This is a fascinating example of long-term ``memory'' exhibited
by the system due to its internal dynamics, even when all explicit
forcing and damping timescales are shorter by orders of magnitude.
This emergent behavior differs strongly from the behavior exhibited by
a linear system, where one expects the response to have identical
frequencies as the forcing.

As a result, it is natural to expect that even though brown dwarfs
have short radiative time constants of $\sim$$10^6\rm\,s$ or less, and
presumably short convective timescales as well, that the atmospheric
circulations on brown dwarfs could nevertheless exhibit
ultra-long-time variability that would manifest in long-term
monitoring surveys.  This provides motivation for continuing to
monitor specific brown dwarfs over periods of years, to search for
multi-annual variability, as they may well have variability on such
long timescales in addition to the shorter-term variability that is
typically the emphasis of current surveys.

\subsection{Waves and role in driving the QBO-like oscillation}
\label{diag}

Here we demonstrate the role of waves in driving the QBO-like
oscillations, characterize the wave population, and investigate the
wave modes specifically responsible for driving the oscillations in
our models.  For concreteness, we focus on one particular model
experiment, namely the one shown in the top row of Figure~\ref{qbo1},
which is also the same one depicted in
Figures~\ref{qbo-merid-u}--\ref{qbo3}.

%%%%%%%%%%%%%%%%%%%%%%%%%%%%%%%%%%%%%%%%%%%%%%%%%%%%%%%%%%%%%%%%%%%%%%%%
% FIGURE 11: Eddy acceleration overlain on zonal wind (meridional slice)
%%%%%%%%%%%%%%%%%%%%%%%%%%%%%%%%%%%%%%%%%%%%%%%%%%%%%%%%%%%%%%%%%%%%%%%%
\begin{figure*}
\centering
\includegraphics[scale=0.4, angle=0]{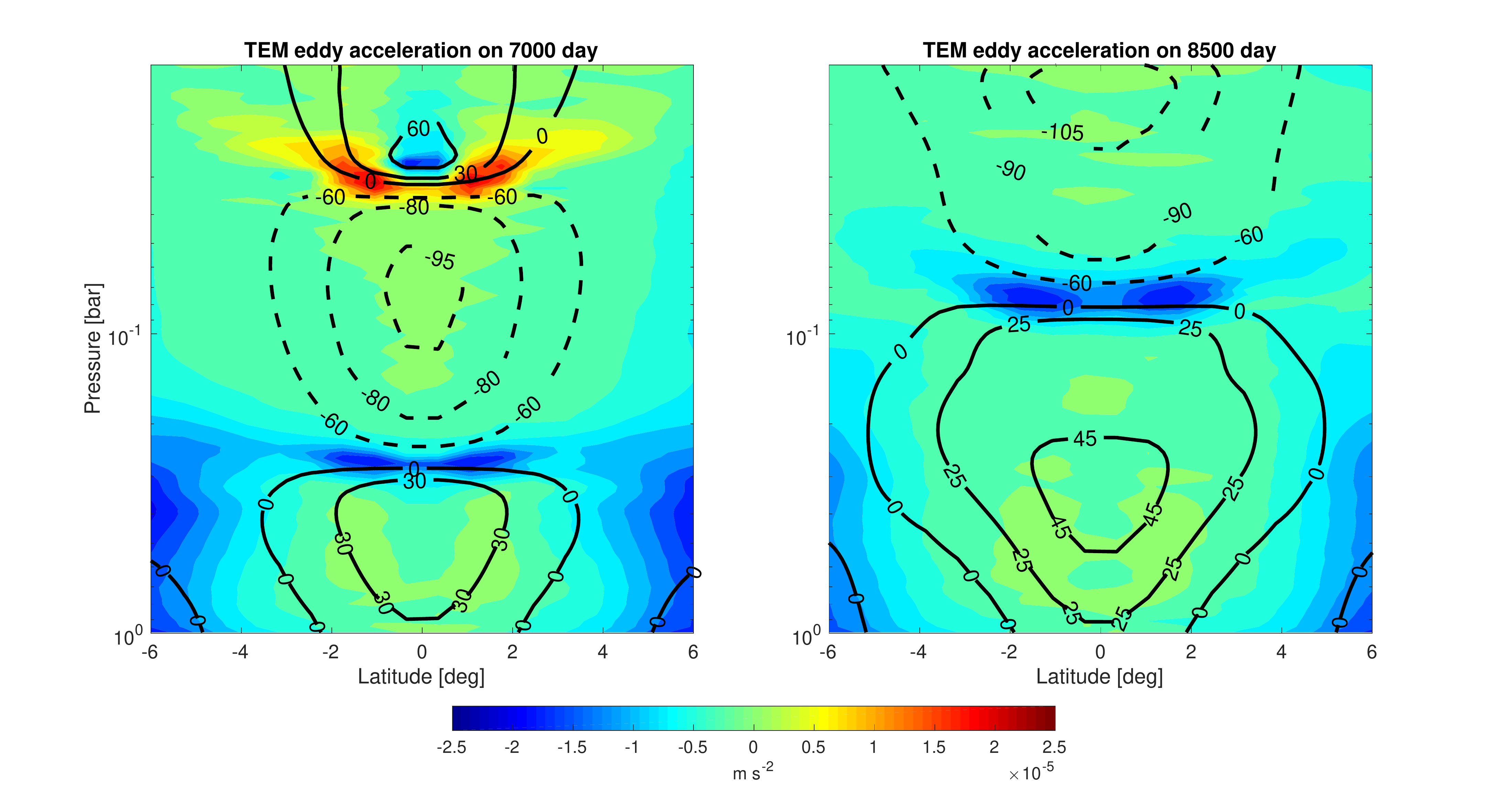}
%\begin{minipage}[c]{0.5\textwidth}
%\includegraphics[scale=0.29, angle=0]{eddyaccel+zonalu-7000day.pdf}
%\put(-235.,205.){(a)}
%\end{minipage}
%\begin{minipage}[c]{0.45\textwidth}
%\includegraphics[scale=0.4, angle=0]{eddyaccel+zonalu-8500day.pdf}
%\put(-235.,205.){(b)}
%\end{minipage}
\caption{Zonal-mean wave-induced acceleration (colorscale,
  $\rm\,m\,s^{-2}$) versus latitude and pressure at two times during
  the cycle of a QBO-like oscillation, at 7000 and 8500 days on the left
  and right, respectively.  Zonal-mean zonal wind is depicted in thick
  black contours (labeled in $\rm m\,s^{-1}$); eastward is positive
  (solid) and westward is negative (dashed).  Note that the
  wave-induced accelerations of a given sign occur primarily near the
  base of the zonal jet of that same sign---specifically, at both
  times, the waves induce westward acceleration near the base of the
  westward jet and eastward acceleration near the base of the eastward
  jet.  This has the effect of causing the jet to migrate downward
  over time and indicates that the QBO-like oscillation is wave
  driven.  This is the same model as shown in the top row of
  Figure~\ref{qbo1}.}
\label{waves1}
\end{figure*}

We first determine the forcing of the zonal-mean zonal flow
caused by waves, and show that it has the structure necessary to
drive a QBO-like oscillation.
The role of waves and eddies in driving the mean flow can be quantified
using the framework of the ``Transformed Eulerian Mean'' or TEM
equations, which provide a representation for the evolution of the
zonal-mean zonal wind and meridional circulation in response to
waves, radiation, and other effects
\citep[e.g.][]{andrews-etal-1983, andrews-etal-1987}.  The TEM
zonal-momentum equation in pressure coordinates is
\begin{eqnarray}
\nonumber
{\partial\overline{u}\over\partial t}=\left[-\left({1\over a\cos\phi}
{\partial (\overline{u}\cos\phi)\over \partial \phi} - f\right)\overline{v}^*
- {\partial \overline{u}\over\partial p}\overline{\omega}^*\right]\\
+ {1\over a\cos\phi}\nabla\cdot {\bf F} + \overline{X}.
\label{tem}
\end{eqnarray}
Here, all variables have been decomposed into their zonal means,
denoted by an overbar, and the deviations therefrom, denoted by a
prime, such that for any quantity $A$, we have $A=\overline{A}+A'$.
So, $\overline{u}$ is the zonal-mean zonal wind, and $\overline{X}$
represents the zonal-mean frictional drag (if any).  In the above
equation, $\overline{v}^*$ and $\overline{\omega}^*$ give the
``residual-mean'' meridional and vertical velocities, respectively,
which typically provide a much better approximation of the
Lagrangian-mean circulation and tracer transport than do the
conventional Eulerian-mean velocities $\overline{v}$ and
$\overline{\omega}$ \citep[e.g.][]{andrews-etal-1987}.  In the above
equation, the zonal acceleration of the zonal-mean flow caused by
waves is given by the divergence of the Eliassen-Palm flux ${\bf F}$
in the meridional plane.\footnote{In pressure coordinates on the
  sphere, the Eliassen-Palm flux, ${\bf F}=(F_\phi,F_p)$, is
  \citep[e.g.,][]{andrews-etal-1983}
\begin{multline}
{\bf F}=a\cos\phi \Bigg\{ -\overline{u'v'} 
+ {\overline{v'\theta'}\over  \partial\overline \theta/\partial p}
{\partial\overline{u}\over\partial p}, \\
  -\overline{u'\omega'} - {\overline{v'\theta'}\over  \partial\overline \theta/\partial p}\left[{1\over a\cos\phi}{\partial(\overline{u}\cos\phi)\over
\partial \phi} - f\right]\Bigg\}
\label{ep-flux}
\end{multline}
and the divergence of ${\bf F}$ is 
\begin{equation}
\nabla\cdot {\bf F}= {1\over a\cos\phi}{\partial (F_\phi \cos\phi)
\over\partial \phi} + {\partial F_p\over\partial p}.
\label{ep-flux-div}
\end{equation}}

%Our results demonstrate that atmospheric waves induce the
%accelerations that in turn cause the QBO-like oscillations to migrate
%downward over time.  

Our results show that, at low latitudes, the wave driving induces
vertically stacked layers of eastward and westward acceleration, with
the peak accelerations shifted downward from the peak of the zonal
jet, in just the manner required to cause a QBO-like oscillation.
Figure~\ref{waves1} illustrates this phenomenon; it shows meridional
cross sections of the zonal-mean zonal wind (thick black contours) and
Eliassen-Palm flux divergence (colorscale) versus latitude and
pressure at two different phases of the oscillation, at 7000 and 8500
Earth days (the long-term evolution of the equatorial-jet structure
for this same model can be seen in Figure~\ref{qbo3}).  At both
timeframes, the peak eastward acceleration caused by the waves occurs
on the lower flank of the eastward equatorial jet, whereas the peak
westward acceleration caused by the waves occurs on the lower flank of
the westward equatorial jet.  At 7000 days (Figure~\ref{waves1},
left), the zonal jet structure comprises an eastward jet extending
from $\sim$1~bar to $\sim$0.3~bar, a westward jet from 0.3~bar to
0.03~bar, and then an eastward jet from 0.03~bar to the top of the
model.  Strong westward acceleration occurs in a layer between
0.2--0.3 bars---at the base of the westward jet---and strong eastward
acceleration occurs in a layer near 0.03 bars---at the base of the
uppermost eastward jet.  At 8500 days, the uppermost eastward jet has
migrated downward to the bottom of the model; its top reaches
$\sim$0.08~bars, above which is a newly formed westward jet that
extends from 0.08~bars to the top of the model.  At this time, strong
westward acceleration occurs right at the base of the westward jet at
a pressure of $\sim$0.06--0.08~bars.  As described earlier, this basic
acceleration pattern---namely that the peak accelerations of a given
sign occur primarily near the base of the zonal jet of that same
sign---causes the zonal jets to migrate downward over time.
Figure~\ref{waves1} demonstrates that this acceleration is caused by
waves.

%%%%%%%%%%%%%%%%%%%%%%%%%%%%%%%%%%%%%%%%%%%%%%%%%%%%%%%%%%%%%%%%%%%%%%%%
% FIGURE 12: equatorial slice of eddy fields showing waves
%%%%%%%%%%%%%%%%%%%%%%%%%%%%%%%%%%%%%%%%%%%%%%%%%%%%%%%%%%%%%%%%%%%%%%%%
\begin{figure*}
\centering
\includegraphics[scale=0.45, angle=0]{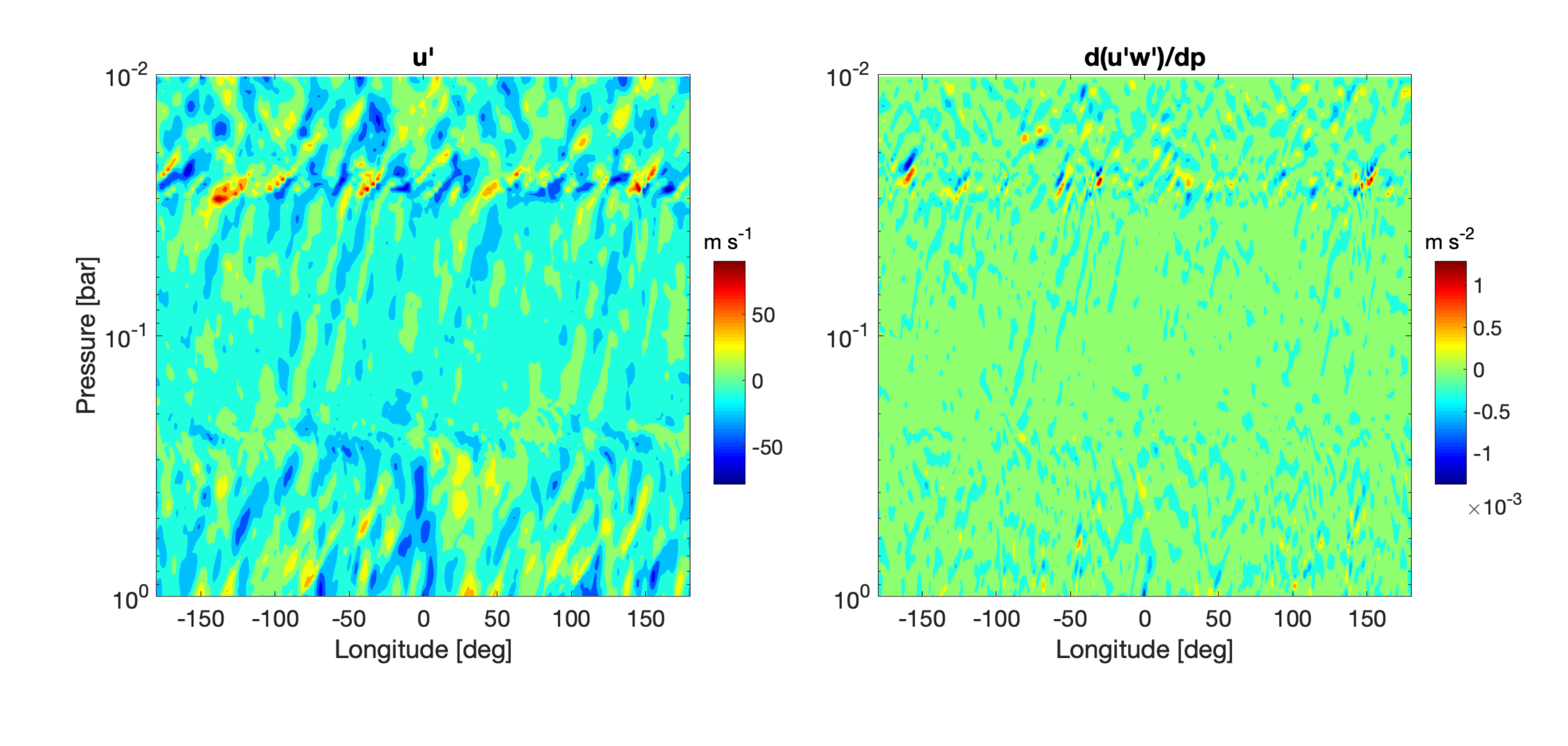}
%\begin{minipage}[c]{0.5\textwidth}
%\includegraphics[scale=0.4, angle=0]{u'.png}
%\put(-220.,160.){(a)}
%\end{minipage}
%\begin{minipage}[c]{0.45\textwidth}
%\includegraphics[scale=0.4, angle=0]{d(u'w')-dp.png}
%\put(-220.,160.){(b)}
%\end{minipage}
\caption{Spatial structure of typical waves at the equator in a
  simulation generating a QBO-like oscillation.  (a) Zonal wind
  anomalies $u'$ in $\rm m\,s^{-1}$ (that is, deviations of zonal wind
  from the zonal-mean at that pressure) shown at the equator versus
  longitude and pressure.  Waves of typical zonal wavenumber
  $\sim$10--20 can be seen, tilting both upward to the right and
  upward to the left, reminiscent of Kelvin waves and equatorially
  trapped Rossby and/or mixed Rossby-gravity waves, respectively.  The
  wave structure undergoes transitions at $\sim$0.2-0.3 and 0.03~bars
  where the zonal-mean zonal wind changes sign.  Although this shows
  perturbations of zonal wind $u'$, the patterns look qualitatively
  similar in $v'$ and $T'$.  (b).  The wave-induced acceleration
  $\partial(u'\omega')/\partial p$, which is one component of the EP
  flux divergence, again shown versus longitude and pressure at the
  equator.  The accelerations are strongest at a pressure
  $\sim$0.03~bars, near the transition between an eastward jet above
  and a westward jet below.  Both panels are snapshots at 7000 days in
  the same model shown in Figure~\ref{waves1}.  }
\label{waves2}
\end{figure*}

We next turn to characterize the types of waves that are relevant,
starting with an examination of the wave spatial structure.
Figure~\ref{waves2}a shows a snapshot of the wave perturbations versus
longitude and pressure in the equatorial plane.  
%This specifically
%shows $u'$, but the patterns would look qualitatively similar for $T'$
%or $v'$.  Several features are prominent.  First, many superposed
%modes exist, including waves whose phases tilt upward to the east, and
%others that tilt upward to the west.
Many superposed modes are present, including waves whose phases tilt
upward to the east and waves whose phases tilt upward to the west.  A
count of wave crests and troughs suggests that modes exhibiting zonal
wavenumbers $k\sim10$--20 are prominent.  Moreover, the character of
the waves changes significantly with altitude, suggesting a strong
interaction between the waves and the background zonal flow.  This
snapshot is at 7000 Earth days in the same model discussed previously,
with eastward jets at the top and bottom and a westward jet in
between, and there are strong transitions in the wave structure at
pressures of $\sim$0.2--0.3 bars and $\sim$0.03~bars, exactly where
the zonal jet changes sign.  The phase tilts evident in
Figure~\ref{waves2}a suggest the existence of Kelvin waves, which
exhibit phase tilting upward to the east, as well as equatorially
trapped Rossby and/or mixed Rossby-gravity waves, which exhibit phase
tilting upward to the west \citep[see phase relations in][]
{andrews-etal-1987}.  The waves attain velocity amplitudes of tens of
$\rm m\,s^{-1}$ or more---which approaches the zonal-mean zonal wind
speeds (Figure~\ref{waves1}).  This fact suggests that the
wave-mean-flow interactions are probably not in a regime where the
waves can be considered small perturbations relative to the mean flow.
Figure~\ref{waves2}b shows the structure of the term
$\partial(u'\omega')/\partial p$, which is one component of the EP
flux divergence.  Consistent with Figure~\ref{waves1}a, its peak
values occur at a pressure of $\sim$0.03~bars; here, however, no zonal
mean has been performed, and we can see that the acceleration varies
longitudinally in a highly patchy manner, consistent with wave
acceleration being caused by the absorption of local wave packets, as
they reach their critical levels.

%%%%%%%%%%%%%%%%%%%%%%%%%%%%%%%%%%%%%%%%%%%%%%%%%%%%%%%%%%%%%%%%%%%%%%%%
% FIGURE 13: Hovmoller diagrams
%%%%%%%%%%%%%%%%%%%%%%%%%%%%%%%%%%%%%%%%%%%%%%%%%%%%%%%%%%%%%%%%%%%%%%%%
\begin{figure}
\centering
\includegraphics[scale=0.62, angle=0]{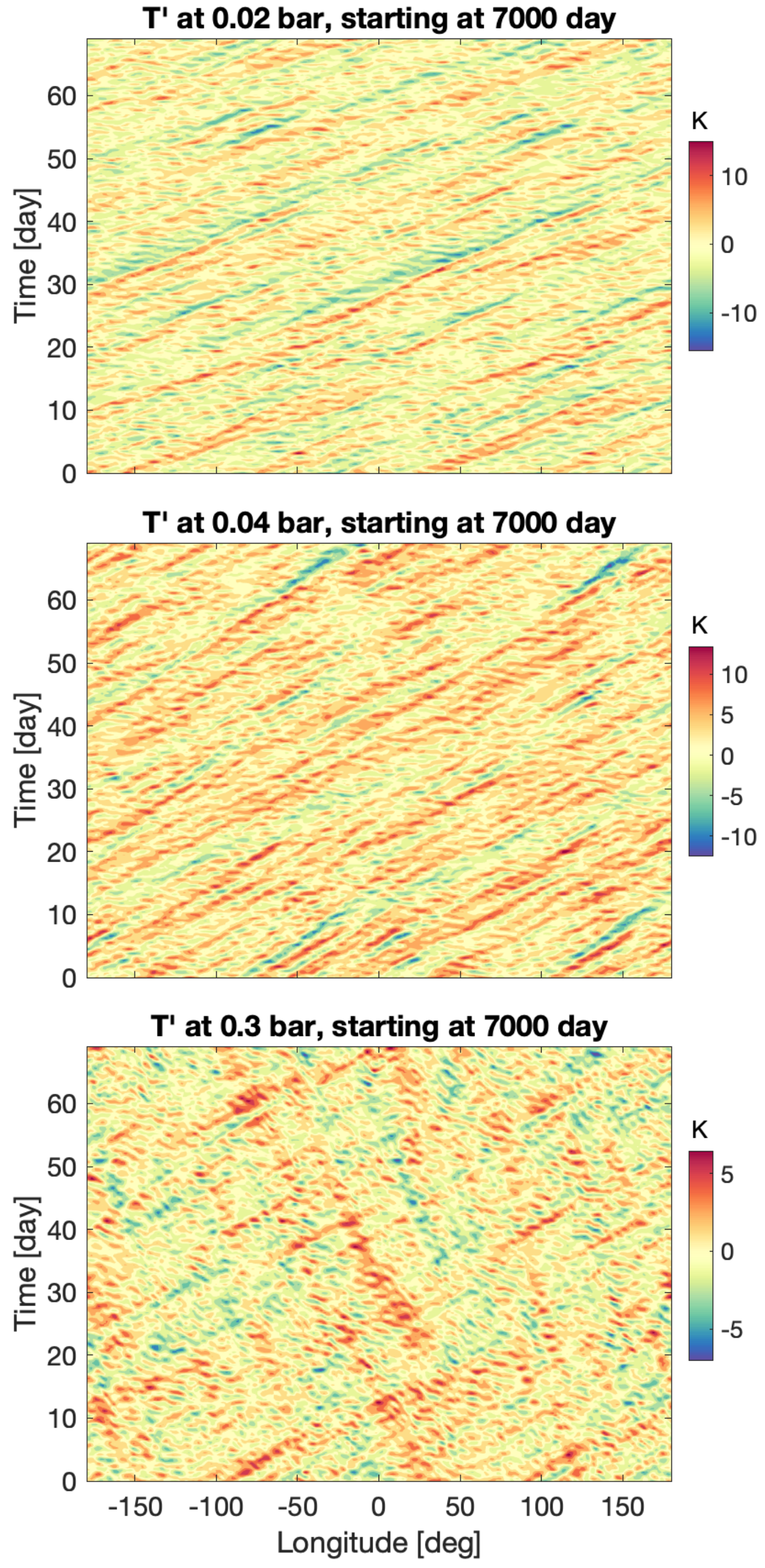}
%\includegraphics[scale=0.4, angle=0]{Hovmoller-Tstar-0-02bar-7000day.pdf}
%\includegraphics[scale=0.39, angle=0]{Hovmoller-Tstar-0-04bar-7000day.pdf}
%\includegraphics[scale=0.39, angle=0]{Hovmoller-Tstar-0-3bar-7000day.pdf}
%\put(-140.,110.){(a)}
%\put(-140.,110.){(b)}
\caption{Hovmoller diagrams, showing the time evolution of temperature
  anomalies at the equator (that is, local temperature minus the zonal
  mean, in K) over a 70-day interval at 7000 days in a model
  exhibiting a QBO-like oscillation.  Pressure levels of 0.3, 0.04,
  and 0.02 bars are shown, lying within vertically stacked eastward,
  westward, and eastward equatorial jets, respectively (see
  Figure~\ref{waves1}, left, for the zonal jet structure at this
  time).  Both eastward and westward-propagating modes can be seen
  within the lowermost eastward jet (0.3 bars), but the most prominent
  slow westward-propagating modes have been filtered out at 0.04 and
  0.02 bars, suggesting absorption of these waves at critical levels.
  This is the same model shown in Figure~\ref{qbo1} (top row),
  \ref{qbo-merid-u}, \ref{qbo-merid-temp}, and \ref{qbo3}.  }
\label{hovmoller}
\end{figure}

Hovmoller diagrams confirm the existence of both eastward and westward
propagating wave modes, constrain their phase speeds, and show that
the wave population changes significantly with altitude.
Figure~\ref{hovmoller} presents such diagrams---specifically,
temperature anomalies $T'$ versus longitude and time are shown at
three different pressure levels at the equator over a 70-day interval
starting at 7000 days.  The equatorial jet structure at this time
comprises a vertically stacked structure with an eastward jet at the
bottom, a westward jet in the middle, and another eastward jet at the
top (see Figure~\ref{waves1}a), and in Figure~\ref{hovmoller} we plot
Hovmoller diagrams within these three jets, at 0.3, 0.04, and 0.02
bars, respectively.  

In Figure~\ref{hovmoller}, eastward-propagating modes tilt upward to
the right, while westward-propagating modes tilt upward to the left.
Both types of modes are prominent at 0.3 bars (Figure~\ref{hovmoller},
bottom panel), which is inside the lowermost eastward jet associated
with the oscillation.  The eastward-propagating modes circumnavigate
the planet's circumference in $\sim$50 days, implying a phase speed of
about $100\rm\,m\,s^{-1}$ eastward.  At least two classes of
westward-propagating modes are evident.  A slow westward-propagating
component is most prominent, and travels about $100^\circ$ longitude
in the 70-day interval, implying a phase speed of $-20\rm\,m\,s^{-1}$
westward.  A faster, less-prominent component traverses the planetary
circumference in about 25--30 days, implying phase speeds approaching
$-200\rm\,m\,s^{-1}$.  Moving upward into the westward jet at
0.04~bars (Figure~\ref{hovmoller}, middle), the eastward-propagating
modes still exist.  However, a key point is that the slowly
propagating westward-propagating modes are now absent.  The top panel,
at 0.02~bars, is similar---lacking the slow westward-propagating modes
yet still retaining the eastward-propagating component.  The most
prominent eastward-propagating modes now traverse the planet's
circumference in about 40 days, implying a speed of about
$130\rm\,m\,s^{-1}$.  Because the westward jet attains peak speeds of
$-90\rm\,m\,s^{-1}$---which exceeds the phase speeds of the prominent,
slowly propagating westward-propagating modes seen in
Figure~\ref{hovmoller}, bottom---the implication is that those modes
have encountered a critical level and been absorbed, explaining their
absence in the top two panels of Figure~\ref{hovmoller}.  The change
in speed of the eastward-propagating modes between the middle and top
panel suggests either that the wave properties change with altitude or
that the slower of the eastward-propagating modes have been removed,
leaving the faster waves behind.  The fast-westward-propagating modes
(mentioned above) exist not only at 0.3 bars but also 0.04 and 0.02
bars as well (Figure~\ref{hovmoller}, middle and top); because the phase
speeds of these waves exceed the wind speed of the westward jet, these
waves do not encounter critical levels, and thus it makes sense that
they would be present at all levels shown.  The key takeaway point is
that the wave properties presented in Figure~\ref{hovmoller}---in
particular the loss of the slow westward-propagating component between
the bottom and middle panels---match expectations for the QBO-driving
mechanism described in the preceding subsection.

\begin{figure}
\centering
\includegraphics[scale=0.57, angle=0]{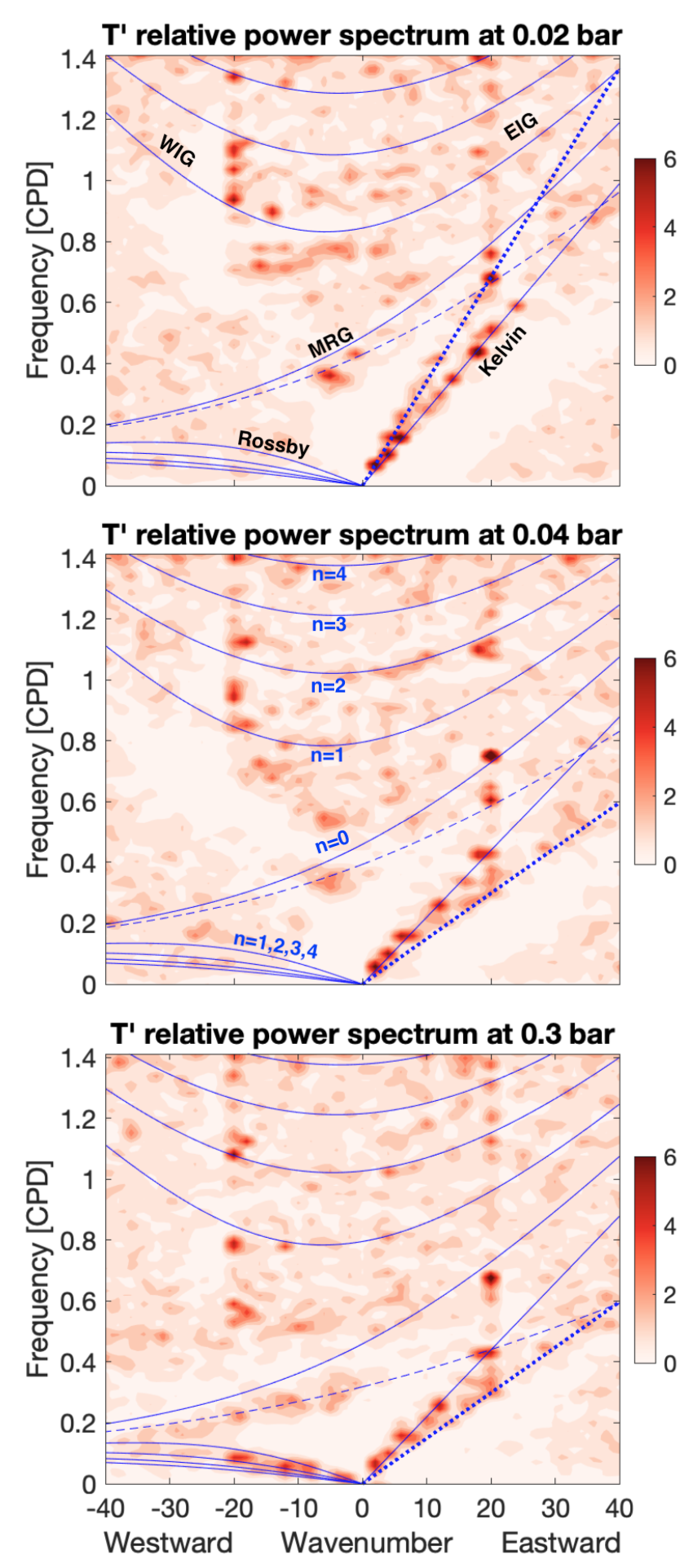}
%\includegraphics[scale=0.11, angle=0]{wkSpacetime-Tstar-0-02bar-7000day.jpg}
%\includegraphics[scale=0.115, angle=0]{wkSpacetime-Tstar-0-04bar-7000day.jpg}
%\includegraphics[scale=0.12, angle=0]{wkSpacetime-Tstar-0-3bar-7000day.jpg}
%\put(-140.,110.){(a)}
%\put(-140.,110.){(b)}
\caption{Wavenumber-frequency spectra (Wheeler-Kiladis diagrams) at
  7000 days in a model exhibiting a QBO-like oscillation.  Pressure
  levels of 0.3, 0.04, and 0.02 bars are shown, lying within
  vertically stacked eastward, westward, and eastward equatorial jets,
  respectively.  Abscissa gives the zonal wavenumber; positive and
  negative represent eastward and westward propagating waves,
  respectively.  Ordinate is frequency in cycles per day (CPD).
  Colorscale indicates wave power density; darker red coloration
  indicates that waves with significant power exist at that wavenumber
  and frequency.  Curves indicate theoretical dispersion relations for
  various wave types; Kelvin waves, equatorially trapped Rossby waves,
  mixed Rossy-gravity (MRG) waves, and eastward and westward
  inertio-gravity waves (EIG and WIG, respectively) are all shown.
  Equivalent depths are chosen to be broadly illustrative of the types
  of waves seen; no attempt has been made to fit the theoretical
  curves to the power densities.  In the upper panel, the solid lines
  all correspond to $h_e=33\rm\,m$, dotted line is a Kelvin mode with
  $h_e=63\rm\,m$, and the dashed line is an MRG mode with
  $h_e=20\rm\,m$.  In the middle panel, the corresponding three values
  are 26, 12, and $14\rm\,m$, respectively, while in the bottom panel
  they are 26, 12, and $6\rm\,m$, respectively.  }
\label{wheeler}
\end{figure}

Additional insight into the wave properties can be obtained by
performing a spectral analysis at the equator in the
wavenumber-frequency domain, similar to that performed by
\citet{wheeler-kiladis-1999}.  To do so, we first obtain the eddy
terms as a function of longitude and time at the equator and at a
given pressure level.  We do the analysis using eddy temperatures,
though qualitatively similar results would be expected for other eddy
components as well. Then we perform two-dimensional fast Fourier
transforms on the data set to obtain the raw Fourier coefficients in
the wavenumber-frequency space. We then perform smoothing by applying
40 times the so-called 1-2-1 filter in wavenumber and frequency to the
raw coefficients, after which we get a background spectrum. Finally we
apply the 1-2-1 filter once to the raw coefficients in wavenumber and
frequency and then divide by the background spectrum. What we finally
obtain is the relative power of the coefficients in
frequency-wavenumber space.  Significant signals will appear to have
values greater than 1.  Note that we have performed the analysis on
the full eddy field, rather than separating it into symmetric and
antisymmetric components about the equator as done by
\citet{wheeler-kiladis-1999}.  The final spectra are plotted in
Figure~\ref{wheeler}.  As is standard practice, the frequency is
plotted in cycles per day (thus, to obtain the angular frequency,
multiply the frequencies by $2\pi$).

%%%%%%%%%%%%%%%%%%%%%%%%%%%%%%%%%%%%%%%%%%%%%%%%%%%%%%%%%%%%%%%%%%%%%%%%
% FIGURE 14: Wheeler-Kiladis w-k diagrams
%%%%%%%%%%%%%%%%%%%%%%%%%%%%%%%%%%%%%%%%%%%%%%%%%%%%%%%%%%%%%%%%%%%%%%%%

Our wavenumber-frequency spectra demonstrate the existence of Kelvin
waves, equatorially trapped Rossby waves, and mixed Rossby-gravity
(MRG) waves, and indicate that some of these modes are preferentially
absorbed with height.  Figure~\ref{wheeler} shows such
wavenumber-frequency spectra for the same three pressure levels and
same time as shown in Figure~\ref{hovmoller}.  In the diagrams,
positive wavenumbers correspond to eastward phase propagation while
negative wavenumbers represent westward phase propagation.  Analytic
dispersion relations for relevant equatorial waves are overplotted
\citep[see, e.g.,][]{andrews-etal-1987}, and are labelled in the
figure.  A brief summary of the primary modes identifiable in our
frequency-wavenumber diagrams is as follows:
\begin{itemize}
\item {\bf Kelvin waves:} The Kelvin wave branch comprise straight
  lines extending from the origin upward to the right, exhibiting only
  eastward phase speeds.  The highest-amplitude Kelvin waves present
  in our model exhibit zonal wavenumbers of 20, but high-amplitude
  Kelvin modes also exist at wavenumbers $\sim$2--13, and lower
  amplitude Kelvin modes manifest across a range of other wavenumbers.
  More than one branch of Kelvin waves are visible, likely due to the
  presence of a range of vertical wave lengths, which would lead to a
  range of frequencies and phase speeds for a given zonal wavenumber.
  Phase speeds $\omega/k$ range from 90 to $200\rm\,m\,s^{-1}$.  
  Some of the modes are present at all three pressure levels, particularly
  the faster modes (e.g., the mode at 0.8~CPD and $k=20$), but it 
  appears that the slower modes are only present at the deepest levels.

\item {\bf Rossby waves:} Rossby waves lie at low frequencies to the
  left of the origin, exhibiting only westward phase speeds
  (Figure~\ref{wheeler}, bottom panel).  Rossby wave amplitudes are
  strongest for zonal wavenumbers of $\sim$3--20, though such modes
  can be identified at wavenumbers as high as 40.  Phase speeds calculated
  for the largest-amplitude Rossby waves in Figure~\ref{wheeler} range
  from $-20$ to $-40\rm\,m\,s^{-1}$.  Crucially, the Rossby waves
  only exist in the bottom panel of Figure~\ref{wheeler}, suggesting
  that they have been absorbed before they can propagate to the pressures
  shown in the top two panels.  As mentioned previously, at the time
  of this snapshot, any westward-propagating wave with a phase speed
  amplitude between 0 and $-90\rm\,m\,s^{-1}$ will encounter a critical level
  before reaching the middle or upper panel of Figure~\ref{wheeler}.

\item {\bf Mixed-Rossby-gravity (MRG) waves:} Interestingly, the
  analysis also indicates the presence of the MRG at the deepest
  levels.  In principle, the MRG can exhibit either eastward or
  westward phase speeds, but the MRG modes with the strongest
  amplitudes in our model are westward and exhibit zonal wavenumbers
  between $\sim$1--40 (Figure~\ref{wheeler}, bottom panel).  The phase
  speeds for the most prominent modes exhibit a wide range, from $-20$
  to nearly $-400\rm\,m\,s^{-1}$.  Just as for Rossby waves, most
  of these MRG modes, particularly the ones with slower phase speeds,
  are absent at lower pressures (upper two panels of Figure~\ref{wheeler}).
  Again, this is consistent with their loss via absorption at critical
  levels encountered at the base of the westward equatorial jet.

\item {\bf Inertio-gravity waves:} Finally, the analysis indicates the
  presence of fast inertio-gravity waves propagating both east and
  west.   Figure~\ref{wheeler} shows that the power in these IG modes
  remains focused at wavenumbers comparable to the forcing wavenumber
  ($k=20$), suggesting that nonlinear interactions are too weak to
  cause significant transfer of power to other wavenumbers (unlike the
  Kelvin, Rossby, and MRG waves, where the presence of power over a
  broad range of wavenumbers suggest that such interactions are
  important).  Nevertheless, the frequencies of the most prominent
  IG modes exhibit modest changes as a function of altitude, suggesting
  possible effects of the stratification and/or the background mean
  flow.  The phase speeds for the dominant IG modes in
  Figure~\ref{wheeler} are fast, exceeding an amplitude of
  $200\rm\,m\,s^{-1}$, so no critical level absorption is expected,
  and this is consistent with the fact that these wave modes exist at
  all three levels shown.
\end{itemize}
In summary, this Wheeler-Kiladis analysis indicates that the
convective forcing generates a wide range of equatorially trapped
waves, and that some of these wave modes disappear at progressively
higher altitudes, in a manner consistent with critical level
absorption.

To seek ``smoking gun'' evidence for the role of waves in driving the
QBO-like oscillation, we now calculate the phase speed spectra of the
wave modes present near the equator, and show how these phase speed
spectra vary with pressure.  To analyze the eddy flux (for example
$u'\omega'$) in $\omega$-$k$ space we need to do cross spectral
analysis. The basic procedure is analogous to that described in
\citet{randel-held-1991}. First we calculate the cross-spectral power
density (CSPD) in $\omega$-$k$ space for propagating atmospheric waves
following the method laid out by \citet{hayashi-1971}. This is done
for all relevant eddy flux including $u'\omega'$, $u'v'$, and
$v'T'$. Then we convert the CSPD to wavenumber-phasespeed space
following \citet{randel-held-1991}. The Eliassen-Palm flux in
wavenumber-phasespeed space is calculated by multiplying the constant
coefficients and grouping all relevant terms as shown in
Equation~(\ref{ep-flux}). The TEM eddy acceleration in
wavenumber-phasespeed space is calculated by taking the appropriate
derivatives as in Equation~(\ref{ep-flux-div}). Finally, the TEM eddy
acceleration in the phasespeed domain is calculated by integrating the
function over the wavenumber domain.

\begin{figure*}
\centering
\includegraphics[scale=0.5, angle=0]{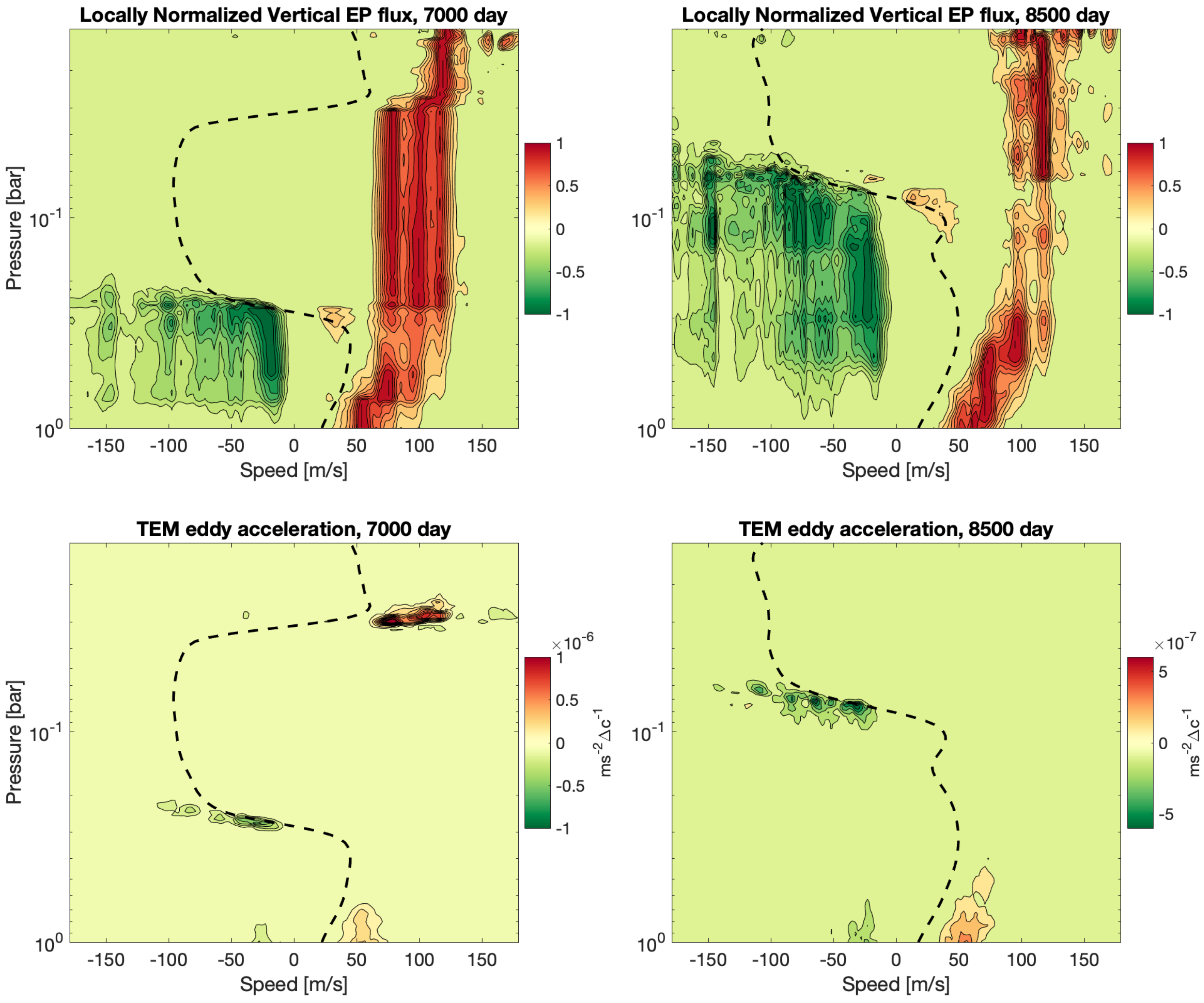}
%\begin{minipage}[c]{0.5\textwidth}
%\includegraphics[scale=0.37, angle=0]{uwflux_normalized_7000day.pdf}
%\includegraphics[scale=0.34, angle=0]{phasespeed_powerdensity_TEMeddy_7000day.pdf}
%\end{minipage}
%\begin{minipage}[c]{0.45\textwidth}
%\includegraphics[scale=0.37, angle=0]{uwflux_normalized_8500day.pdf}
%\includegraphics[scale=0.34, angle=0]{phasespeed_powerdensity_TEMeddy_8500day.pdf}
%\end{minipage}
\caption{Phasespeed spectra of the Eliassen-Palm flux (top) and
  Eliassen-Palm flux divergence (bottom) at 7000 and 8500 Earth days
  on the left and right, respectively, showing the absorption of waves
  at critical levels in a model with a QBO-like oscillation.  The
  equatorial zonal-mean zonal wind profiles at the corresponding times
  are overplotted in thick dashed curves.  In the top panels, the
  spectra are of the vertical component of the Eliassen-Palm flux
  vector, and the spectra are normalized at every pressure by the
  maximum value at that pressure.  The spectra comprise eastward
  propagating modes with speeds of $50$ to $\sim$$140\rm\,m\,s^{-1}$
  and westward propagating modes with speeds of about $-10$ to
  $-150\rm\,m\,s^{-1}$, which propagate upward from near the bottom
  (pressures of around a bar).  The westward propagating modes shown
  here (which as shown in Figure~\ref{wheeler} are primarily Rossby
  and mixed Rossby-gravity waves) are absorbed near a critical level
  when they reach the base of the westward zonal jet (at 0.02--0.03
  bars at 7000 days, and $\sim$0.06-0.08 bars at 8500 days). Some of
  the eastward-propagating modes (which as shown in
  Figure~\ref{wheeler} are primarily Kelvin waves) are absorbed at
  critical levels, though some of them are able to propagate upward
  throughout the plotted region without being absorbed.  The bottom
  panels show that the zonal accelerations caused by the wave
  absorption are eastward (orange/red colors) on the lower flank of
  the eastward jet, and westward (green) on the lower flank of the
  westward jet, as necessary to cause the downward zonal-jet migration
  occurring in the QBO-like oscillation.  This figure provides strong
  evidence that the QBO-like oscillation in the numerical experiment
  is driven by critical-level absorption of vertically propagating
  atmospheric waves, as described in the text.}
\label{phasespeed-spectra}
\end{figure*}

Figure~\ref{phasespeed-spectra} shows the resulting phase-speed
spectra versus pressure for the same two timeframes during a QBO-like
oscillation as are shown in Figure~\ref{waves1}---namely, 7000 and
8500 days, on the left and right, respectively.  First examine the top
row, which shows the phasespeed spectra of the vertical component of
the Eliassen-Palm flux.  At any given pressure, the amplitudes at
various wavenumbers represent a spectrum giving the range of zonal
phase speeds exhibited by the waves (with the plotted amplitudes
normalized by the maximum value at that pressure), and comparing
different pressures shows how the mode population varies with
altitude.  The thick dashed curve shows the zonal-mean zonal wind at
the time of this analysis.  At both time intervals, the deepest
regions shown exhibit prominent waves with westward phase speeds
ranging from $-10$ to $-180\rm m\,s^{-1}$ and eastward phase speeds
from $50$ to $150\rm\,m\,s^{-1}$.  At 7000 days, it can be seen that
the modes with westward phase speeds continue until they reach
pressures of $\sim$0.2--0.3 bars, above which they disappear.  At 8500
days, the pattern is similar, except that the westward-propagating
modes reach pressures as low as 0.06--0.07 before they disappear.  It
is clear that many of these westward-propagating waves have reached a
critical level, because their amplitude sharply decreases with
altitude at precisely the altitude where their zonal phase speed
matches that of the zonal-mean zonal wind.  These waves are primarily
Rossby and mixed Rossby-gravity waves, as Figure~\ref{wheeler} makes
clear.  Considering now the eastward-propagating modes, the slowest
modes (phase speeds $\lesssim 60\rm\,m\,s^{-1}$) disappear at
pressures less than $\sim$0.3~bars.  At 7000 days, Kelvin waves with
eastward phase speeds of $70$--$100\rm\,m\,s^{-1}$ disappear on the
lower flank of the eastward jet (although the Kelvin wave with phase
speed $\sim$$120\rm\,m\,s^{-1}$ continues unimpeded to the top of the
plot).  In contrast, at 8500 days, there is no strong eastward jet at
the top of the model, and most of the fast Kelvin waves propagate
upward unimpeded.

The bottom row of Figure~\ref{phasespeed-spectra} show the phasespeed
spectra of the TEM acceleration versus pressure.  Essentially, this is
just a decomposition of $\nabla\cdot{\bf F}$ (see Equation~\ref{tem})
showing the phase speeds of the waves that contribute to the TEM
acceleration; positive values indicate an eastward acceleration and
negative values indicate a westward acceleration.  It can be seen that
absorption of westward-propagating waves near the critical levels at
the base of the westard jet causes westward acceleration, while the
absorption of eastward-propagating waves on the critical level near
the base of the eastward jet causes eastward acceleration.  These
zonal accelerations caused by critical-level absorption are just those
shown in Figure~\ref{waves1}, and which are responsible for the
downward migration of the QBO-like oscillation in our experiments.

To summarize, Figure~\ref{phasespeed-spectra} provides decisive evidence
that the wave-driving mechanism, as understood for the terrestrial
QBO and as outlined in the preceding subsection, causes the QBO-like
oscillations in our models.  

Nevertheless, a puzzling aspect emerging from
Figure~\ref{phasespeed-spectra} is that some of the
westward-propagating waves absorbed on the lower flank of the westward
jet, and some of the eastward-propagating waves absorbed on the lower
flank of the eastward jet, have phase speeds {\it faster} than those
of the zonal-mean zonal wind, implying that they do not appear to have
reached critical levels.\footnote{At 7000 days, this is particularly
  true for westward-propagating waves with phase speed amplitudes
  faster than $90\rm\,m\,s^{-1}$ at $\sim$0.3~bars, and for
  easward-propagating waves with speeds of $70$--$100\rm\,m\,s^{-1}$
  at $\sim$0.03~bars.  At 8500 days, it is primarily true for
  westard-propagating waves of speeds exceeding $100\rm\,m\,s^{-1}$ at
  $\sim$0.06~bars, and eastward-propagating modes of speeds
  $\sim$70--$80\rm\,m\,s^{-1}$ that are absorbed near 0.3~bars.}
Stated another way, Figure~\ref{phasespeed-spectra} seems to indicate
that the waves {\it act} as if they have reached critical levels for a
zonal-mean zonal-wind profile that is {\it faster} than the actual
zonal-mean zonal-wind profile.  This behavior is not predicted by
classical wave-mean-flow interaction theory, which applies to small
amplitude waves and adopts the WKB approximation, under which the
background zonal-flow properties change over length scales long
compared to the vertical wavelength of the waves.

We suggest two hypotheses that could potentially explain the
unexpected nature of the critical-level dynamics in our simulations.
First, the WKB approximation is not well satisifed in our models; the
wavelengths of the convectively generated waves in our models are not
generally short compared to the vertical scale over which the zonal
wind varies (compare the wave vertical wavelengths in
Figure~\ref{waves2} with the zonal-mean zonal-wind structure in
Figures~\ref{waves1} and \ref{phasespeed-spectra}).  We speculate that
this effect could change the nature of the wave-mean-flow
interactions, perhaps helping to explain the behavior seen in
Figure~\ref{phasespeed-spectra}.  The second hypothesis is that the
phenomenon results from a nonlinearity associated with the large wave
amplitudes: as mentioned previously and shown in Figure~\ref{waves2},
the local wave amplitudes $u'$ attain speeds approaching that of the
local zonal flow, and thus quasi-linear wave-mean-flow interaction
theory may not apply.  While the precise mechanism remains to be
investigated, we speculate that the large wave amplitude causes the
wave to ``sense'' an effective zonal-jet speed greater than the actual
value of $\overline{u}$.  For example, Figure~\ref{waves2} indicates
that at 0.03~bars---where many eastward-propagating waves are
absorbed---the wave amplitudes are $50\rm\,m\,s^{-1}$ at most
longitudes, and reach twice this value in a few local regions.  This
level and time corresponds to the base of an eastward jet, whose
zonal-mean zonal wind speeds $\overline{u}$ reach $50\rm\,m\,s^{-1}$
(Figure~\ref{phasespeed-spectra}, left).  This implies that, {\it
  locally}, at the longitudes where the wave anomalies are eastward,
the zonal wind reaches speeds up to $150\rm\,m\,s^{-1}$.  We speculate,
then, that the waves of phase speeds $70$--$100\rm\,m\,s^{-1}$ (which
Figure~\ref{phasespeed-spectra} indicates are absorbed) act as though
they are reaching a critical level, due to the fact that the local
zonal wind speed at certain longitudes matches the wave's phase
speed---even though the {\it zonal-mean} zonal wind never reaches such
high values.  Detailed investigation of this hypothesis and other
aspects are left for the future.

%%%%%%%%%%%%%%%%%%%%%%%%%%%%%%%%%%%%%%%%%%%%%%%%%%%%%%%%%%%%%%%%%%%%%%%%
% FIGURE 15: Phase speed spectra of wave amplitude and TEM acceleration
% (similar to Randel & Held 1991 type plots)
%%%%%%%%%%%%%%%%%%%%%%%%%%%%%%%%%%%%%%%%%%%%%%%%%%%%%%%%%%%%%%%%%%%%%%%%

\subsection{Implications for Jupiter and Saturn}

Our idealized model with isotropically imposed thermal perturbations
holds promise for understanding the general circulation of Jupiter and
Saturn.  In particular, some of our models exhibit zonal-jet
structures very similar to those on Jupiter and Saturn, including
multiple off-equatorial (mid-to-high latitude) zonal jets and a broad,
stable, equatorial eastward jet---i.e., equatorial
superrotation. Figure~\ref{zonal-winds-vorticity} illustrates an
example, showing the zonal-wind and vorticity at 1.5~bars pressure,
which is in the troposphere, deeper than the level where the QBO-type
oscillations occur.  Thus, the equatorial superrotation shown in
Figure~\ref{zonal-winds-vorticity} is a stable, long-term feature, in
contrast to the oscillating eastward and westward jets seen in the
QBO-type oscillations in other models.  In
Figure~\ref{zonal-winds-vorticity}, the zonal-wind speed in the
mid-to-high-latitude and equatorial jets reach 50--$100\rm\,m\,^{-1}$
and $\sim$$300\rm\,m\,s^{-1}$, respectively. These speeds are
intermediate between the jet speeds on Jupiter and Saturn, though
closer to the latter.  Although the equatorial jet in this model is a
stable feature, even in this model, there exists---at lower pressure,
at $\sim$0.02--1 bar, within the model's stratosphere and upper
troposphere---a QBO-type oscillation {\it within} this robust
equatorial superrotation, in a similar way that the QQO and SAO on
Jupiter and Saturn represent stratospheric perturbations around a
long-lived, stable equatorial superrotating jet that is rooted in the
troposphere.

Although the zonal jets in this model are rather zonally symmetric at
low latitudes, they experience greater meanders at high latitudes,
which resemble a large-scale, polygonal structure when viewed from
over the pole (Figure~\ref{vorticity-globes}, upper right panel; a
quasi-hexagonal structure can be seen in the yellow, circumpolar ring
of postive relative vorticity at $\sim$65--$70^\circ$ latitude).  This
feature is similar to---and relevant for understanding---the polar
hexagon in Saturn's northern hemisphere
\citep[e.g.][]{baines-etal-2009, morales-juberias-etal-2015}.  The
hexagonal pattern seen in Figure~\ref{vorticity-globes} represents
equatorward/poleward deflections of the prominent zonal jet at
$\sim$$65^\circ$, which is the closest eastward zonal jet to the pole
in this model (see Figure~\ref{zonal-winds-vorticity}a).
Nevertheless, the hexagonal structure in this simulation is less
prominent than Saturn's hexagon, both in that the vertices are not as
sharp, and the sides are not all equal in length.  The hexagon on
Saturn is also at higher latitude ($\sim$$76^\circ$) than that shown
here---presumably a simple result of the latitude at which the most
poleward jet occurs.  Within the hexagon, there exist several
cyclonic and anticyclonic vortices as well as turbulent filaments
likely associated with vorticity mixing.  Further studies of the 
properties of such polar polygons in this class of model would 
be useful.

The QBO-type oscillations in the models shown in this paper exhibit
{\tt periods} that bracket those observed on in the Jovian QQO and
Saturnian SAO.  The QBO-type oscillations for the two simulations in
Figure~\ref{qbo1} exhibit periods of 12 and 19 years, which bracket
the 15-year period for Saturn's SAO \citep{orton-etal-2008,
  fouchet-etal-2008, guerlet-etal-2011}.  The QBO-type oscillation for
the model in Figure~\ref{zonal-winds-vorticity} has a period of
$\sim$400--500 days, which is several times shorter than that of
Jupiter's QQO.

{\tt A detailed comparison of the QBO-like oscillations in our models
  to Jupiter's QQO and Saturn's SAO reveals similarities and
  differences.  The observed temperature anomalies on both planets
  exhibit temperature extrema at the equator that are anticorrelated
  to those immediately north and south \citep{fletcher-etal-2016,
    cosentino-etal-2017, guerlet-etal-2011, guerlet-etal-2018}, a
  feature also shared by our simulations
  (Figure~\ref{qbo-merid-temp}).  The temperature anomalies observed
  on Jupiter exhibit amplitudes of order $\pm 5\rm\,K$, similar to
  those in our model depicted in Figure~\ref{qbo-merid-temp}, whereas
  the temperature anomalies of Saturn's SAO are roughly twice as
  large, about $\pm 10\rm\,K$.  A significant discrepancy between our
  models and the actual QQO and SAO is that the temperature anomalies
  (including both the central temperature extrema at the equator and
  the anticorrelated extrema on either side) extend to latitude
  20--25$^\circ$ on Jupiter and Saturn but only to $15^\circ$ in our
  models (Figure~\ref{qbo-merid-temp}).  This difference probably
  results in large part from the faster rotation rates of our models,
  which confines the equatorial waveguide closer to the equator.  The
  equatorial Rossby deformation radius---which represents the
  half-width of the waveguide---is $L_{\rm eq}=\sqrt{NH/\beta}$.
  Our models adopt a rotation rate twice that of Jupiter, which, everything
  else being equal, implies an equatorial deformation radius 70\% that
  on Jupiter, and should lead to equatorial oscillations with a meridional
  width that is narrower by a similar factor. 

   Another significant difference from Jupiter and Saturn is that the
   QBO-like oscillations in our models occur at greater pressures than
   Jupiter's QQO and especially Saturn's SAO.  In our model, the
   oscillation occurs primarily between 1~bar and the top of the model
   at 10~mbar.  The QQO extends from $\sim$1--100~mbar
   \citep{fletcher-etal-2016, cosentino-etal-2017} while the Saturn
   SAO occurs primarily between 0.01--10~mbar.  An obvious future step
   would be to extend the model top to lower pressures and investigate
   whether the additional ``room'' promotes a natural extention of the
   oscillation to lower pressures.  One also expects the oscillation
   to be affected by the vertical stratification profile. The greater
   gravities of brown dwarfs shifts the tropopause (defined here as
   the deepest level where the stratification becomes significant) to
   deeper pressures than on Jupiter and Saturn\footnote{For a constant
     specific absorption coefficient $k_{\rm abs}$ in the IR, one
     expects the pressure at optical-depth unity to be approximately
     $g/k_{\rm abs}$.  Because the tropopause usually occurs near this
     level, one expects the tropopause pressure to scale with gravity
     in a similar manner.  A typical brown dwarf has a gravity dozens
     of times larger than Jupiter's, and thus can have a tropopause
     that is much deeper than on Jupiter and Saturn.} and this fact
   motivated our choice of background temperature profile
   (Figure~\ref{setup}).  We speculate that a model whose background
   temperature profile is chosen to match that of Jupiter or Saturn
   may naturally exhibit oscillations that are shifted to lower
   pressures.  Regardless, the question of what controls the vertical
   pressure range over which the oscillation occurs---and the extent
   to which the modeled range matches that observed on Jupiter and
   Saturn---is a critical question that needs to be addressed in
   future models.}

Although we have captured many elements of the circulations on Jupiter
and Saturn in some simulations---numerous east-west zonal jets,
equatorial superrotation, stratospheric QQO and SAO-type oscillations,
and tendencies toward polar polygons---our overall emphasis has been
on understanding the overall dynamical behavior across a wide range of
conditions relevant to giant planets and brown dwarfs generally.  As
such, we have not attempted to ``tune'' models to match precisely the
observed properties of Jupiter and Saturn.  It would be valuable to
perform follow-on studies aimed at determining the conditions under
which the particular details of the circulations on these
planets---jet speeds and profiles, temperature perturbations, and
periods, amplitudes, and meridional structure of the QQO and SAO---can
be matched.

In the context of the present simulations, the temperature profile
primarily exerts an influence via the height-dependence of the
Brunt-Vaisala frequency that it implies, rather than via the absolute
temperature itself.\footnote{The radiative time constant strongly affects the
results, of course, but this is an independent parameter in our
setup.}  We thus might expect that otherwise similar simulations that
replace the temperature profile in Figure~\ref{setup} with one more
appropriate for Jupiter or Saturn would yield qualitatively similar
results.  To test this, we performed a few simulations where we
adopted a temperature profile similar to that observed on Jupiter,
with an isothermal temperature $T_{\rm iso}=110\rm\,K$ at the top,
transitioning to an adiabatic temperature profile with a constant
potential temperature of $\theta_{\rm ad}=165\rm \,K$ in the interior.
We also lessened the forcing amplitude in these models to be more
appropriate for the cooler temperatures.  As expected, these models
produce behavior qualitatively similar to that described in
Sections~\ref{flow-regime}--\ref{qbo-like}, including the generation
of multiple zonal jets and, in some models, QBO- or QQO-like
oscillations.  We defer for the future a focused investigation in the
exact parameter regime of Jupiter and Saturn.

\section{Conclusions and Discussion}
\label{conclusions}

We presented idealized 3D simulations of brown dwarfs and Jupiter and
Saturn-like giant planets to test the hypothesis that interaction of
convection with an overlying stratified atmosphere can lead to a
vigorous atmospheric circulation consisting of zonal jets and
turbulence, long-term variability, and stratospheric oscillations, and
to ascertain how the properties of the circulation vary over a wide
range of parameters relevant to brown dwarfs and Jupiter-like planets.
Convection was parameterized by introducing small-scale, random,
horizontally isotropic thermal perturbations near the bottom of the
domain, which represent the effect of convective plumes in
perturbing the radiative-convective boundary at the base of the
atmosphere.  

Our primary results are as follows:
\begin{itemize}
\item{We showed that under the rapidly rotating conditions relevant to
  brown dwarfs and Jupiter-like planets, zonal jets are a robust,
  ubiquitous outcome of the dynamics.  Under forcing and damping
  conditions relevant to brown dwarfs and giant planets, wind speeds
  typically range from tens to hundreds of $\rm m\,s^{-1}$ and
  horizontal temperature perturbations on isobars are typically
  several to tens of K.  These ranges agree with the wind speeds and
  temperature perturbation amplitudes predicted in an analytical scaling
  theory by \citet{showman-kaspi-2013}.   As expected, stronger forcing
  leads to stronger jets, while stronger radiative and/or frictional
  damping leads to weaker jets.  Generally, for similar wind speeds,
  models with weak forcing and damping exhibit a more zonally symmetric
  pattern than models with strong forcing and damping.  In models with
  strong jet formation, the potential vorticity (PV) tends to be homogenized
  in strips, leading to a staircase pattern of PV with latitude.}

\item{Our simulations show that under conditions of weak radiative and
  frictional damping, the zonally banded pattern---and zonal
  jets---occur over a wide latitude range from the equator to near the
  poles.  When the radiative or frictional damping are strong,
  however, the zonal banding becomes confined to low latitudes, with
  the higher latitudes dominated primarily by wave dynamics.  This
  behavior naturally results from the fact that the ability of the
  convective perturbations to generate jets is stronger near the
  equator and weaker near the poles, due to a combination of factors,
  including the latitudinal variation of both the $\beta$ effect and
  the ability of the convective perturbations to generate Rossby waves,
  which are critical in jet formation.}

\item{Under appropriate conditions, our models produce long-term
  oscillations in the stratospheric jet structure, in which vertically
  stacked eastward and westward zonal jets migrate downward, analogous
  to the terrestrial QBO, Jovian QQO, and Saturnian SAO.  Our
  simulations are the first demonstration of a QBO-like oscillation in
  full 3D numerical simulations a giant planet.  The ranges of periods
  and other properties seen in our oscillations are similar to, and
  bracket, those of the observed QQO and SAO.  The possibility of such
  phenomena on brown dwarfs suggests the possibility of very long term
  (multi annual) variability, which could be monitored in long-term
  groundbased surveys.}

\item{\tt Detailed diagnostics show that the convective forcing drives
  a broad population of equatorial waves, including Kelvin, Rossby,
  mixed Rossby-gravity, and inertio-gravity waves, which propagate
  vertically upward from near the radiative-convective boundary.  We
  showed in detail that the QBO-like oscillations in our models are
  driven by the interaction of these waves with the mean flow; in
  particular, absorption of waves at critical levels on the lower
  flanks of the stacked eastward and westward equatorial stratospheric
  jets causes the jets to migrate downward over time.}

\item{Some of our models produce zonal-jet profiles very similar to
  those on Jupiter and Saturn, including stable, long-lived equatorial
  superrotation in the troposphere, numerous high-latitude jets,
  and hints of polar cyclones that resemble Saturn's hexagon.}
\end{itemize}

Our results support a picture wherein the convective forcing triggers
a population of Rossby waves, the latitudinally preferential breaking
of which leads to a coherent zonal jet structure with eastward eddy 
acceleration in the eastward jets and westward eddy acceleration in
the westward jets.  As proposed by \citet{showman-kaspi-2013}, such
jet accelerations cause an overturning circulation in the meridional
plane.  The vertical motion associated with this circulation transports
entropy vertically and leads to horizontal temperature perturbations on
isobars, even in the absense of any externally imposed irradiation
gradients.  In a model with clouds, the vertical motions associated
with these overturning circulations would lead to patchy clouds,
which will help to explain the light curve variability observed
on a wide range of brown dwarfs.

{\tt Our models show that the atmosphere of brown dwarfs and giant
  planets can exhibit the emergence of oscillations having timescales
  orders of magnitude longer than any of the explicit forcing and
  damping timescales.  The implication is that brown dwarfs are likely
  to be variable not only on short timescales (as has already been
  extensively observed) but on longer timescales of years or even
  decades as well, despite the fact that the radiative and convective
  timescales are likely short on these objects.}

The existence of banded flow patterns in all our models---including
those with very strong radiative damping---differs from the results
reported by \citet{zhang-showman-2014}, who found using a one-layer
shallow-water model that, when the forcing was weak and/or the damping
was strong, the flow transitioned to an isotropic state dominated by
turbulent eddies with no (statistically) preferred directionality.
The differing behavior results from the fact that in 3D models,
radiative damping can remove horizontal temperature variations and
therefore (via the thermal-wind equation) vertical wind shears, but it
cannot damp the barotropic mode---that is, the pressure-independent
component of the wind, which is not associated with any horizontal
temperature gradients.  Essentially, by assuming a quiescent interior
underlying their one-layer atmosphere, \citet{zhang-showman-2014}
assumed that there is no barotropic mode.  In our 3D models, the
barotropic mode can be damped by the friction imposed near the base of
our model.  In the present context, then, the setup of
\citet{zhang-showman-2014} would best be represented by the limit of
very strong drag at the base of the model.

Our model setup---with idealized convective forcing and radiative
damping---represents a useful platform for further dynamical studies
of the Jovian QQO, Saturnian SAO, and potentially even Earth's QBO.
Inclusion of gravity-wave-drag parameterizations and further
exploration of the large model parameter space may yield a more
complete understanding of the conditions under which these
oscillations occur, including the amplitude, wavenumber, and
latitudinal profile of convective wave forcing that yield oscillations
in agreement with observations.  Moreover, it would be straightforward
to add a simple seasonal cycle, relevant to Saturn and Earth, which
would allow an investigation of how QBO-like oscillations interact with
a seasonally varying stratospheric meridional circulation.  This is
relevant to understanding the tendency of the QBO to ``lock onto'' 
the seasonal cycle \citep[e.g.,][]{rajendran-etal-2016,
  rajendran-etal-2018} and may be important for understanding why
Saturn's SAO has a period close to half a Saturn year.  Moreover,
both the terrestrial QBO and Saturn's SAO have recently experienced
disruptions \citep{fletcher-etal-2017, newman-etal-2016, 
dunkerton-2016}, a phenomenon which could be explored in our
model framework by introducing perturbations to the convective
forcing or seasonal cycle to determine how they affect the
QBO-like oscillation properties.

Our models lack any representation of cloud feedbacks, which are
likely important for the atmospheric dynamics on many brown dwarfs
\citep[e.g.,][]{tan-showman-2018}.  The heating and cooling associated
with time-variable patchy clouds could lead to significant horizontal
temperature variations due purely to radiative effects, which in turn
will be important in driving the atmospheric circulation, particularly
on L dwarfs, which have relatively opaque clouds and extremely high
heat fluxes due to their high temperatures.  Including such cloud
feedbacks in 3D models of this type is an important avenue for future
research.  Nevertheless, the current idealized cloud free models
provide a critical foundation for understanding more complex
scenarios.  Moreover, our cloud-free models are likely directly
relevant to a wide range of brown dwarfs where cloud effects are
relatively weak, such as the mid-to-late T dwarfs, which exhibit
relatively cloud-free atmospheres, or cooler giant planets (including
Y dwarfs as well as Jupiter and Saturn themselves) where the lower
temperatures imply that the cloud radiative forcing is likely to be
weaker.

%%%%%%%%%%%%%%%%%%%%%%%%%%%%%%%%%%%%%%%%%%%%%%%%%%%%%%%%
\appendix
\section{Forcing}
\label{append}
{\tt
%Our primitive equation (PE) model of the atmospheric circulation on
%brown dwarfs adopts a simple forcing scheme that represents the
%perturbing effects of convection impinging on the radiative-convective
%boundary (RCB).  
Here, we physically motivate plausible forcing amplitudes that may be
relevant to brown dwarfs of specified heat flux and other parameters.
We imagine material surfaces just above the RCB being perturbed up and
down by buoyant, rising convective plumes overshooting into the
stratified atmosphere (or of descending plumes dripping off the bottom
of the RCB into the interior).  To represent these effects, thermal
forcing is added to the thermodynamic-energy equation in a forcing
layer in the bottom scale height of the domain, with the aim of
deflecting material surfaces up and down in a way analogous
to convection.  Here, we estimate the expected amplitudes this forcing
should have.

Specifically, the question is: how do we select appropriate values of
$f_{\rm amp}$, if we want to represent convection from a brown dwarf
of some given heat flux?  We break this into three steps.  First,
let's calculate the typical overshoot distance of plumes, $\Delta z$,
given some vertical velocity at the RCB, which we could estimate from
mixing length theory.  Second, let's estimate, in our forcing scheme,
what characteristic horizontal variations of temperature would be
needed in the bottom scale height of our model to generate
material-surface displacements of this magnitude, and third, we will
determine what values of $f_{\rm amp}$ are needed to generate those
necessary horizontal temperature variations.

\subsection{Overshoot distance of convective plumes at RCB}

We first calculate the overshoot distance, $\Delta z$, for a rising
convective plume of velocity $w$ that overshoots across the RCB, given
a specified background potential temperature gradient immediately
above the RCB of $\partial\theta/\partial z$. If at some time $t$ the
parcel has ascended a height $\Delta z'$, then the instantaneous
difference in potential temperature between the parcel and its
surroundings will be $\Delta\theta=-(\partial\theta/ \partial z)\Delta
z'$ and the negative buoyancy acceleration will be
$-g\,\Delta\theta/\theta$.  Using parcel theory, we simply write the
equation of motion for the parcel as
\begin{equation}
{d^2\Delta z'\over dt^2}= {\Delta\theta\over\theta}g = -{g\over\theta}
{\partial\theta\over \partial z}\Delta z'
\end{equation}
which has solution $\Delta z'=\Delta z \sin Nt,$ where $N$ is the
Brunt-Vaisala frequency and $\Delta z$ is the maximum amplitude of the
displacement (i.e., the height above the RCB at which the parcel
reaches its maximum altitude and begins descending).  The time-varying
parcel velocity is just the time derivative of this solution,
$d \Delta z'/ dt = N \Delta z \cos Nt$,
which has a maximum velocity amplitude of $N\Delta z$.  Since this 
is just our assumed convective velocity, we can write that, given
some velocity $w$ at the RCB, the overshoot distance is
\begin{equation}
\Delta z \sim {w\over N}.
\label{dz1}
\end{equation}

%Note that it's possible to obtain this same estimate from a couple of
%simpler, order-of-magnitude arguments.  The most direct way is to
%simply say that the kinetic energy per unit mass of the overshooting
%air parcel, $w^2$, is transferred to the potential energy per mass of
%the air column, which is acceleration times distance travelled
%(basically work done in increasing the atmospheric potential energy
%equals force times distance), i.e., just equating the two energies,
%\begin{equation}
%w^2 \sim {g \Delta\theta\over\theta}\Delta z,
%\end{equation}
%and given that $\Delta \theta \sim \Delta z \partial\theta/\partial z$, we
%obtain 
%\begin{equation}
%w \sim \left({g\over \theta} {\partial\theta\over\partial z}\right)^{1/2}
%\Delta z
%\end{equation}
%which in turn immediately yields Equation~(\ref{dz1}).  An alternate
%way is to say that the time for an overshooting parcel to slow to a
%stop equals velocity over acceleration, i.e. $\tau_{\rm os} \sim
%w/(g\Delta \theta/\theta) \sim w\theta/(g \Delta z \partial
%\theta/\partial z)$.  Likewise, the distance traveled is velocity
%times the overshoot time, $\Delta z \sim w \tau_{\rm os}$, and when we
%plug in the expression for $\tau_{\rm os}$ into the latter expression
%for distance traveled, we again obtain Equation~(\ref{dz1}).

In Equation~(\ref{dz1}), it's not obvious {\it a priori} what value of
$N$ would be appropriate, since it depends both on the stratification
above the RCB, as well as the gravity and other parameters, all of
which might vary widely on brown dwarfs.  Therefore, it's convenient
to rewrite expression (\ref{dz1}) using a simple dimensionless
parameter that characterizes the subadiabaticity of the thermal
profile.  Following \citet{zhang-showman-2014}, we define a
``subadiabaticity parameter'' $\gamma = 1 + (c_p/g) \partial
T/\partial z = 1 - \kappa^{-1}\partial \ln T/\partial \ln p$, where
$\kappa=R/c_p$ is the ratio of the specific gas constant to the
specific heat at constant pressure.  Using this definition, the
Brunt-Vaisala frequency can be expressed as $N^2 = {g^2\over T c_p}
\gamma,$ where $\gamma=1$ for an isothermal atmosphere and $\gamma=0$
for an adiabatic atmosphere.  What is an appropriate value of
$\gamma$?  Although the atmospheres of brown dwarfs should be quite
stratified aloft (approaching an isotherm at high altitude according
to 1D models; e.g., \citealt{burrows-etal-2006}), the stratification
should be weak just above the RCB, with only minimal deviations from
an adiabat.  \citet{zhang-showman-2014} suggested $\gamma=0.1$ as
appropriate.  The value will depend on how sharply the thermal
gradient transitions from adiabatic to isothermal above the RCB, and
on how this distance compares to the overshoot distance.

It is convenient to express the overshoot distance in fractions of a
scale height, i.e., from Equation~(\ref{dz1}), $\Delta z/H \sim
w/NH$, where $H$ is scale height. Within factors of order unity, $NH$
is the phase speed of fast horizontal gravity waves (those with long
vertical wavelengths).  It can be shown that
\begin{equation}
NH = \sqrt{RT}\left[\kappa - {\partial \ln T\over \partial \ln p}\right]^{1/2}
= \sqrt{R T \kappa}\left[1 - {1\over \kappa}{\partial \ln T\over\partial \ln p}
\right]^{1/2}\equiv \sqrt{\gamma R T \kappa}.
\label{nh}
\end{equation}
Combining Equations~(\ref{dz1}) and (\ref{nh}), one obtains a final
expression for overshoot distance,
\begin{equation}
{\Delta z \over H} \sim {w\over \sqrt{\gamma R T \kappa}}.
\label{dz3}
\end{equation}

From the point of view of the above equations, $w$ is an independently
specified parameter, but we can constrain it using mixing length
theory for convection, which implies
\begin{equation}
w\sim \left({\alpha F g l \over \rho c_p}\right)^{1/3},
\end{equation}
where $\alpha$ is thermal expansivity, $F$ is transported heat flux,
$l$ is mixing length, and $\rho$ is density.  Assuming ideal gas
($\alpha=1/T$ and $\rho=p/RT$) and that the mixing length is a scale
height, this becomes
\begin{equation}
w \sim \left({F R^2 T\over p c_p}\right)^{1/3}
\label{w}
\end{equation}
which, for brown dwarf parameters ($F\sim 10^4$--$10^5\rm\,W\,m^{-2}$,
$R=3700\rm \,J\,kg^{-1}\,K^{-1}$, $T\sim10^3\rm\,K$,
$p\sim10^6\rm\,Pa$ for the RCB, and $c_p=10^4\rm\,J\,kg^{-1}\,K^{-1}$)
yields $w\sim 20$--$50\rm\,m\,s^{-1}$.  For a planet with Jupiter's heat flux
($\sim$$10\rm\,W\,m^{-2}$), we obtain $w\sim1$--$2\rm\,m\,s^{-1}$.

Combining Equations~(\ref{dz3}) and (\ref{w}), we obtain overshoot
distances
\begin{equation}
{\Delta z\over H} \sim \left({F\over R p}\right)^{1/3} \left({c_p\over T}\right)^{1/6}
{1\over \gamma^{1/2}}.
\label{dz4}
\end{equation}

We are now in a position to obtain numerical estimates of the overshoot
distance from Equation~(\ref{dz3}) or (\ref{dz4}).  Adopting
$\gamma=0.1$, $R=3700\rm\,J\,kg^{-1}\,K^{-1}$, $T=1500\rm\,K$, and
$\kappa=2/7=0.2857$ appropriate for a diatomic H$_2$ atmosphere, and
the same range of heat fluxes as used when estimating the
velocities, we obtain ${\Delta z/ H} \sim 0.05-0.12$.  In other words,
mixing length theory suggests an overshoot distance of order 10\% of a
scale height, varying weakly as the heat flux to the one-third power.
For a planet with Jupiter's heat flux, the appropriate value is order
$\Delta z/H\sim 0.01$.

\subsection{Required temperature perturbations in the forced layer}

Our goal is for the parameterization to vertically perturb material
surfaces by amplitudes comparable to that given in
Equation~(\ref{dz4}).  We now address the question: if we accomplish
these displacements by imposing horizontally varying heating in the
bottom $\sim$scale height of the model domain (following
Equation~\ref{thermo}), then what amplitude of horizontal temperature
variations in this ``parameterized forcing'' layer are needed in order
to generate these deflections?

This can straightforwardly be estimated from hydrostatic balance
applied to some local region, i.e.,
\begin{equation}
{\partial\Phi\over\partial \ln p} = -RT,
\end{equation}
and this implies that the increment in height $\delta z$ over some range
of log pressures is given by
\begin{equation}
\delta \Phi = g \delta z \approx - R T \delta \ln p
\end{equation}
where here, $\delta$ is a vertical difference across some finite slab,
such that $\delta\ln p$ is the range of log-pressures across the forced
layer, and $\delta z$ is the height (thickness) of the forced layer.
Now imagine heating or cooling causes a change in temperature of the
gas inside the forcing layer, such that we start with an initial
temperature $T_i$ and end with a final temperature $T_f$.  This will
change the thickness of the forcing layer following $g\delta z_i
\approx - R T_i \delta \ln p$ and $g\delta z_f \approx -R T_f \delta
\ln p$, where $\delta z_i$ and $\delta z_f$ are the initial and final
vertical thicknesses of the forcing layer.  (The
pressure range of the forced layer does not change, so we adopt the
same $\delta \ln p$ for the two states.)  The difference in the
thickness of the initial and final forcing layer is
\begin{equation}
\delta z_f - \delta z_i \approx -{R(T_f - T_i)\over g} \delta \ln p.
\end{equation}
In the limit of zero horizontal divergence, the entire overlying
column will move up or down by this amount, in other words, this is
the amplitude of vertical deflections of material surfaces in the air
above the forced layer.\footnote{In principle there could be subtle
  interactions with the dynamics, in that there will be horizontal
  divergence in regions where the forcing layer expands vertically,
  and horizontal convergence in regions where the forcing layer
  contracts vertically, which might conceivably lessen the actual
  vertical deflections of material surfaces in the atmosphere
  overlying the forcing layer, as compared to the case where we
  essentially suppress dynamics and imagine zero horizontal divergence.
  Still, this limit of zero horizonatal divergence should be roughly
  correct for the case---relevant to our simulations---where the
  horizontal forcing scale is comparable to or larger than the Rossby
  deformation radius, because the rotation will tend to supress the
  ability of the fluid to expand/contract horizontally.}  Thus, we 
expect that the deflections of material surfaces caused by temperature
perturbations $\Delta T_{\rm force}$ that are applied over a vertical
thickness of one scale height ($\delta\ln p=1$) are of order
\begin{equation}
\Delta z  \sim {R \Delta T_{\rm force}\over g}
\label{dz5}
\end{equation}
Using the definition of scale height, this equation is simply
$\Delta z/H \sim \Delta  T_{\rm force}/T$.  Adopting our earlier expression
for vertical deflections, we thus require fractional temperature
differences in the forcing layer of
\begin{equation}
{\Delta T_{\rm force}\over T} \sim {w\over \sqrt{\gamma R T \kappa}},
\end{equation}
which, as estimated previously, should have values of $\sim$0.05--0.12
for a typical range of conditions encountered on brown dwarfs, and as
low as 0.01 for a Jupiter-like planet.  For a typical brown
dwarf temperature of 1500~K above the RCB, and rounding, implies that
we need temperature perturbations in the forcing layer of order
50--200~K, or even lower for a planet with a heat flux comparable
to Jupiter.

\subsection{Required forcing amplitudes}

We now wish to determine the characteristic forcing amplitudes $f_{\rm
  amp}$ that are needed to achieve the desired $\Delta T_{\rm force}$
calculated above.  We expect that because the two terms on the
righthand side of Equation~(\ref{markov}) are uncorrelated, the
forcing will statistically equilibrate its amplitude such that
\begin{equation}
|S_h| \approx |\hat F|
\label{sh}
\end{equation}
where $|\hat F|$ is the characteristic amplitude of $F$, the random
modifier of the forcing (see Equation~\ref{markov}), and $|S_h|$ is
the characteristic amplitude of the forcing $S_h$.  In
Equation~(\ref{random}), the normalized Legendre polynomials each have
characteristic values of order unity, as do the cosines multiplying
them.  A standard result is that adding $N$ one-dimensional sinusoids
of equal wavenumber, amplitude=1, but random phases leads to a
characteristic summed amplitude $\sqrt{N}$.  Analogously, we expect
that adding up $n_f$ modes of equal total wavenumber but random phases
should lead to a characteristic amplitude of order $\sqrt{n_f}$,
implying from Equation~(\ref{random}) that $|\hat F| \sim f_{\rm
  amp}\sqrt{n_f}$.  Thus,
\begin{equation}
|S_h| \approx f_{\rm amp} \sqrt{n_f}.
\label{sh}
\end{equation}

This forcing will cause the characteristic horizontal temperature
contrasts to increase with time in the forcing layer, although not
necessarily linearly with time, since at any given location the sign
of $S_h$ can change from positive to negative over characteristic
timescales $\tau_{\rm for}$, implying a degree of cancellation in how
the forcing integrates in time.  This growth of $\Delta T_{\rm force}$
will be resisted by the radiative damping in the system, and
potentially also by horizontal advection if the horizontal winds in
the forcing layer become sufficiently strong, causing $\Delta T_{\rm
  force}$ to statistically equilibrate.  In the case where radiative
timescales are sufficiently short for the radiation to comprise the
dominant damping process, we could write
\begin{equation}
{\Delta T_{\rm force}\over \tau_{\rm rad}} \sim f_{\rm amp} \sqrt{n_f},
\label{forcing-balance}
\end{equation}
implying that the equilibrated temperature variations in the forcing
layer have amplitude $\Delta T_{\rm force}\sim \tau_{\rm rad}f_{\rm
  amp}\sqrt{n_f}$.  The model analyzed in Section~\ref{diag}, for
example, has $\tau_{\rm rad}=10^6\rm\,s$, $f_{\rm
  amp}=5\times10^{-6}\rm\,K\,s^{-1}$, and $n_f=20$, and the above
equation predicts $\Delta T_{\rm force}\approx 25\rm\,K$, similar to
the actual horizontal temperature contrasts at the bottom of the
forcing layer, which fluctuate between 15--$25\rm\,K$ once the forcing
reaches a statistical equilibrium.  

As a final step, we can rewrite the balance as an expression for the
desired forcing amplitude, which, when the balance
(\ref{forcing-balance}) dominates, yields
\begin{equation}
f_{\rm amp}  \sim {\Delta T_{\rm force} \over \tau_{\rm rad}\sqrt{n_f}}
\sim {\Delta z\over H} {T\over \tau_{\rm rad} \sqrt{n_f}}
\end{equation}
where $\Delta z/H$ could be specified from Equation~(\ref{dz4}), and
in the latter expression, we have used the fact that $\Delta T_{\rm
  force}/T \sim \Delta z/H$.  Plugging in $\Delta z/H\sim 0.1$
appropriate for typical brown dwarfs, $\tau_{\rm rad}\sim10^6\rm\,s$,
$T=1500\rm\,K$ and $n_f=20$ yields $f_{\rm amp}\sim
3\times10^{-5}\rm\,K\,s^{-1}$, similar to our highly forced models.
Adopting $\Delta z/H\sim0.01$ relevant to a low, Jupiter-like heat
flux instead yields $f_{\rm amp}\sim 3\times10^{-6}\rm\,K\,s^{-1}$,
similar to our weaker-forced models, including the model exhibiting a
QBO-like oscillation analyzed in Section~\ref{diag}.  Therefore, the forcing
amplitudes adopted in this paper are appropriate for brown
dwarfs and giant planets.

However, we emphasize that these estimates linking $f_{\rm amp}$ to
objects of a particular heat flux should be taken as only a very rough
guide, since there are many uncertainties, not only in the convective
velocities and vertical thermal profile near the RCB (which controls $\gamma$) but
in the processes that cause the convection to self organize, which
will determine whether the overshoot distance $\Delta z$ relevant on
the convective scale is also relevant on the scale of large-scale
convective organization, as we have implicitly assumed.  Moreover, the
framework of thermal perturbations that we have adopted is only one
approach; in the future, it may also be interesting to investigate
alternate forcing schemes, such as the introduction of momentum
forcing near the RCB, which might parameterize the effect of
convective Reynolds stresses acting at the base of the stratified
atmosphere.

}% end \tt

%%%%%%%%%%%%%%%%%%%%%
% End document body %
%%%%%%%%%%%%%%%%%%%%%

% Acknoledgements section
\acknowledgements APS thanks the NSF (grant AST1313444) and
Peking University for support.

% References
\def\icarus{Icarus}

\bibliographystyle{apj}
\bibliography{showman-bib}

%\begin{thebibliography}{30}

%\end{thebibliography}

\end{document}